%% file: main.tex
\documentclass[a4paper,onecolumn,11pt,accepted=2024-03-25]{quantumarticle}
\pdfoutput=1
\input{packages.tex}

\RequirePackage{etex} 

\usepackage[cal = pxtx, scr = dutchcal]{mathalpha}

\usepackage{xcolor}
\usepackage{multirow}
\usepackage{graphics}
\usepackage{complexity}
\usepackage{xparse}

\usepackage{pdflscape}

\usepackage{index}

\usepackage{nomencl}

 \usepackage{multirow}
 \usepackage{colortbl}

\usepackage[baseline]{euflag}

\usepackage{nameref}
\input{adam.tex}
\usepackage[utf8]{inputenc}

\usepackage{tikz}
\usetikzlibrary{backgrounds,graphs,quotes,positioning}
\usepackage{keycommand}
\usepackage{ifthen}

\usepackage{caption}
\usepackage[boxsize=2em]{ytableau}

\allowdisplaybreaks

\usepackage{autobreak}

\usepackage{scrextend}
\usepackage{float}

\newcommand{\cG}{\mathcal{G}}
\newcommand{\cV}{\mathcal{V}}

\newcommand{\bunderline}[1]{\underline{#1\mkern-4mu}\mkern4mu }

\newcommand{\cB}{\mathcal B}

\def\ben{\begin{enumerate}}
\def\een{\end{enumerate}}

\def\RR{\mathbb{R}}
\def\N{\mathbb{N}}

\def\cM{{\mathcal M}}

\newcommand{\free}[2]{#1\langle#2\rangle}

\def\bs{\bigskip}

\def\sec{\section}
\def\ssec{\subsection}
\def\sssec{\subsubsection}

\def\C{\mathbb{C}}

\def\V{\mathrm{V}}
\def\E{\mathrm{E}}

\def\cH{\mathcal{H}}

\newcommand{\spec}{\mathrm{eigs}}
\newcommand{\eig}{\mathrm{eig}}
\newcommand{\ssymidl}{\tilde{\mathcal{I}}^\mathrm{swap}}
\newcommand{\salg}{\mathcal{A}^\mathrm{swap}}
\newcommand{\ialg}[1]{\mathcal A^{#1}}
\newcommand{\sidl}{\mathcal{I}^\mathrm{swap}}
\newcommand{\cIp}{\mathcal{I}^\textrm{Pauli}}
\newcommand{\ham}[1]{H_{#1}}
\newcommand{\hampol}[1]{h_{#1}}
\newcommand{\hamirrep}[2]{H_{#1}^{#2}}
\newcommand{\irrp}[1]{\mathrm{Irrep}[#1]}

\def\starn{{\bigstar_{n}}}
\newcommand{\stars}[2]{{\bigstar_{#1}^{#2}}}
\newcommand{\hamclint}[1]{{\eta_{#1}}}
\newcommand{\hamclinth}[1]{{\widehat{\eta}_{#1}}}
\newcommand{\clique}[1]{K_{#1}}
\newcommand{\completion}[1]{\clique{}(#1)}
\newcommand{\irrpsym}{\lambda}
\newcommand{\altirrpsym}{\xi}
\newcommand{\rep}[1]{\rho_{#1}}
\newcommand{\chara}[2]{{\chi_{#1}(#2)}}
\nomenclature{$\hamclint{n-k,k}$}{hamclint\{n-k,k\}}
\newcommand{\partn}[2]{\mathrm{Par}(#1,#2)}
\nomenclature{$\partn{k,m}$}{partn\{k,m\}}
\nomenclature{$\spec$}{spec}

\def\exactn{numerically exact\xspace}

\newcommand{\mswap}{Swap Matrix Algebra\xspace}
\newcommand{\sswap}{Symbolic Swap Algebra\xspace}

\newcommand{\sign}[1]{\mathrm{sgn}\left(#1\right)}
\nomenclature{$S^{\lambda}$}{specht\{$\lambda$\}}

\DeclareMathOperator{\sw}{Swap}
\newcommand{\swa}{M^{\mathrm{swap}}}

\DeclareMathOperator{\spn}{span}
\DeclareMathOperator{\SOS}{SOS}
\DeclareMathOperator{\LT}{LT}
\DeclareMathOperator{\LC}{LC}
\DeclareMathOperator{\End}{End}
\DeclareMathOperator{\Hom}{Hom}

\def\symgpn{S_n}
\def\perm{\pi}
\def\nkmin{{n-k,k-1}}
\def\nmink{{n-k-1,k}}
\def\nk{{n-k,k}}
\newcommand{\genlingp}[1]{\GL(#1)}
\newcommand{\genlingpmat}[1]{\GL_{#1}(\mathbb{C})}
\def\qmaxcut{Quantum Max Cut\xspace}
\newcommand{\qmc}{QMC\xspace}
\newcommand{\irrpmod}[1]{V_{#1}}
\newcommand{\irrpmodgl}[1]{L_{#1}}
\DeclareMathOperator{\GL}{GL}
\NewDocumentCommand{\repgl}{o}{%
    \ensuremath{\zeta}
    \IfValueT{#1}{\ensuremath{_#1}}
}

\makeatletter
\def\amsbb{\use@mathgroup \M@U \symAMSb}
\makeatother

\newcommand{\rmax}[1]{#1\mathrm{-max}}
\newcommand{\rmin}[1]{#1\mathrm{-min}}
\newcommand{\maxeig}{\eig_{\mathrm{max}}}
\newcommand{\mineig}{\eig_{\mathrm{min}}}

\makeindex
\makenomenclature

\nomenclature{$\salg$}{salg}
\nomenclature{$\sidl$}{sidl}
\nomenclature{$\cIp$}{cIp}
\nomenclature{$\ham{G}$}{ham\{G\}}
\nomenclature{$\hampol{G}$}{hampol\{G\}}
\nomenclature{$\hamirrep{G}{n-k,k}$}{hamirrep\{G\}\{n-k,k\}}
\nomenclature{$\irrp{\nk}$}{irrp\{n - k,k\}}
\nomenclature{$\nk$}{nk}
\nomenclature{$\nkmin$}{nkmin}
\nomenclature{$\nmink$}{nmink}
\nomenclature{$\bigstar$}{bigstar}
\nomenclature{$\starn$}{starn}
\nomenclature{$\stars{k}{j}$}{stars\{k\}\{j\}}
\nomenclature{$S_n$}{symgpn}
\nomenclature{$\genlingp{V}$}{genlingp\{V\}}
\nomenclature{$\genlingpmat{n}$}{genlingp\{n\}}
\nomenclature{$\clique{n}$}{clique\{n\}}
\nomenclature{$\irrpsym$}{irrpsym}
\nomenclature{$\rep{\irrpsym}$}{rep\{irrpsym\}}
\nomenclature{$\altirrpsym$}{altirrpsym}
\nomenclature{$\sw$}{sw}
\nomenclature{$\swa$}{swa}
\nomenclature{\mswap}{mswap}
\nomenclature{\sswap}{sswap}
\nomenclature{\qmc}{qmc\xspace}
\nomenclature{$\irrpmod{[\nk]}$}{irrpmod{[nk]}}
\nomenclature{$\irrpmod{\lambda}$}{irrpmod{lambda}}
\nomenclature{$\irrpmodgl{\lambda}$}{irrpmodgl{lambda}}
\nomenclature{$\ssymidl$}{ssymidl}

\nomenclature{$\maxeig$}{maxeig}
\nomenclature{$\mineig$}{mineig}

\nomenclature{$\spn$}{spn}

\nomenclature{$\chara{\irrpsym}{x}$}{chara\{irrpsym\}\{x\}}
\nomenclature{$\repgl[n]$}{repgl[n]}

\nomenclature{$\LT$}{LT}
\nomenclature{$\LC$}{LC}
\nomenclature{Quantum Max Cut}{qmaxcut}

\nomenclature{$\ialg{\lambda}$}{ialg\{$\lambda$\}}

\nomenclature{$\SOS$}{SOS}

\begin{document}

\title{Relaxations and Exact Solutions to Quantum Max Cut via the Algebraic Structure of Swap Operators}
\author{Adam Bene Watts}
\affiliation{University of Waterloo} \email{adam.benewatts1@uwaterloo.ca}
\author{Anirban Chowdhury}
\affiliation{University of Waterloo} \email{anirban.chnarayanchowdhury@uwaterloo.ca}
\author{Aidan Epperly}
\affiliation{University of California San Diego} 
\email{aepperly@ucsd.edu}
\author{J. William Helton}
\affiliation{University of California San Diego}
\email{helton@math.ucsd.edu}
\author{Igor Klep}
\affiliation{Faculty of Mathematics and Physics, University of Ljubljana}
\affiliation{Institute of Mathematics, Physics and Mechanics, Ljubljana, Slovenia}
\email{igor.klep@fmf.uni-lj.si}
\maketitle

\begin{abstract}

The \qmaxcut (\qmc) problem has emerged as a test-problem for designing approximation algorithms for local Hamiltonian problems. In this paper we attack this problem using the algebraic structure of \qmc, in particular the relationship between the quantum max cut Hamiltonian and the representation theory of the symmetric group.\looseness=-1

The first major contribution of this paper is an extension of non-commutative Sum of Squares (ncSoS) optimization techniques to give a new hierarchy of relaxations to Quantum Max Cut. 
The hierarchy we present is based on optimizations over polynomials in the qubit swap operators. This is in contrast to the ``standard'' quantum Lasserre Hierarchy, which is based on polynomials expressed in terms of the Pauli matrices. To prove correctness of this hierarchy, we exploit a finite presentation of the algebra generated by the qubit swap operators. This presentation allows for the use of computer algebraic techniques to manipulate and simplify polynomials written in terms of the swap operators, and may be of independent interest. Surprisingly, we find that level-2 of this new hierarchy is \exactn (up to tolerance $10^{-7}$) on all QMC instances with uniform edge weights on graphs with at most 8 vertices. 

The second major contribution of this paper is a polynomial-time algorithm that computes (in exact arithmetic) the maximum eigenvalue of the \qmc Hamiltonian for certain graphs, including graphs that can be ``decomposed'' as a signed combination of cliques. A special case of the latter are complete bipartite graphs with uniform edge-weights, for which exact solutions are known from the work of Lieb and Mattis \cite{lieb1962ordering}. Our methods, which use representation theory of the symmetric group, can be seen as a generalization of the Lieb-Mattis result. 

\end{abstract}

\newpage

\tableofcontents

\sec{Intro}

The local Hamiltonian problem plays a central role in quantum complexity theory. Given as input a $2^n \times 2^n $ Hermitian matrix $H$, the goal of the problem is to compute its maximum (or minimum) eigenvalue. The matrix acts on $n$ qubits and can be concisely specified as a sum of \emph{local} terms, $$H = \sum_{
\substack{
S \subseteq \{1,\ldots,n\}\\[1mm] |S| =k} } H_S,$$ 
where each $H_S$ acts non-trivially on at most a constant $k$ number of qubits. We call such an $H$ a $k$-local Hamiltonian. 

The $k$-local Hamiltonian problem for any $k \geq 2$ is known to be hard for the complexity class \QMA{}
 \cite{kitaev2002classical, kempe2006complexity}, the quantum analog of \NP. Thus, it is unlikely to be efficiently solvable in general. However, specific instances of the local Hamiltonian problem may be easy, in the sense that they admit exact  arithmetic solutions \cite{lieb1962ordering}, or polynomial-time algorithms to approximate the maximum eigenvalue to high precision \cite{landau2015polynomial}. In other cases, it is interesting to study the best approximation to the maximum eigenvalue attainable in polynomial time. A number of works have, for example, provided algorithms to obtain constant-factor approximations to the maximum eigenvalue of local Hamiltonians \cite{gharibian2012approximation,brandao2013product,bravyi2019approximation,Harrow2017extremaleigenvalues}. 
These can be viewed as quantum analogues of  results about approximation algorithms for \NP-hard problems and the hardness of approximation.\looseness=-1

Particularly relevant to us are approaches that  address the local Hamiltonian problem using the tools of semidefinite programming (SDP) relaxations and noncommutative polynomial optimization. Indeed the local Hamiltonian problem can be formulated as a maximization problem, where the goal is to compute $\eig_{\max}(H) = \max_{\rho} \trace\left(\rho H \right)$ subject to $\rho \succeq 0$, $\trace(\rho)=1$.
This defines an SDP involving exponentially large matrices and thus becomes intractable as system size grows. However, one can define a hierarchy of efficiently computable SDP \emph{relaxations} which provides upper bounds to the maximum eigenvalue \cite{brandao2013product,bravyi2019approximation,gharibian2019almost,parekh2021application,hastings2022optimizing}. This is in contrast to variational methods using Ans\"atze 
such as tensor networks which give lower bounds by approximating the maximum-eigenvalue eigenstate from within a subset of the set of quantum states \cite{cirac2021matrixproduct,Bridgeman2017}. SDP relaxations, on the other hand, give \emph{outer} approximations to quantum optimization problems -- their solutions may not correspond to actual quantum states. These relaxations are efficiently computable in the sense that at any fixed level of the hierarchy one can compute the SDP value in polynomial time.

\paragraph*{\qmaxcut} We will focus on the local Hamiltonian problem defined on a specific family of 2-local Hamiltonians called Quantum Max Cut (QMC) Hamiltonians. (We sometimes simply call this problem Quantum Max Cut.) Each \qmaxcut Hamiltonian is defined with respect to a graph $G$ on $n$ vertices with edge set $\E(G)$ and for weights $w_{ij}$ as
\begin{align}
     \ham{G} = \sum_{(i,j)\in \E(G)} w_{ij}\left(I - \sigma^i_{X}\sigma^j_{X} - \sigma^i_{Y} \sigma^j_{Y} - \sigma^i_{Z}\sigma^j_{Z}\right),
\end{align}
where $\sigma_{X}$, $\sigma_{Y}$ and $\sigma_{Z}$ are the Pauli matrices, and $\sigma^i_{W} = \bigotimes_{j=1}^{i-1} I \otimes \sigma_{W} \otimes\bigotimes_{j=i+1}^{n} I $ for $W \in \{X,Y,Z\}$. 
Equivalently, $\ham{G}$ can be written in terms of qubit SWAP operators, $\sw_{ij} = \tfrac{1}{2}\left(I+ \sigma^i_{X}\sigma^j_{X} + \sigma^i_{Y} \sigma^j_{Y} + \sigma^i_{Z}\sigma^j_{Z}\right)$, as
\begin{align}
    \ham{G} = \sum_{(i,j) \in E(G)} 2w_{ij} (I-\sw_{ij}).
\end{align}
\qmaxcut was introduced in the context of approximation algorithms by Gharibian and Parekh \cite{gharibian2019almost} as a quantum generalization of the well-known Max Cut problem. The Hamiltonian $\ham{G}$ has also been studied extensively in physics as the antiferromagnetic quantum Heisenberg model. Determining the maximum eigenvalue of $\ham{G}$ for an arbitrary graph $G$ is known to be the \QMA-hard~\cite{piddock2017complexity}, and the Quantum Max Cut problem has received significant attention as a test-bed for designing approximation algorithms to solve \QMA-hard problems \cite{anshu2020beyond,parekh2021application,parekh2022optimal,king2022improved,lee2022optimizing}. Additionally, exact  arithmetic solutions are known in a few cases, such as for complete bipartite graphs \cite{lieb1962ordering} and one-dimensional chains \cite{levkovich2016bethe}. 

\paragraph*{Max Cut and Sum-of-Squares relaxations}

As suggested by the name, \qmaxcut is closely related to the classical problem of Max Cut where, given a graph $G$ with edge-set $E(G)$ and weights $w_{ij}>0$, the goal is to maximize the objective $\sum_{(i,j)\in \E(G)} w_{ij}\frac{1 - z_i z_j}{2}$ over assignments of the variables $z_i\in \{+1,-1\} \text{ for } i=1,\ldots,n$. 
Determining the optimum value of the Max Cut objective is \NP-hard, but nontrivial upper-bounds on it can be obtained from the Lasserre or Sum-of-Squares (SOS) hierarchy of SDP relaxations. Level $d$ of this hierarchy involves optimizing over ``pseudo-expectations'' -- linear functionals $\mu$ which are defined on polynomials of degree at most $2d$ in the $z_i$ variables and which satisfy $\mu(1) = 1$ and $\mu(f^2) \ge 0$ for any polynomial $f$ with degree at most $d$.\footnote{There is some inconsistency in the literature between $d$ and $2d$ when discussing levels of the SoS hierarchy. Here we use the convention that level $d$ of this hierarchy involves optimizing over linear functionals which are positive on squares of degree $d$ polynomials; note these squares may include terms of degree up to $2d$.} The set of pseudo-expectations contains the set of ``actual'' expectation values which can be obtained from probability distributions assigning values to the variables $z_i$, and converges to this set as $d \rightarrow \infty$. For this reason, the SoS hierarchy gives a converging series of upper bounds on the true value of the  Max Cut problem. Additionally, for any constant $d$, the level $d$ SoS upper bound can be found in polynomial time, though in practice this optimization quickly becomes infeasible as $d$ grows large.\looseness=-1

The best-known approximation algorithm (i.e., constructive lower bound algorithm) for Max Cut is due to Goemans and Williamson and can be viewed as an algorithm which ``rounds'' the pseudo-expectation produced by a level 1 Lasserre relaxation to an actual expectation~\cite{goemans1995improved}. In general, Lasserre relaxations in conjunction with rounding have been powerful techniques for upper bounding and approximating hard constraint satisfaction problems (CSPs) \cite{barak2011rounding}. In fact, assuming the Unique Games Conjecture (UGC) of Khot~\cite{khot2002power}, Raghavendra~\cite{raghavendra2008optimal,raghavendra2009approximating} showed that it is NP-hard to outperform a canonical approximation algorithm based on rounding from the first level of the SoS hierarchy  for any classical constraint satisfaction problem .

\paragraph*{Quantum Lasserre hierarchy}

The success of SDP relaxations for classical CSPs has motivated their use in approximations to local Hamiltonian problems \cite{brandao2013product,bravyi2019approximation,hastings2022optimizing} and in particular \qmaxcut \cite{gharibian2019almost,parekh2021application,parekh2022optimal,king2022improved,kim2018synchronous}. Formulating these quantum relaxations is non-trivial and builds on results in non-commutative polynomial optimization \cite{HM04} and its application to problems in quantum information \cite{doherty2008quantum,navascues2008convergent}.\looseness=-1

The quantum Lasserre relaxation for \qmaxcut \cite{gharibian2019almost,parekh2022optimal} was defined with respect to the $n$-qubit Pauli algebra. A solution to the level-$d$  relaxation assigns pseudo-expectation values to operators corresponding to Pauli polynomials, i.e., polynomials of degree at most $2d$ in the non-commuting variables $\{X_i, Y_i, Z_i\}$ for $i=1,\ldots,n$. In analogy with the classical situation, the pseudo-expectation of the identity is fixed to be 1 and, for any Pauli polynomial $P$ with degree at most $d$, the pseudo-expectation of its square must be non-negative. The relaxation can be viewed as relaxing from operator algebraic states (positive linear functionals) on the $n$-qubit Pauli algebra to pseudo-states that are positive only with respect to squares of degree-$d$ polynomials. 

In contrast to the classical case, the relationship between the quantum Lasserre Hierarchy and approximation algorithms for local Hamiltonian problems is much less well understood. In~\cite{gharibian2019almost}, Gharibian and Parekh give an approximation algorithm for Quantum Max Cut based on a rounding from the first level of the quantum Lasserre hierarchy. Later works then gave algorithms which rounded from the second level of the hierarchy and lead to better approximation ratios \cite{parekh2021application,parekh2022optimal}. In~\cite{lee2022optimizing}, a modified Lasserre relaxation was introduced, which considered only a subset of degree-2 polynomials sufficient to recover the \qmaxcut objective. Assuming the Unique Games Conjecture along with a technical conjecture known as the vector valued Borel's inequality, the authors of~\cite{hwang2022unique} showed the second level of the quantum Lasserre hierarchy (and therefore also approximation algorithms building off the second level) could strictly outperform the first level. 
Indeed, the second level of the Lasserre hierarchy has been shown to be exact for certain graphs such as the unweighted star, and capture physically meaningful properties of quantum states such as monogamy of entanglement inequalities \cite{parekh2021application}. These properties have been used to improve the analysis of rounding algorithms \cite{parekh2021application, parekh2022optimal}.

\subsection{Results}

The first major contribution of this paper is to extend non-commutative Sum of Squares (ncSoS) optimization techniques to give a new hierarchy of relaxations to the problem of computing the maximum eigenvalue of the \qmc Hamiltonian. 
This hierarchy is a non-commutative sum-of-squares hierarchy (ncSoS) which captures the algebraic structure arising from the swap operators $\sw_{ij}$ which define the \qmc Hamiltonian objective. 

A critical ingredient in our construction of the SDP hierarchy is the ability to manipulate polynomials in the swap operators and enforce polynomial identities. Towards this end, we give a fully algebraic charcterization (i.e., finite presentation) of the swap operators. We note that this presentation follows quickly as a special case of results in representation theory, cf.~\cite[§6.1 Theorem]{Procesi}\footnote{\cite[§6.1]{Procesi} applies to swaps on qudits, though here we only work with qubits.} but provide a standalone proof for the sake of completeness.
This characterization is necessary for the use of ``dimension free'' computer algebraic techniques (e.g.  Gr\"obner bases) for manipulating polynomials in swap operators. That is to say, we do not need to use the matrix representations of the swap operators which could be exponentially large. We develop swap algebra theory including replacement rules useful for identifying polynomial identities in the swaps. These results may be of independent interest, beyond their use in constructing higher levels of the SDP hierarchy in swaps.

This new hierarchy of relaxations to \qmaxcut is distinct from, and may sometimes outperform, the usual quantum Lasserre relaxations over Paulis. 
 We show that this hierarchy is exact on an $n$-vertex graph at the level $\lceil \frac{n}{2} \rceil$, no higher levels are needed -- a consequence of the previously discussed identities for simplifying swap polynomials.
For instance, the first level of this hierarchy enforces 
relations among the QMC terms which are not enforced by the first level of the quantum Lasserre relaxation \cite{parekh2021application}. Moreover, tests on small instances indicate that our hierarchy can provide tighter bounds to \qmaxcut compared to the level-1 and level-2 quantum Lasserre relaxations over Paulis. Specifically, we find that level-2 of this new hierarchy 
gives a numerically exact
upper bound on the QMC problem for all $\leq8$ vertex graphs with uniform edge weights. Of course, these small instances can be solved by diagonalization in practice. But our numerical results suggest that the swap hierarchy can give fairly tight upper bounds on the value of the Quantum Max Cut problem.
Thus, it is possible that a better theoretical understanding of the performance of the swap hierarchy can lead to improved approximation guarantees for Quantum Max Cut. We leave such an analysis for future work.  
\looseness=-1

The second major contribution of this paper is a polynomial-time algorithm that computes in exact arithmetic the maximum eigenvalue of the \qmc Hamiltonian for certain graphs, including graphs that can be ``decomposed'' as a signed combination of cliques. A special case of the latter are complete bipartite graphs with uniform edge-weights, for which exact arithmetic solutions are known from the work of Lieb and Mattis \cite{lieb1962ordering}. The Lieb-Mattis solutions use angular momentum algebra to identify invariant subspaces of the \qmc Hamiltonian. Our methods, which use representation theory of the symmetric group, can be seen as a generalization of that of Lieb and Mattis.

In closing, we alert the reader that some care has been taken in this paper to distinguish between various notions of exactness. 
When discussing numerical results we use the phrase \df{\exactn} to mean $10^{-7}$ accuracy.
When discussing theoretical results we distinguish algorithms which are \textit{exact in theory} (meaning they give a provably correct output up to floating point precision) from the stricter notion of algorithms which provide as output an \textit{exact arithmetic}  representation of the solution. 

\paragraph*{Note} While preparing this manuscript, we became aware of contemporaneous work \cite{jun2023hierarchy} which establishes related results. The main commonality is that both papers construct hierarchies of semidefinite programs based on the swap matrices which give convergent series of upper bounds to the maximum eigenvalue of the \qmc Hamiltonian. However, there are significant technical differences between the hierarchies constructed here and in \cite{jun2023hierarchy}, cf.~\cref{app:trz}. More importantly, we focus on developing the algebraic machinery to construct a hierarchy in the swaps, whereas \cite{jun2023hierarchy} focuses on understanding the performance of low levels of SDP hierarchies for \qmaxcut and a number of other many-body physics problems. 
While both papers  in addition touch upon exact solvability of \qmaxcut, the results are quite distinct. We provide exact  arithmetic solutions using representation theory of the symmetric group whereas \cite{jun2023hierarchy} theoretically prove exactness of SDP relaxations for certain graphs. We refer the reader to \cref{app:trz} for a detailed comparison between the two papers.\looseness=-1

\subsection{Overview of Techniques} \label{ssec:overview_techniques}

We prove both of these results by exploiting a connection between the QMC Hamiltonian and the representation theory of the symmetric group. More formally, we use the fact that the QMC Hamiltonian can be written as a sum of swap matrices, which can, in turn, be viewed as representations of the group algebra of the symmetric group. This representation can be understood (in particular, the irreducible representations, or \df{irreps}, appearing in this representation can be characterized) via Schur-Weyl duality, and plays a key role in deriving several important theoretical results in quantum information~\cite{Harrow2013TheCO}.

To construct a hierarchy of semidefinite programming relaxations specific to these swap matrices, we introduce an abstract $*$-algebra, which we call the \df{\sswap}. We give a finite presentation for this algebra, and then, using the characterization of the swap matrices given by Schur-Weyl, we explain that this algebra is isomorphic to the one generated by the swap matrices, thereby establishing a special case of \cite[Theorem, §6.1]{Procesi}.
Informally, this means that the \sswap captures the full behavior of the swap matrix algebra. Finally, we show how standard non-commutative optimization techniques (in particular, the non commutative sum of squares, or ncSoS hierarchy) can be specified to this $*$-algebra to bound the eigenvalues of Quantum Max Cut Hamiltonians.

An important complexity in this process comes from the necessity of finding all linear relationships between polynomials in the symbolic swap algebra (or equivalently, finding a system for simplifying expressions written in terms of swap matrices). 
We solve this problem in a two ways, depending on the degree of the polynomials considered. First, for any integer $n$, we show how to construct a set of linearly independent polynomials which span all polynomials of degree $\leq 4$ in the $n$-qubit symbolic swap algebra. 
This lets us simplify all terms appearing when running the ncSoS hierarchy up to level two in the symbolic swap algebra. To simplify terms appearing in higher levels of the hierarchy, we connect to the theory of non-commutative Gr\"obner bases and show that this lets us find, in principle, systems of rewrite rules for any swap matrix polynomials of fixed finite degree in polynomial time. A byproduct of our simplification theory 
is that to any polynomial $p$ we can associate a 
``simpler'' polynomial $q$ with the same values on swap 
matrices.
\looseness=-1

The polynomial time algorithm for solving  in exact arithmetic some QMC Hamiltonians (and simplifying others) is obtained by building on standard techniques for studying representations of the symmetric group. One important concept we introduce to connect to these techniques is that of the \df{Quantum Max Cut Irrep Hamiltonian} -- defined to be the matrix obtained by replacing the swap matrices appearing the the standard Quantum Max Cut Hamiltonian with the representation of equivalent permutations inside some irrep of the symmetric group. For any irrep, we then show how  an exact arithmetic representation of the eigenvalues of the QMC Irrep Hamiltonian associated with a clique can be found in polynomial time using Schur's Lemma and the Murnaghan-Nakayama rule. (This, combined with the characterization of the swap matrix algebra irreps given by Schur-Weyl, lets us reproduce a well known characterization of eigenvalues of the QMC Hamiltonian associated with a clique.)

Then we consider graphs whose associated QMC Hamiltonians decompose as a signed sum of cliques, and show how the previously discussed  exact arithmetic solutions to the clique irrep Hamiltonians coupled with Young's branching rule can be used to find the $k$ max and min eigenvalues of these Hamiltonians. When graphs do not decompose entirely into a signed sum of cliques, we show similar techniques can be used to bound the max and min eigenvalues of these graphs in terms of the max and min eigenvalues of smaller ``residual" graphs. 

\subsection{Reader's Guide}

In \Cref{sec:QMC_and_swaps_background} we introduce some basic definitions related to the Quantum Max Cut problem, swap matrices, and the representation theory of the symmetric group algebra. We then introduce the concept of Quantum Max Cut Irrep Hamiltonians and give some basic results concerning them, including an exact  arithmetic formula for the  single (integer) eigenvalue of any Quantum Max Cut Irrep Hamiltonian associated with a clique. 

In \Cref{sec:the_symbolic_swap_algebra} we introduce the symbolic swap algebra, give a presentation for it, and show it is isomorphic to the swap matrix algebra. 
We then prove some basic properties about polynomials in this algebra, to include how one produces a ``simple"
polynomial $q$ equivalent to a given polynomial $p$
evaluated on swaps.

In \Cref{sec:swap_ncSoS} we discuss how the non-commutative Sum of Squares (ncSoS) algorithm can be applied to the symbolic swap algebra to produce a semidefinite programming hierarchy analogous to the Quantum Lasserre Hierarchy. We also construct an explicit linear algebra basis for polynomials of degree $\leq 4$ in the symbolic swap algebra. This shows the first two levels of the swap-ncSoS hierarchy can be run in polynomial time. 

Then, in \Cref{sec:GB}, we discuss the theory of Gr\"obner bases and show it is possible to find rewrite rules for polynomials of any fixed degree in the swap variables. This lets us conclude that any finite level of the swap-ncSoS hierarchy can run in, in principle, polynomial time. 

In \Cref{sec:eig_via_clique_decomp} we discuss graphs for which we can compute, in exact arithmetic, any constant number of max and min eigenvalues  associated to the QMC Hamiltonian using representation theoretic techniques. We give an algorithm which identifies these graphs, then we show how to compute their eigenvalues (or bound the eigenvalues of other graphs). Finally, we give some simple examples of these algorithms in practice.

There are 3 online appendices.
\Cref{app:swap3} contains  relationships on Swaps, all used for proofs in the paper.
\Cref{app:GB34} gives some examples and properties of Gr\"obner bases for 
ideals involving Swaps. Finally, in
\Cref{app:trz} 
we compare and contrast this paper with the independent and simultaneously released paper \cite{jun2023hierarchy}.

\subsection*{Acknowledgements}

The authors thank Claudio Procesi 
for bringing 
the notion of a ``good permutation'' to their attention,
and for sharing his expertise in invariant and representation theory. IK thanks  Jurij Vol\v ci\v c for discussions.
ABW and AC thank Ojas Parekh for discussions and in particular pointing out the connection between quantum max cut and swap matrices. AC acknowledges discussions with Kevin Thompson. ABW and AC also thank William Slofstra and David Gosset for helpful discussions. Additionally, the authors wish to thank Brendon Rhoades for discussions and creating an alternative proof of \Cref{thm:swap}.  The authors also thank anonymous referees for helpful and insightful comments.
ABW and AC acknowledge the support of the Natural Sciences and Engineering Research Council of Canada through grant number RGPIN-2019-04198.
IK was supported by the Slovenian Research Agency program P1-0222 and grants 
J1-50002, J1-2453, N1-0217, J1-3004.
IK's work was performed within the project COMPUTE, funded within the QuantERA II Programme that has received funding from the EU's H2020 research and innovation programme under the GA No 101017733 {\normalsize\euflag}.

 \subsection*{Code Availability}

\noindent The code used to obtain the numerical results presented in this paper is available on request. 

\section{Quantum Max Cut and The Swap Matrix Algebra}
\label{sec:QMC_and_swaps_background}

In this section we explore a connection between the the Quantum Max Cut Hamiltonian discussed in the introduction and the representation theory of the symmetric group. We begin with a brief review of representation theory, particularly concerning irreducible representations (irreps) of the symmetric group. Then, we consider the algebra generated by the swap matrices, which we call the \mswap, and which turns out to be a representation of the symmetric group algebra. We note that (as has been pointed out previously \cite{osborne2006statics, parekh2021application}) the \qmc Hamiltonian lies inside this algebra. Connecting these ideas leads to the notation of the \qmc Hamiltonian ``inside'' of an irrep. Finally, we show how combining all these ideas with some standard results in representation theory can be used to give a closed form expression for \textit{all} eigenvalues of the \qmc Hamiltonian on a clique with uniform edge weights.

\subsection{Pauli Matrices and the \qmaxcut Hamiltonian}

Recall the three Pauli matrices,
\beq\label{eq:pauli}
\sigma_X=\begin{pmatrix}0&1\\ 1&0\end{pmatrix},
\quad
\sigma_Y=\begin{pmatrix}0&-i\\ i&0\end{pmatrix},
\quad
\sigma_Z=\begin{pmatrix}1&0\\0 &-1\end{pmatrix}.
\eeq
Together with $\sigma_I:=I$ these form a basis for $M_2(\C)$. They satisfy 
the following relations:
\beq\label{eq:pauliRel}
\sigma_X^2=\sigma_Y^2=\sigma_Z^2=I,\quad
\sigma_X\sigma_Y=i\sigma_Z, \quad
\sigma_Y\sigma_X=-i\sigma_Z. \quad
\eeq

Fix $n\in\N$. For 
$W \in\{I,X,Y,Z\}$ we shall also consider the matrices
\beq\label{eq:pauliTensor}
\sigma^j_{W}:= \underbrace{I \otimes \cdots \otimes I }_{j-1}{} \otimes \sigma_{W} \otimes I \otimes \cdots \otimes I \in M_{2^n}(\C).
\eeq
Observe that
\beq\label{eq:pauliBasis}
\{\sigma^1_{W_1} \sigma^2_{W_2}\cdots \sigma^n_{W_n} \mid W_j \in \{I,X,Y,Z\},\ j  = 1,\dots, n  \}
\eeq
is a basis of $M_{2^n}(\C)$.
Further, given $i\neq j$,
\[
[ \sigma^i_{W_i}, \sigma^j_{W_j}] =0,
\]
where $[\cdot, \cdot]$ denotes the additive commutator
\[
[a,b]:=ab-ba.
\]

Let $\mathbb{G}^n$ denote the set of all graphs on $n$ vertices. 
    Let $\E(G)$ denote its set of edges, and
    let $\V(G)$ denote its set  vertices.
\begin{defn}\label{def:qmaxcut-Paulis}
    Given a graph $G$ with vertex set $\V(G)$, edge set $\E(G)$ and edge weights $\{w_{ij}>0 \mid (i,j)\in \E(G) \} $, the \qmaxcut (QMC) Hamiltonian is defined to be
        \begin{align}\label{eq:qmaxcutHamiltonian}
            \ham{G} = \sum_{(i,j)\in \E(G)} w_{ij}  \left(I - \sigma^i_{X}\sigma^j_{X} - \sigma^i_{Y} \sigma^j_{Y} - \sigma^i_{Z}\sigma^j_{Z}\right) .
        \end{align}
\end{defn}
While the right-hand side of \cref{eq:qmaxcutHamiltonian} clearly depends on the weights $w_{ij}$, we suppress this dependence in the notation $H_G$.

\subsection{Representations of the Symmetric Group} \label{sec:representationssymgpn}

In this section we review standard facts about representations of the symmetric group which will be necessary to prove our results.

\subsubsection{Preliminary definitions}

For any finite-dimensional vector space $V$, we use $\genlingp{V}$ to denote the group of invertible linear transformations from $V$ to itself.
We use $\symgpn$ to denote the symmetric group, the group of permutations of $n$ objects. We will specify elements group using cycle notation, with $e$ denoting the identity element. 
Recall that a representation of $\symgpn$ is a group homomorphism $\rho: \symgpn \rightarrow \genlingp{V}$. 
The vector space $V$ is also referred to as an \df{$\symgpn$-module} or simply a module.

The \df{group algebra} $\C[\symgpn]$ can be defined as an $\symgpn$-module as follows. Promote the elements $\{\perm_1, \perm_2, \dots, \perm_{n!}\} \in \symgpn$ to  basis vectors $\mathbf{\perm_1}, \mathbf{\perm_2}, \dots, \mathbf{\perm_{n!}} $ with the multiplication rule $\mathbf{\perm_i}\mathbf{\perm_j} = \mathbf{\perm_k}$ if $\perm_i \perm_j = \perm_k$. Then $\C[\symgpn]$ is given by
\begin{align}\label{eq:groupalg}
    \C[\symgpn] = \{c_1\mathbf{\perm_1} + c_2\mathbf{\perm_2} + \dots + c_{n!}\mathbf{\perm_{n!}}\mid c_j\in\C\}
\end{align}
where $c_i \in \C$ for all $i$, and $\symgpn$ acts on $\C[\symgpn]$ by left-multiplication.
A representation $(\rho, V)$ of $\symgpn$ also gives rise to a \emph{representation of the algebra} $\C[\symgpn]$ via the homomorphism $\tilde{\rho} : \C[\symgpn] \rightarrow \C[\genlingp{V}]$ defined by its action 
\begin{align}
    \tilde{\rho} \left(\sum_{i=1}^{n!} c_i \mathbf{\perm_i} \right) = \sum_{i=1}^{n!} c_i \rho(\perm_i).
\end{align}
It is common to use $\rho$ to refer to the representation of both the group and its group algebra.

A $\symgpn$-module in general decomposes into a number of \df{$\symgpn$-submodules}. A module $V$ is \df{irreducible} if the only submodules of $V$ are itself and the trivial module $\{0\}$. For a representation $(\rho, V)$ of any finite group, it follows from Maschke's Theorem that there exists a decomposition of $V$ as
\begin{align}\label{eq:isotypicmodule}
    V = \bigoplus_{\irrpsym} \irrpmod{\irrpsym}
\end{align}
where each $\irrpmod{\irrpsym}$ is an irreducible module. 

We will be interested in matrix representations $\rho$ of $\symgpn$ where $V$ is $(\C^2)^{\otimes n}$, which is to say that $\rho(\pi)$ is a $2^n \times 2^n$ matrix for each $\pi\in\symgpn$. 
Maschke's Theorem then implies that there is an invertible matrix $U$ such that 
\begin{align}\label{eq:irrepmatrixrep}
    \rho(\perm) = U\left[\bigoplus_{\irrpsym} \rep{\irrpsym}(\pi) \right]U^{-1}, \quad \perm\in \symgpn,
\end{align} 
where $\rep{\irrpsym}$ are irreducible representations or irreps of $\symgpn$. 
 The decomposition of \cref{eq:isotypicmodule} into irreducible modules can thus be seen as a \df{block-diagonalization} of matrices in $\rho (\C[\symgpn])\subseteq M_{2^n}(\C)$.

\subsubsection{Irreducible representations of \texorpdfstring{$\symgpn$}{Sn}} \label{ssubsec:irrepsympgpn}
Irreducible representations of $\symgpn$ are in one-to-one correspondence with integer partitions of $n$,\looseness=-1
\begin{align}
    \irrpsym = [\irrpsym_1, \irrpsym_2,\dots,\irrpsym_k],~~~~\irrpsym_1 \ge \irrpsym_2 \ge\dots\ge \irrpsym_k > 0, \quad
    \sum_{i=1}^k\irrpsym_i=n
\end{align}
which we will denote as $\irrpsym \vdash n$. 
It is convenient to associate the partition $\irrpsym$ with the \emph{Young diagram} of shape $\irrpsym$, which consists of $k$ rows indexed top to bottom such that the $i$-th row contains $\irrpsym_i$ boxes. As an example, the partition $[3,2]$ corresponds to the shape 
\begin{align}
    \ydiagram{3,2}
\end{align}
Of crucial importance to us later in the paper will be those irreps $\rep{[\nk]}$ that correspond to two row Young diagrams and the respective modules $\irrpmod{[\nk]}$, defined for $k =0, \dots, \lfloor \frac n 2 \rfloor$. 

\subsection{Swap Matrices and Permutations}\label{ssec:swap}
In order to analyze the QMC Hamiltonian, we introduce the \df{swap matrices} $\sw_{ij}$ as permutations on $n$-qubit states. 
\begin{defn}\label{def:swapmatrices}
    The swap matrices $\sw_{ij}$ are defined by their action on tensor products of $\ket{\psi_1},\ket{\psi_2},\dots,\ket{\psi_n} \in \C^2$ as follows:
    \begin{align}\label{eq:swapqubits}
        \sw_{ij} \Big( \ket{\psi_1} \otimes \dots \otimes \ket{\psi_i} \otimes &\dots \otimes \ket{\psi_j} \otimes \dots \otimes \ket{\psi_n} \Big) \nonumber \\ &= \ket{\psi_1} \otimes \dots \otimes \ket{\psi_j}\otimes \dots \otimes \ket{\psi_i} \otimes \dots \otimes \ket{\psi_n}.
    \end{align}
\end{defn}
The connection to Quantum Max Cut is made by noticing that the $\sw_{ij}$ is the following linear combination of the Pauli matrices
\beq\label{eq:swap}
\sw_{ij}= \frac12\Big( I + \sigma^i_X\sigma^j_X+\sigma^i_Y\sigma^j_Y+\sigma^i_Z\sigma^j_Z\Big).
\eeq
\begin{prop}
    The QMC Hamiltonian $\ham{G}$ of \cref{eq:qmaxcutHamiltonian} in terms of the $\sw_{ij}$'s defined in \cref{eq:swapqubits} is given by
    \begin{align}\label{eq:Hamiltonian-swap}
        \ham{G} = \sum_{(i,j)\in \E(G)} 2w_{ij}\left( I - \sw_{ij} \right)
    \end{align}
\end{prop}

Now, observe that we can write the right-hand side of \cref{eq:swapqubits} as
\begin{align}
    \ket{\psi_{(i\ j)^{-1}(1)}} \otimes \dots \otimes \ket{\psi_{(i\ j)^{-1}(i)}}\otimes \ket{\psi_{(i\ j)^{-1}(j)}} \otimes \dots \otimes \ket{\psi_{(i\ j)^{-1}(n)}},
\end{align}
where $(i\ j) \in \symgpn $ is the transposition of objects $i$ and $j$. 
This is in fact a representation of $\symgpn$ on $(\C^2)^{\otimes n}$ defined by
\begin{equation}\label{eq:symgpntensorrep}
    \rep{n}(\pi)\left(\ket{\psi_1} \otimes \dots \otimes \ket{\psi_n}\right) = \ket{\psi_{\pi^{-1}(1)}} \otimes \dots \otimes \ket{\psi_{\pi^{-1}(n)}}
\end{equation}
\begin{defn}
    The algebra $\rho_n(\C[\symgpn])$, for the representation $\left(\rep{n}, (\C^2)^{\otimes n} \right)$ in \cref{eq:symgpntensorrep}, is called the \mswap $\swa_n$.
\end{defn}

\begin{ex}
For small $n$, the swap algebra $\swa_n$ can be explicitly determined. Firstly, since the swap matrices $\sw_{ij}$ 
are defined in terms of a representation of $\symgpn$,
there is a $*$-homomorphism 
\beq\label{eq:canonicalProj}
\C[\symgpn]\to\swa_n, 
\quad
(i\ j)\mapsto \sw_{ij},
\eeq
whence
$\swa_n$ is isomorphic to a semisimple quotient of the group algebra $\C[\symgpn]$ of the symmetric group $\symgpn$. 
Thus representation theory of $\symgpn$ can be used to study $\swa_n$.
For instance,
$S_3$ has two one-dimensional representations (the trivial one and the signature), and one two-dimensional one, hence\looseness=-1 
\[
\C[S_3]\cong \C\oplus\C\oplus M_2(\C).
\]
Since $\dim \swa_3=5$, this immediately yields
\[
\swa_3\cong \C\oplus M_2(\C).
\]
Similarly, classifying irreducible representations for $S_4$ gives
\[
\C[S_4]\cong \C\oplus\C\oplus M_2(\C)\oplus M_3(\C)\oplus M_3(\C),
\]
which
together with $\dim \swa_4=14$ implies that
\[
\swa_4\cong \C\oplus M_2(\C) \oplus M_3(\C).
\]
\end{ex}

The general characterization of the \mswap follows from Schur-Weyl duality 
 \cite{etingof}
of $\symgpn$ and $\genlingpmat{2}$, the group of invertible $2\times 2$ complex matrices. 
The natural representation of $\genlingpmat{2}$ on $ (\C^2)^{\otimes n}$ is defined by the diagonal action of the group elements on tensor products of $\ket{\psi_1},\ket{\psi_2},\dots \ket{\psi_n}\in \C^2$,
\begin{align}\label{eq:tensorrepGL2C}
    \repgl[n](g)\left(\ket{\psi_1}\otimes \ket{\psi_2} \otimes \dots \otimes \ket{\psi_n} \right) = g\ket{\psi_1}\otimes g\ket{\psi_2} \otimes \dots \otimes g\ket{\psi_n}, \quad g \in \genlingpmat{2}.
\end{align}
The irreducible modules of $\genlingpmat{2}$ are indexed by two row Young diagrams with an unbounded number of boxes. We denote the former by $\irrpmodgl{[\nk]}$ for $n \in \mathbb{N}$ and $k =0, \dots, \lfloor \frac n 2 \rfloor$. 
In fact,
$\irrpmodgl{[\nk]}$ is the space of all linear maps from $\irrpmod{[\nk]}$ to $(\C^2)^{\otimes n}$ that commute with the action of $\symgpn$, that is
\begin{align}
\irrpmodgl{[\nk]} = \Hom_{\symgpn}(\irrpmod{[\nk]},(\C^2)^{\otimes n}).
\end{align}
The following lemma is essentially a restatement of Schur-Weyl duality for $\symgpn$ and $\genlingpmat{2}$.
\begin{lem}\label{lem:schur-weyl}
    The algebras $\swa_n$ and $\repgl[n]\left(\C[\genlingpmat{2}]\right)$ are centralizers of each other inside $\End((\C^2)^{\otimes n})=M_{2^n}(\C)$.
    Moreover, the space $(\C^2)^{\otimes n}$ decomposes under the action of the direct product $ \genlingpmat{2}\times \symgpn $ as\looseness=-1
    \begin{align}\label{eq:schur-weyl}
        (\C^2)^{\otimes n}\cong \bigoplus_{k=0}^{\lfloor \frac n 2 \rfloor} \irrpmodgl{[\nk]} \otimes \irrpmod{[\nk]}.
    \end{align}
\end{lem}

\begin{proof}
     See, e.g., \cite[Sec.~5.19]{etingof} or \cite[§6.1]{Procesi}.
\end{proof}

Since the action of $\symgpn$ on $\irrpmodgl{[\nk]}$ is trivial,
as an $\symgpn$-module, the space $(\C^2)^{\otimes n}$
decomposes 
by \Cref{lem:schur-weyl}
into irreducible modules 
$\irrpmod{[\nk]}$ with multiplicities
as follows:
\begin{align}\label{eq:schurW}
(\C^2)^{\otimes n} = \bigoplus_{k=0}^{\lfloor \frac n 2 \rfloor} 
(\irrpmod{[\nk]})^{\dim(\irrpmodgl{[\nk]})}.
\end{align}
The Weyl character formula \cite[Theorem 5.22.1]{etingof}
gives an explicit formula for $\dim(\irrpmodgl{[\nk]})$.

Further,
\Cref{lem:schur-weyl} immediately leads to the following characterization of $\swa_n$ in terms of the irreducible representations of $\C[\symgpn]$; cf.~\cite[(1.12)]{Procesi2021}.

\begin{thm} \label{thm:mswap_decomp}
    The \mswap decomposes into the direct sum 
    of simple algebras generated by the two row irreps of the symmetric group. That is, we have 
    \begin{align}
        \swa_n \cong \bigoplus_{k=0}^{\lfloor \frac n2\rfloor}
        \rep{[\nk]}(\mathbb{C}[S_n]).
    \end{align}
\end{thm}
\begin{proof}
This is immediate from 
\Cref{lem:schur-weyl} and \cref{eq:schurW}.
\end{proof}

\ssec{Quantum Max Cut and Irreps}
\label{sec:QMC_and_irreps}

Now we discuss how the decomposition of the \mswap into irreps described in the previous section can be applied to calculations related to the eigenvalues of Quantum Max Cut Hamiltonians. 
Among other benefits the calculations in this section
are essential to \Cref{sec:eig_via_clique_decomp}.

\begin{defn}
Let $\irrpsym\vdash n$ be any partition labeling an irrep of $S_n$, and let $G$ be an $n$ vertex graph with edge set $\E(G)$ and edge weights $w_{ij}$. Then define the \df{\qmc irrep Hamiltonian} $\hamirrep{G}{\lambda}$ by\looseness=-1
\begin{align}
\hamirrep{G}{\lambda} = \rep{\irrpsym}\left(\sum_{(i,j)\in \E(G)} 2w_{ij}\left( I - (i\ j) \right)\right)
\end{align}
\end{defn}

We will frequently use the phrase ``all two row irrep Hamiltonians of $G$'' to refer to the set of irrep Hamiltonians 
\begin{align}
   \{ \hamirrep{G}{[\nk]} : 1 \leq k \leq \lfloor n / 2 \rfloor \}. 
\end{align}
Now we state a straightforward corollary of \Cref{thm:mswap_decomp} which makes the significance of the two row irrep Hamiltonians clear.  
\begin{cor} \label{cor:spec_decomp}
    The spectrum of the \qmc Hamiltonian of $G$ is given by the union of the spectra of all two row irrep Hamiltonians of $G$. That is 
    \begin{align}
        \spec(\ham{G}) = \bigcup_{k=0}^{\lfloor \frac n2\rfloor}
        \spec(\hamirrep{G}{[\nk]})
    \end{align}
    In particular, we have 
    \begin{align}
        \maxeig(\ham{G}) = \max_{k=0}^{\lfloor \frac n2\rfloor}\left(
        \maxeig(\hamirrep{G}{[\nk]})\right)
    \end{align}
\end{cor}

\begin{proof}
    Immediate from \Cref{thm:mswap_decomp}. 
\end{proof}

\Cref{cor:spec_decomp} suggests an immediate technique for computing the max eigenvalue of a \qmc Hamiltonian -- rather than computing the maximum eigenvalue of the matrix $\ham{G}$ directly we can instead compute the maximum eigenvalue of each of the $\hamirrep{G}{[\nk]}$ matrices. When combined with brute force computation this yields a modest computational advantage, since the dimension of each $\hamirrep{G}{[\nk]}$ matrix is smaller than that of $\ham{G}$.\footnote{ This approach also requires computing  the form of the $\hamirrep{G}{[\nk]}$ matrix in some basis. This can be done, for example, by explicitly computing matrix representations of transpositions in the $[\nk]$ irrep via their action on the Specht basis (see~\cite{james2006representation} for details), then summing these matrices to obtain $\hamirrep{G}{[\nk]}$. While involved, all these operations have runtime polynomial in the dimension of the $[\nk]$ irrep. In practice, these computations can be also be performed using a convenient software package, for example \cite{groupmath}, which was the approach used in this paper.  }
In the next lemma we calculate these dimensions.
It should be pointed out that the dimension of these irreps still scales exponentially with $n$, meaning this approach does not give a polynomial time algorithm for computing the max eigenvalue of $\ham{G}$. 
In \Cref{sec:irrepMax} we give the results of some simple experiments regarding which irreps provide the maximizing eigenvalue for \qmc Hamiltonians. 

We calculate the dimension of each irrep by computing the character (or trace) of the identity element in the irrep. When working with the symmetric group, we will use $e \in S_n$ denote the identity and $\chi_{[\nk]} : \symgpn \to \C$ denote the character of $\rho_{[\nk]}$, so $\chara{[\nk]} \pi := \Tr (\rho_{[\nk]}(\pi))$ for all $\pi \in \symgpn$. Then, in particular,
\begin{align}
\chara{[n-k,k]} e = \Tr (\rho_{[\nk]} (e) )
\end{align}
is the dimension of the irrep $\rho_{[\nk]}$ of $\symgpn$.

\begin{lem} \label{lem:irrep_dim}
\begin{align}
    \chara{[n-k,k]}e=\frac{n-2k+1}{n-k+1}\frac{n_{[k]}}{k!}= \frac{n-2k+1}{n-k+1} {n \choose k} .
\end{align}
Here 
we used the falling factorial notation $n_{[k]}$, defined by 
\begin{align}\label{eq:ffrac}
    n_{[k]} := \frac{n!}{(n-k)!} = n(n-1)\cdots(n-k+1),
\end{align}
and $n_{[0]} = 0_{[k]} = 1$ by definition. 
\end{lem}

\begin{proof}
The dimension of the irrep of the partition $[\nk]$ is equal to the trace, or character, of the identity element in this irrep, that is, $\chara{[\nk]}{e}$. We compute this quantity using the well-known hook length formula. Straightforward inspection gives that the product of the hook lengths of this irrep is given by: 
\begin{align}
    \prod_{ij}\mathrm{hook}_\irrpsym(i,j) = (n-k+1)^{[k]}(n-2k)!k! = \frac{(n-k+1)!k!}{(n-2k+1)}.
\end{align}
Then the dimension of the $[\nk]$ irrep is given by 
\[
    \chara{[\nk]}{e} = \frac{n!}{\prod_{ij}\mathrm{hook}_\irrpsym(i,j)} = \frac{n-2k+1}{n-k+1} \frac{n_{[k]}}{k!} = \frac{n-2k+1}{n-k+1} \binom{n}{k}.\qedhere
\]
\end{proof}

From \Cref{lem:irrep_dim} we see that, for any constant $k$, the irreps $\rep{[\nk]}$  of the symmetric group $\symgpn$ have dimension polynomial in $k$. Then the maximum eigenvalue (and indeed, entire spectrum) of the $\hamirrep{G}{[\nk]}$ matrices can be computed in time polynomial in $n$ for constant $k$.

\sssec{Maximizing irreps for 7 and 8 vertex graph Hamiltonians}
\label{sec:irrepMax}

The following tables give the results of some numerical experiments comparing the maximum eigenvalue of irrep Hamiltonians and the maximum eigenvalues of the \qmc Hamiltonian for small graphs. In \Cref{tab:irrepMax7} we focus on 7 vertex graphs, $n=7$, and the associated irreps of $S_7$.
In \Cref{tab:irrepMax8} we focus on 8 vertex graphs,
$n=8$, and the associated irreps of $S_8$.

The third column of both of these tables gives the number of graphs for which the maximum eigenvalue of the \qmc Hamiltonian $\ham{G}$ is equal to the maximum eigenvalue of the irrep Hamiltonian $\hamirrep{G}{\irrpsym}$ for the appropriate irrep $\irrpsym$. The fourth column gives the minimum over all graphs of the ratio between the max eigenvalue of the irrep Hamiltonian and the max eigenvalue of the \qmc Hamiltonian, that is it is equal to 
$$
\min_{\parbox{4em}  {\centering \tiny
n-vertex connected graphs~G}} 
\frac
{\maxeig{\hamirrep{G}{\irrpsym}}}{\maxeig{\ham{G}}}. 
$$
\begin{table}[ht]
    \centering
    \begin{tabular}{|c|c|c|c|c|}
        \hline
        Irrep & Irrep Dim 
        & \# Irrep Max Eig=Swap Max Eig & Irrep Max Eig / Swap Max Eig \\
        \hline
        $[6, 1]$ & 6    &    1      & 0.413 \\
        $[5, 2]$ & 14   &    32     & 0.766 \\
        $[4, 3]$ & 14   &    824    & 0.714 \\
        \hline
    \end{tabular}
    \caption{Summary data for all 853 connected graphs 
    on $7$ vertices. }
    \label{tab:irrepMax7}
\end{table}

\begin{table}[ht]
    \centering
    \begin{tabular}{|c|c|c|c|c|}
        \hline
        Irrep & Irrep Dim 
        & \# Irrep Max Eig=Swap Max Eig & Irrep Max Eig / Swap Max Eig \\
        \hline
        $[7, 1]$ & 7      &      1      & 0.354 \\
        $[6, 2]$ & 20      &    45     & 0.665 \\
        $[5, 3]$ & 28   &    1445    & 0.75 \\
        $[4, 4]$ & 14 &    9114      & 0.625 \\
        \hline
    \end{tabular}
    \caption{Summary data for all 11,117 connected graphs 
    on $8$ vertices. 
}
    \label{tab:irrepMax8}
\end{table}

One might wonder why the sum in column 3
does not equal the total number of graphs.
This for two reasons. One: two different irreps 
can have the same maximum eigenvalue.
Two: equality of eigenvalues was checked 
inexactly (to a tolerance of $10^{-13}$).

\bs

\ssec{Exact Arithmetic Solution for Clique Hamiltonians}
\label{ssec:exact_arithmetic_solns_for_cliques}
Now we consider the \qmc Hamiltonian $\ham{\clique{n}}$ corresponding to the clique $\clique{n}$ on $n$ vertices. We will show that it is possible to compute the maximum eigenvalue (and indeed -- entire spectrum) of this Hamiltonian efficiently by using the irreducible representations of \mswap. Note that this fact is known in the physics literature, see, e.g., \cite[Lemma 36]{cubitt2018universal} and also \cite{osborne2006statics}.

Our proof proceeds in two parts: in \Cref{lem:clique_ham_identity} we show the corresponding \qmc Irrep Hamiltonians $\hamirrep{\clique{n}}{\irrpsym}$ are constant multiples of the identity for each irrep $\irrpsym$, that is 
\begin{align}
    \hamirrep{\clique{n}}{\irrpsym}= \hamclint{\irrpsym} I.
\end{align} 
Then, in  \Cref{lem:irrpsym_value} we compute the value of these constants $\hamclint{\irrpsym}$. 

\begin{lem} \label{lem:clique_ham_identity}
    For any integer $n$ and irrep $\irrpsym$ of $\symgpn$, we have 
    \begin{align}\label{eq:etalambda}
    \hamirrep{\clique{n}}{\irrpsym} = \hamclint{\irrpsym} I
    \end{align}
    where $\hamclint{\irrpsym}$ is some scalar depending only on the irrep $\irrpsym$ and $I$ is an identity matrix of the appropriate dimension. 
\end{lem}

\begin{proof}

We begin by showing something stronger: let 
\begin{align}
q_{\clique{n}} = \sum_{(i,j) \in \E(\clique{n})} (i\ j)\in\C[\symgpn]
\end{align}
denote the sum over all transpositions in $\symgpn$. Then, for any permutation $\pi \in \symgpn$, we have 
\begin{align}
    \pi q_{\clique{n}} \pi^{-1} = \pi \sum_{i,j = 1,\dots, n } (i\ j) \pi^{-1} = \sum_{i,j = 1,\dots, n } (\pi(i)\ \pi(j)) = q_{\clique{n}}
\end{align}
from which it follows that the element $q_{\clique{n}}$ is central in the group algebra $\C[\symgpn]$. In particular, for any irrep $\lambda = [\nk]$, we see that $\rep{\lambda} (q_{\clique{n}})$ is central in the irreducible algebra of matrices $\rep{\lambda}(\C[\symgpn])$ acting on the vector space $V_\lambda$ as defined in \cref{eq:isotypicmodule}.

Similarly for any such $\rep{\lambda}$ we see $\rep{\lambda}(2\binom{n}{2}e - 2q_{\clique{n}}) = \hamirrep{\clique{n}}{\irrpsym}$ is an $\symgpn$-linear map from the irreducible module $V_\lambda$ to itself and hence, by Schur's lemma, is a scalar multiple of the identity matrix. This completes the proof. 
\end{proof}

With a little bit of extra work, we can also compute the scalar multiple $\hamclint{\irrpsym}$ from \cref{eq:etalambda} associated with Hamiltonian of the clique in each irrep. We do this next.

\begin{lem}\label{lem:irrpsym_value}
Let $\hamclint{\irrpsym}$ be as in \Cref{lem:clique_ham_identity}. Then, for any integer $n$ and  irrep $\rep{[\nk]}$ we have
\begin{align}\label{eq:traceHKn}
    \hamclint{[\nk]} = 2 k (n + 1) - 2 k^2
\end{align}
\end{lem}

\begin{proof}
We can find the value of $\hamclint{\irrpsym}$ by computing two quantities: the dimension of the $\lambda$ irrep (i.e., the trace of $\rep{\irrpsym}(e)$) and the trace of $ \hamirrep{\clique{n}}{\irrpsym}$. It is possible to compute both of these quantities with the Frobenius trace formula, but we shall save some effort by instead using the hook length formula and Murnaghan-Nakayama rule. To deal with the many partial factorials appearing in this calculation, recall the falling factorial notation $n_{[k]}$
from \cref{eq:ffrac}. We shall also use the dimension of the $[\nk]$ irrep given in \Cref{lem:irrep_dim}.
For ease of future use we also write this formula as
\begin{align}
    \chara{[s,t]}{e} = \frac{s-t+1}{s+1} \binom{s+t}{t} = \frac{(s-t+1)(s+t)_{[t-1]}}{t!}. \label{eq:id_chara}
\end{align}

Next we compute the trace of $ \hamirrep{\clique{n}}{\irrpsym}$. We can do this by computing the character $\chara{\irrpsym}{(i\ j)}$ of a single transposition $(i\ j)$ in the $\lambda$ irrep. Since all transpositions belong to the same conjugacy class, they will also have the same character. We now compute this character. The Murnaghan-Nakayama rule relates the character of a permutation in some irrep to the character of smaller permutations in irreps of the symmetric group on fewer elements. In the special case considered here, it gives 
\begin{align}
    \chara{\irrpsym}{(i\ j)} = \sum_{\xi} (-1)^{h(\altirrpsym)} \chara{\irrpsym / \altirrpsym}{e} \label{eq:MN-Rule}
\end{align}
where the sum runs over all ways of removing two adjacent boxes from the irrep $\irrpsym$ while leaving a valid Young diagram; $h(\altirrpsym)$ is one if the boxes removed are stacked vertically, zero otherwise; and $\irrpsym / \altirrpsym$ is the resulting partition when the boxes are removed from $\irrpsym$. More explicitly, we write $\irrpsym =[{\nk}]$ 
and then, for any $n > 3$: 
\begin{align}
\chara{[\nk]}{(i\ j)} = \begin{cases}
\chara{[n-k-2,k]}{e} &\text{if } k = 1 \\
\chara{[n-k-2,k]}{e} + \chara{[n-k,k-2]}{e} &\text{if } 2 \leq k \leq n/2-1 \\
\chara{[n-k,k-2]}{e} &\text{if } k = (n-1)/2 \\
\chara{[n-k,k-2]}{e} - \chara{[n-k-1,k-1]}{e} &\text{if } k = n/2. \label{eq:transposition_characters_MN}
\end{cases}
\end{align}

By \cref{eq:Hamiltonian-swap} the Hamiltonian for a clique $K_n$ is 
\begin{align}\label{eq:Hamiltonian-swap2}
    \ham{K_n} &=  2 \sum_{(i,j)\in \E(K_n)} I - 2 \sum_{(i,j)\in \E(K_n)} \sw_{ij}  \\ 
    &= 2  \binom{n}{2}    I 
    - 2 \sum_{(i,j)\in \E(K_n)} 
    \rho_{[n-k,k]}((i \ j )),
\end{align}
and taking the trace of this expression gives
\begin{align}
\Tr[\hamirrep{\clique{n}}{[\nk]}]
&= 
       2    \binom{n}{2}   \chara{[n-k,k]}{e}
        - 2 \sum_{(i,j)\in \E(G)} \chara{[n-k,k]}{(i \ j )} \\
        &= 2 \binom{n}{2} \big(\chara{[\nk]}{e} - \chara{[\nk]}{(1 \ 2)} \big).
\end{align}
But we also have 
\begin{align}
\Tr[\hamirrep{\clique{n}}{[\nk]}] = \Tr[\hamclint{[\nk]} I] = \hamclint{[\nk]} \chara{[\nk]}{e}
\end{align}
by definition of the trace.
Putting these together gives
\begin{align}
\hamclint{[\nk]} = \frac{2 \binom{n}{2} (\chara{[\nk]}{e} - \chara{[\nk]}{(1 \ 2)})}{\chara{[\nk]}{e}} = 2\binom{n}{2} - 2\hamclinth{[\nk]}
\label{eq:hamclint_characters}
\end{align}
where 
\begin{align}
    \hamclinth{[\nk]} := \frac{\sum_{(i,j)\in \E(K_n)} 
    \chara{[n-k,k]}{(i \ j )}}{\chara{[\nk]}{e}} 
    = \binom{n}{2}\frac{\chara{[\nk]}{(1 \ 2)}}{\chara{[\nk]}{e}} .
\end{align}

Finally, combining \cref{eq:hamclint_characters,eq:id_chara,eq:transposition_characters_MN} gives a closed form expression for $\hamclint{[\nk]}$. We do that below and simplify, assuming that $n > 5$ and treating several cases separately.\\

\noindent Case ${k = 1}$:
\begin{align*}
\hamclinth{[\nk]} &= \binom{n}{2} \chara{[n-k-2,k]}{e} / \chara{[\nk]}{e} \\
&= \left(\frac{n(n-1)}{2}\right) \left(\frac{(n - 2k - 1)(n-2)_{[k-1]}}{k!}\right)\left(\frac{k!}{(n-2k+1)n_{[k-1]}}\right) \\
&= \frac{n(n - 3)}{2} 
\end{align*}

\noindent Case ${k = 2}$:
\begin{align*}
\hamclinth{[\nk]} &= \binom{n}{2} (\chara{[n-k-2,k]}{e} +\chara{[n-k,k-2]}{e}) / \chara{[\nk]}{e} \\
&= \left(\frac{n(n-1)}{2}\right) \left(\frac{(n - 2k - 1)(n-2)_{[k-1]}}{k!}\right)\left(\frac{k!}{(n-2k+1)n_{[k-1]}}\right) \\
&\hspace{10pt}+ \left(\frac{n(n-1)}{2}\right) \left(\frac{k!}{(n-2k+1)(n)^{[k-1]}}\right) \\ 
&= \frac{(n-1)(n-5)(n-2)}{2(n-3)} + \frac{(n-1)}{(n-3)}\\
&= \frac{n(n-5)}{2} + 2
\end{align*}

\noindent Case ${3 \leq k \leq (n-1)/2}$:
\begin{align*}
\hamclinth{[\nk]} &= \binom{n}{2} (\chara{[n-k-2,k]}{e} +\chara{[n-k,k-2]}{e}) / \chara{[\nk]}{e} \\
&= \left(\frac{n(n-1)}{2}\right) \left(\frac{(n - 2k - 1)(n-2)_{[k-1]}}{k!}\right)\left(\frac{k!}{(n-2k+1)(n)^{[k-1]}}\right) \\
&\hspace{10pt}+ \left(\frac{n(n-1)}{2}\right) \left(\frac{(n - 2k + 3)(n-2)^{[k-3]}}{(k-2)!}\right)\left(\frac{k!}{(n-2k+1)(n)^{[k-1]}}\right) \\ 
&= \frac{(n - 2k - 1)(n-k+1)(n-k)}{2(n-2k+1)} + \frac{(n - 2k + 3)k(k-1)}{2(n - 2k + 1)} \\
&= \frac{n(n-1)}{2} + k^2 - k(n+1)
\end{align*}

\noindent Case ${k = (n-1)/2}$:
\begin{align*}
\hamclinth{[\nk]} &= \binom{n}{2} \chara{[n-k,k-2]}{e} / \chara{[\nk]}{e} \\
&= \left(\frac{n(n-1)}{2}\right) \left(\frac{(n - 2k + 3)(n-2)^{[k-3]}}{(k-2)!}\right)\left(\frac{k!}{(n-2k+1)(n)^{[k-1]}}\right) \\ 
&= \frac{(n - 2k + 3)k(k-1)}{2(n - 2k + 1)} \\
&= \frac{(n-1)(n-3)}{4}
\end{align*}

\noindent Case ${k = n/2}$:
\begin{align*}
\hamclinth{[\nk]} &= \binom{n}{2} (\chara{[n-k,k-2]}{e} - \chara{[n-k-1,k-1]}{e})/ \chara{[\nk]}{e} \\
&= \left(\frac{n(n-1)}{2}\right) \left(\frac{(n - 2k + 3)(n-2)_{[k-3]}}{(k-2)!}\right)\left(\frac{k!}{(n-2k+1)n_{[k-1]}}\right) \\
&\hspace{20pt} - \left(\frac{n(n-1)}{2}\right) \left(\frac{(n - 2k + 1)(n-2)_{[k-2]}}{(k-1)!}\right)\left(\frac{k!}{(n-2k+1)n_{[k-1]}}\right) \\
&= \frac{(n - 2k + 3)k(k-1)}{2(n - 2k + 1)} - \frac{(n-k+1)k}{2} \\
&= \frac{3n(n-2)}{8} - \frac{n(n+2)}{8} \\
&= \frac{n(n-4)}{4}
\end{align*}

Straightforward calculation then shows that all of these cases are consistent with the  formula 
\begin{align}
    \hamclinth{[\nk]} = \binom{n}{2} + k^2 - k(n+1) \label{eq:eta_hat_value}
\end{align} 
from the $3 \leq k \leq (n-1)/2$ case. Hence $\hamclint{[\nk]} = 2 k(n+1) - 2k^2$ and we are done. 
\end{proof}

\Cref{lem:irrpsym_value} gives us a straightforward way to compute the max eigenvalue (and in fact, \textit{all} eigenvalues) of the Hamiltonian $\ham{\clique{n}}$.  In \Cref{subsec:Star_Graph} we show how, with a bit more work, we can also use this lemma to compute the eigenvalues of graphs other than the complete graph.\looseness=-1

\subsection{Identities Satisfied by Swap Matrices}
\label{subsec:swap_matrix_identities}

A key tool that will be developed in the subsequent section and then used in the remainder of this paper is a fully algebraic characterization of the swap matrices. In preparation for this, we observe some simple algebraic facts about the swap matrices. 

Since the transpositions $(i\ j)$ are generators of $\symgpn$, it follows that $\swa_n$ is generated by the swap matrices $\sw_{ij}$. 
The latter must satisfy the relations
\begin{subequations}\label{eq:Gp}
\begin{align}
    \sw_{ij}^2 & =I,\label{eq:Gp1} \\
    \sw_{ij}\sw_{jk} & = \sw_{ik}\sw_{ij},\label{eq:TrianglePair} \\ \sw_{ij}\sw_{kl} & = \sw_{kl}\sw_{ij},\label{eq:CommuteReln} 
\end{align}
\end{subequations}
for all $i,j,k,l$ distinct, arising from the relations satisfied by the transpositions $(i\ j)$. One can verify using \cref{eq:swap} that, in addition to the above relations, the swap matrices satisfy
\begin{align} 
    \sw_{ij} \sw_{jk} + \sw_{jk} \sw_{ij} &= \sw_{ij}+\sw_{jk}+\sw_{ik} - I,\label{eq:Btri}\tag{$\triangle$}
\end{align}
which is in general not true for transpositions $(i\ j),(j\ k)\in\C[\symgpn]$. For instance, under the signature representation $\pi\mapsto \sign{\pi}$ we have 
$\sign{(i\ j)}\sign{(j\ k)}+\sign{(j\ k)}\sign{(i\ j)} = 2$ and $\sign{(i\ j)}+\sign{(j\ k)}+\sign{(i\ k)} - 1 = -4$.

\begin{rmk}\label{rem:redundancy}
The commutation relations \cref{eq:CommuteReln} 
are (somewhat surprisingly) implied by 
\cref{eq:Gp1,eq:TrianglePair,eq:Btri}. To prove this, note \cref{eq:app:commute} below contains a list of
19 triples $(f_r,g_r,\ell_r)$ of noncommutative polynomials in the variables $s_{ij}$, $1\leq i,j\leq 4$ such that
\beq\label{eq:commuteAPP}
[s_{12},s_{34}]=\sum_{r=1}^{19} f_r\, g_r\, \ell_r.
\eeq
Further, when the variables $s_{ij}$ are specialized to the matrices $\sw_{ij}$, each $g_k(\sw)$ becomes
an expression from either \cref{eq:Gp} or
\cref{eq:Btri}. Thus
\beq\label{eq:precommuteREDUNDANT}
[\sw_{12},\sw_{34}]=\sum_k f_k(\sw)\, g_k(\sw)\, \ell_k(\sw) =0 .
\eeq
Of course, by reindexing indices in \cref{eq:precommuteREDUNDANT} we then obtain \cref{eq:CommuteReln}.
{\small
\begin{align}\label{eq:app:commute}
\begin{autobreak}
\phantom{=}
(1,\  -s_{13} s_{23} + s_{23} s_{12},\  1),\;  (-s_{24},\  -s_{13} s_{23} + s_{23} s_{12},\  1),\;  
 (1,\  -s_{12} s_{13} + s_{13} s_{23},\  1),\;  (-s_{24},\  -s_{12} s_{13} + s_{13} s_{23},\  1),\;  
 (-1,\  s_{23} s_{34} - s_{24} s_{23},\  s_{12}),\;  (-1,\  -s_{24} s_{34} + s_{34} s_{23},\  s_{12}),\;  
 (-1,\  -s_{23} s_{24} + s_{24} s_{34},\  s_{12}),\;  (1,\  1 - s_{23} - s_{24} - s_{34} + s_{23} s_{34} + s_{34} s_{23},\  s_{12}),\;  
 (1,\  -s_{14} s_{24} + s_{24} s_{12},\  1),\;  (-s_{23},\  -s_{14} s_{24} + s_{24} s_{12},\  1),\;  
 (1,\  -s_{12} s_{14} + s_{14} s_{24},\  1),\;  (-s_{23},\  -s_{12} s_{14} + s_{14} s_{24},\  1),\;  
 (-1,\  -s_{14} s_{24} + s_{24} s_{12},\  s_{13}),\;  (-1,\  -s_{12} s_{14} + s_{14} s_{24},\  s_{13}),\;  
 (-s_{12},\  s_{14} s_{13} - s_{34} s_{14},\  1),\;  (-s_{12},\  1 - s_{13} - s_{14} - s_{34} + s_{14} s_{34} + s_{34} s_{14},\  1),\;  
 (s_{12},\  -s_{13} s_{14} + s_{14} s_{34},\  1),\;  (-1,\  -s_{13} s_{23} + s_{23} s_{12},\  s_{14}),\;  
 (-1,\  -s_{12} s_{13} + s_{13} s_{23},\  s_{14})
\end{autobreak}
\end{align}
} \qed
\end{rmk}

\begin{rmk}
There are several sets of relations that could be used to define the symmetric group. 
One option is to define this group as being generated by the transpositions $(i \ j)$ satisfying relations \Cref{eq:Gp1,eq:TrianglePair,eq:CommuteReln}. Alternately, 
this group can defined with generators $(i \ i+1)$ (that is, only the adjacent elements) and relations \cref{eq:Gp1,eq:CommuteReln} along with the relations
\begin{align}
    \big(\sw_{i, i+1}\sw_{i+1, i+2}\big)^3&=I\label{eq:Gp2}.
\end{align} \qed
\end{rmk}

\sec{The \sswap}
\label{sec:the_symbolic_swap_algebra}

In this section we introduce an abstract $*$-algebra generated by formal variables which satisfy the same relations as the swap matrices introduced in the previous section. We call this algebra the \sswap , and then show that it is isomorphic to the \mswap. The main reason for introducing the \sswap comes from its use in 
 constructing the \df{Non-Commutative Sum of Squares (ncSoS) Hierarchy} for  \qmaxcut , discussed in \Cref{sec:swap_ncSoS}.  Specifically, this allows us to  formulate
rewrite rules for swap matrix polynomials (see \Cref{ssec:sswap_further_properties} and \Cref{sec:GB})  and express polynomial identities without explicitly using the matrix representations, which may be of independent interest.

An alternate approach to giving a fully algebraic characterization (i.e., a presentation) of the swap matrix algebra would be to build directly on the results of \Cref{subsec:swap_matrix_identities}. There we identified generators for $\swa_n$ along with a set of relations (i.e., identities) that those generators had to satisfy. To prove that these generators and relations give a presentation of the swap matrix algebra all that remains is to show that \textit{all} identities in the swap matrix algebra follow from these relations. Indeed, we will see in just a few pages that this is the case. We prefer the longer and slightly more mathematically involved approach described in the previous paragraph because it allows us to clearly distinguish elements of the swap matrix algebra $\swa_n$, which we view as matrices, from elements of the isomorphic \sswap, which we view as formal non-commuting variables. 

\looseness=-1

\subsection{A Presentation of the \sswap}
\label{ssec:the_sym_swap_alg}

\begin{defn}
\label{defn:sswap1}
Define the $n$th \sswap, denoted $\salg_n$, to be the $*$-algebra over $\mathbb{C}$ generated by the set of symmetric elements $\{s_{ij} \mid 1\leq i <  j\leq n\}$ satisfying relations: 
\begin{enumerate}[\rm(1)]
\item \label{eq:sqr_rl} $s_{ij}^2=1$;
\item \label{eq:ftrianglepair}
$s_{ij}  s_{jk} = s_{ik}  s_{ij}$,\quad 
\item 
\label{eq:Brels}
$s_{ij}s_{jk}+s_{jk}s_{ij} = s_{ij}+s_{jk}+s_{ik} - 1$.
\end{enumerate}
for all $1\leq i, j,  k \leq n$ 
 all distinct. In the relations above, for parsimony of notation, we have sometimes used $s_{ij}$ with $i>j$ to mean $s_{ji}$. We will continue to use this $s_{ij}$ shorthand later in the paper. The reason for this abuse of notation is discussed in \Cref{rmk:s_ij_abuse_of_notation}.
\end{defn}

Equivalently, we could have defined the \sswap as a quotient of the free algebra. Consider the free algebra 
$\free{\C}{s_{ij}\mid 1 \leq i\ < j\leq n}$
endowed with the involution $*$ that fixes each $s_{ij}$, and is
complex conjugation on $\C$.
Let $\sidl_n$ be the ideal in this free algebra
generated by
\cref{eq:sqr_rl}, \cref{eq:ftrianglepair}
and \cref{eq:Brels}. Then 
\begin{align}
\salg_n=
\free{\C}{s_{ij}\mid 1 \leq i < j\leq n}
/ \sidl_n
\end{align}
is the $n$th \sswap. We will refer to the elements $s_{ij}$ generating the \sswap as the \df{swap variables}.

\begin{rmk}  
\label{rmk:s_ij_abuse_of_notation}
The abuse of notation used in \Cref{defn:sswap1} above avoids a complication we now describe.
Suppose that we rigidly insist that $s_{ij}$ must have $i<j$.
Then to define the Symbolic Swap Algebra we must use more relations, namely
\ben [\rm(1)]
\item 
$s_{ij}^2-1$
\item \label{rem42:it2}
$
s_{ij}  s_{jk} = s_{ik}  s_{ij},\quad 
 s_{ij}  s_{ik} = s_{jk}  s_{ij},\quad 
 s_{ik}  s_{jk} = s_{jk}  s_{ij},$

\item 
$\big(s_{ij}s_{jk}+s_{jk}s_{ij}\big)  = \big(s_{ij}+s_{jk}+s_{ik} - 1\big), \quad 
\big(s_{ij}s_{ik}+s_{ik}s_{ij}\big)=
\big( s_{ij}+s_{jk}+s_{ik} - 1\big),$
$$
\big(s_{ik}s_{jk}+s_{jk}s_{ik}\big) = \big( s_{ij}+s_{jk}+s_{ik} - 1\big).
$$
\een
To illustrate the issue, we claim that 
$ s_{12}  s_{13} = s_{23}  s_{12}$ is true in the algebra $\salg_n$ but
does not follow from the equations given  in \Cref{defn:sswap1}
had we enforced index ordering $i< j < k$.
However $ s_{12}  s_{13} = s_{23}  s_{12}$ is
of the form of the second equation of 
\Cref{rem42:it2}.

In \Cref{sec:swap_ncSoS,sec:GB} when we turn to more computational aspects of $\salg_n$ 
we will enforce index ordering $i < j$ and encode relations as in \Cref{rmk:s_ij_abuse_of_notation}.
\qed
\end{rmk}

\begin{rmk} 
As in \Cref{rem:redundancy}, 
    we deduce that swaps on disjoint indices commute, i.e.,
    \beq\label{it:commutesymbolically}
    [s_{ij},s_{k\ell}]=0
    \eeq whenever $i,j,k,\ell$ are distinct. Similarly we can deduce the relations 
    \begin{align}
    (s_{ij}s_{jk})^3 = 1
    \end{align}
    given in the standard presentation of the symmetric group. \qed 
\end{rmk}

We can also understand $\salg_n$ as a quotient of the symmetric group algebra  $\mathbb{C}[S_n]$.

\begin{lem}
Let $\ssymidl_n$ be the two-sided ideal of $\mathbb{C}[S_n]$ generated by the elements 
\begin{align} \label{eq:Delta_rels_in_sym_gp}
\big((i\ j) (j\ k)+ (j\ k)(i\ j)\big)  -\big(
(i\ j)+(j\ k)+(i\ k) -1\big) 
\end{align}
for distinct $1\leq i,j,k\leq n$.
Then there is a natural $*$-homomorphism
\[
\C[\symgpn]\to\salg_n,\quad (i\ j)\mapsto s_{ij},
\qquad \text{for }1\leq i<j\leq n,
\]
and the kernel of this homomorphism is given by $\ssymidl_n$.
\end{lem}

\begin{proof}
We begin by noting the group algebra of the symmetric group $\mathbb{C}[\symgpn]$ can be presented as the algebra with generators $\{(i\ j) \mid 1\leq i\neq j\leq n\}$ satisfying relations 
\begin{enumerate}[\rm(1)]
\item $(i\ j)^2=1$;
\item $(i\ j)(j \ k) = (i \ k)(i \ j)$;
\item $(i\ j)(k \ \ell) = (k \ \ell)(i\ j)$ 
\end{enumerate}
for all distinct integers $i,j,k,\ell$.

The \sswap algebra has an equivalent set of generators with the only additional relations being relations being those given in \cref{eq:Delta_rels_in_sym_gp} above, so the proof follows. 
\end{proof}

\ssec{Relationship Between \sswap and Swap Matrix Algebra}
Now we relate the \sswap to both the Swap Matrices and to the irreps of the symmetric group, cf.~\cite[Theorem, §6.1]{Procesi}.

\begin{prop}\label{prop:swap}
The following are characterizations of  $\salg_n$, the \sswap:
\begin{enumerate}[\rm(a)]
\item The elements in $\ssymidl_n$ are precisely those that vanish under all irreps of $\symgpn$ corresponding to at most two rows of the Young tableaux.
\item The polynomials in $\sidl_n$ are exactly the polynomials which annihilate the matrices $\sw_{ij}$.
\end{enumerate}
\end{prop}

\begin{proof}
    For the sake of completeness, we recall additional terminology necessary for analysing the irreducible modules $\irrpmod{\irrpsym}$ of $\symgpn$. 
    A \emph{Young tableau} of shape $\irrpsym \vdash n$ is a filling of the Young diagram $\irrpsym$ by integers $1,2,\dots,n$ such that each box is assigned a unique integer. The action of the group $\symgpn$ on a Young tableau $t$ follows by letting $\pi \in \symgpn$ act on the entries of $t$. 
    Permuting entries within each row of a tableau $t$ gives us an equivalence class of tableaux\,--\,a \emph{tabloid}\looseness=-1
     \begin{align}\label{eq:tabloid}
    \{t_1,t_2,\dots,t_k\}:=\{t\}.
    \end{align}
    The group action extends to tabloids as $\pi \{t\} = \{\pi(t_1),\pi(t_2),\dots,\pi(t_k) \} $. 
    The so-called Specht irreducible module $\irrpmod{\irrpsym}$ of $\C[\symgpn]$ is spanned by \emph{polytabloids}
    \begin{align}
        e_t = \sum_{\pi \in C_t} \sign{\pi} \pi\{t\}
    \end{align}
    where $t$ ranges over Young tableau of shape $\lambda$, and $C_t$ is the set of permutations that permute elements only within the columns of $t$. 

    For a 3-row irrep given by $\lambda = [\lambda_1, \lambda_2, \lambda_3]$, let $T$ be the Young tableau 
    \begin{align}
        T = \begin{ytableau}
                \scriptstyle 1 & \scriptstyle 2 & \none[\scriptstyle \cdots] & \none[\scriptstyle \cdots] & \scriptstyle \lambda_1 \\
                \scriptstyle \lambda_1+1 & \none[\scriptstyle\dots] & \none[\scriptstyle\dots] & \scriptstyle \lambda_2 \\ 
                \scriptstyle \lambda_2+1 & \none[\scriptstyle\dots] & \scriptstyle \lambda_3 .
            \end{ytableau}
    \end{align}
    and $e_T$ the corresponding polytabloid. Consider the transpositions $(1\ \lambda_1+1)$, $(1\ \lambda_2+1)$, $(\lambda_1+1\ \lambda_2+1)$ all of which permute elements within the first column of $T$ and are thus contained in $C_T$. Note that
    \begin{align}\label{eq:37}
        (1\ \lambda_1) e_T = e_{(1\ \lambda_1)T}
    \end{align}
    and on expanding the right-hand side of \cref{eq:37} we see that $T$ appears with coefficient $\sign{(1\ \lambda_1+1)} = -1$. By the exact same reasoning, the coefficients of $T$ in $(1\ \lambda_2) e_T$ and $(\lambda_1\ \lambda_2) e_T$ also equal $-1$. 
    Now, each of $(1\ \lambda_1+1) (\lambda_1+1\ \lambda_2+1) = (1\ \lambda_2+1\ \lambda_l+1)$ and $(\lambda_1+1\ \lambda_2+1) (1\ \lambda_1+1) = (1\ \lambda_1+1\ \lambda_2+1)$ is an even permutation contained in $C_T$. Repeating the above for $(1\ \lambda_1+1) (\lambda_1+1\ \lambda_2+1) e_T$ and $(\lambda_1+1\ \lambda_2+1) (1\ \lambda_1+1)e_T $, we see that the coefficient of $T$ in these must be +1. 
    It follows that 
    \begin{align}
        \left[(1\ \lambda_1+1) + (1\ \lambda_2+1) + (\lambda_1+1\ \lambda_2+1) -1 \right] e_T \neq \{(1\ \lambda_1+1), (\lambda_1+1\ \lambda_2+1)\}e_T
    \end{align}
    in the irrep specified by $\lambda$, and more generally in any irrep with 3 or more rows. That is, $s_{ij}s_{jk}+s_{jk}s_{ij}  - (s_{ij} + s_{jk} + s_{ik} -1)$ does not vanish under the evaluation $ s_{ij}=\rep{\irrpsym} (i\ j)$ for the above irrep $\irrpsym$. 
    Thus we have proved that the swap relations are incompatible with a $\geq3$ row Young Tableaux.\looseness=-1

For the converse, consider distinct indices $i,j,k$ and a tabloid $T$ of shape $[\nk]$. 
Let $\pi\in C_T$.
We  distinguish two cases:
\begin{enumerate}[\rm(a)]
\item
$i,j,k$ lie in the same row of $\pi\{T\}$.\\
Then each of the terms $s_{ij},s_{jk},s_{ik}$ acts as the identity on $\pi\{T\}$, whence 
\beq\label{eq:eT0}
\big(s_{ij}s_{jk}+s_{jk}s_{ij}  - (s_{ij} + s_{jk} + s_{ik} -1)\big)(\pi\{T\})=0.
\eeq
\item
$i,j$ lie in the same row of $\pi\{T\}$, but $k$ does not.\\
Then, 
\[
\begin{aligned} 
s_{ij}(\pi\{T\})&=\pi\{T\}, 
& s_{ij}s_{jk}(\pi\{T\}) & = s_{ik}(\pi\{T\}), 
& s_{jk}s_{ij}(\pi\{T\}) & = s_{jk}(\pi\{T\}).
\end{aligned}
\]
As before, this implies \cref{eq:eT0}.
\end{enumerate}

This shows that for each $\pi\in C_T$, \cref{eq:eT0} holds, whence
$s_{ij}s_{jk}+s_{jk}s_{ij}  - (s_{ij} + s_{jk} + s_{ik} -1)$ vanishes under the evaluation $ s_{ij}=\rep{\irrpsym} (i\ j)$ for any irrep $\irrpsym=[\nk]$. 
\end{proof}

\begin{thm}[\protect{\cite[Theorem, §6.1]{Procesi}}]
    \label{thm:swap}
The algebras $\salg_n$ and $\swa_n$ are isomorphic.
More precisely, the natural map 
\[
\salg_n\to\swa_n,\quad s_{ij}\mapsto\sw_{ij}
\]
is an isomorphism.

\end{thm}

\begin{proof}
An immediate corollary of \Cref{prop:swap}.
\end{proof}

The isomorphism between the \sswap and \mswap given above lets us identify the QMC Hamiltonian associated with a graph $G$ with an element of the \sswap. We make this clear in the following definition.  

\begin{defn}
For any graph $g$ define the symbolic QMC Hamiltonian $\hampol{G}$ by
\begin{align}
    \hampol{G} := \sum_{(i,j)\in \E(G)} 2w_{ij}\left( I - s_{ij} \right)  \in \salg_{n}
\end{align}
\end{defn}
\bs

We end this section by proving some results about the dimension of the \sswap. 
While the formula for the dimension is not used elsewhere in the paper it does bring to light a potentially interesting connection between the swap matrix algebra and the Catalan numbers. 

\begin{cor}\label{cor:swap}
Recall from \Cref{lem:irrep_dim} that $\chara{[n-k,k]}e$ denotes the dimension of the module $\irrpmod{[\nk]}$ of $\symgpn$. Then
\begin{enumerate}[\rm(a)]
\item\label{it:decompAnswap}
$\displaystyle
\salg_n \cong \bigoplus_{k=0}^{\lfloor \frac n2\rfloor}
M_{\chara{[n-k,k]}e}(\C)$.
\item\label{it:dimAnswap}
$\displaystyle \dim \salg_n = \sum_{k=0}^{\lfloor \frac n2\rfloor} 
\left(\frac{n-2k+1}{n-k+1} {n \choose k}  \right)^2= 
\frac1{n+1}{2n\choose n}
$,\\[1mm]
which is the $n$-th Catalan number $C_n$.
\end{enumerate}
\end{cor}

\begin{proof}
\Cref{it:decompAnswap} follows from \Cref{prop:swap}. 

\Cref{it:dimAnswap}: The first equality simply uses the dimension of the full matrix algebra. 
We shall give two proofs of the second equality.

{\it Algebraic proof.} Observe that
\[
\begin{split}
\frac1{n-k+1}{n \choose k} & = \frac1{n+1}{n+1\choose k} \\
k {n+1\choose k}&=(n+1){n\choose k-1}
\end{split}
\]
which simplifies the summand $\displaystyle\left(\frac{n-2k+1}{n-k+1} {n \choose k}  \right)^2$ into 
\[
 \Big( {n+1\choose k}-2 {n\choose k-1}\Big)^2=
 \Big( {n\choose k}- {n\choose k-1}\Big)^2=
{n\choose k}^2 - 2{n\choose k}{n\choose k-1}+{n\choose k-1}^2
,
\]
where we applied the recurrence relation on the binomial coefficients for  the middle  equality.
Thus 
\beq\label{eq:prechu}
\dim \salg_n = \sum_{k=0}^{\lfloor n/2\rfloor}
{n\choose k}^2 - 2{n\choose k}{n\choose k-1}+{n\choose k-1}^2.
\eeq

By the Chu-Vandermonde identity,
\beq\label{eq:chu}
\sum_{k=0}^n{n\choose k}^2={2n\choose n},
\eeq
whence by symmetry,
\beq\label{eq:chu1}
\sum_{k=0}^{\lfloor n/2\rfloor}{n\choose k}^2=
\begin{cases}
\frac12 {2n\choose n} & n\text{ odd}\\[3mm]
\frac12 \Big({2n\choose n}+ {n\choose n/2}^2\Big)& n\text{ even}.
\end{cases}
\eeq
Likewise, we obtain
\beq\label{eq:chu2}
\sum_{k=0}^{\lfloor n/2\rfloor}{n\choose k-1}^2=
\begin{cases}
\frac12 \Big({2n\choose n}- 2{n\choose (n-1)/2}^2\Big)& n\text{ odd}
\\[3mm]
\frac12 \Big({2n\choose n}-{n \choose n/2}^2\Big) & n\text{ even}.
\end{cases}
\eeq
Similarly, the Chu-Vandermonde identity implies
\[
\sum_{k=0}^n {n\choose k}{n\choose k-1}={2n\choose n+1},
\]
whence again by symmetry,
\beq\label{eq:chu3}
\sum_{k=0}^{\lfloor n/2\rfloor} {n\choose k}{n\choose k-1}=
\begin{cases}
\frac12 \Big({2n\choose n+1} - {n\choose (n-1)/2}^2\Big) & n\text{ odd}\\[3mm]
\frac12 {2n\choose n+1} & n\text{ even}.
\end{cases}
\eeq
Inserting \cref{eq:chu1}, \cref{eq:chu2}, \cref{eq:chu3} into \cref{eq:prechu} yields
\[
\dim \salg_n = {2n\choose n}-{2n \choose n+1}  = {2n\choose n}-\frac{n}{n+1}{2n \choose n}  =  \frac1{n+1}{2n\choose n} = C_n,
\]
as desired.

{\it Combinatorial proof.} We also provide a slicker but less self-contained combinatorial proof. It is well-known that the $n$-th Catalan number $C_n$ counts the number of standard Young tableaux of the shape $[n,n]$ 
\cite[Exercise 6.19.ww]{stanley2}.
Picking out the squares containing $1,\ldots,n$ 
and $n+1,\ldots, 2n$, respectively, 
from such a tableaux,
we obtain two Young tableaux of the same shape $[n-k,k]$ for some $k\in\N$ (see \cref{eq:young1to2} below for a simple example). 
 \begin{align}
\begin{ytableau}
1 & 3 & 4 & 6 \\
2 & 5 & 7 & 8            
\end{ytableau} 
&\quad\rightsquigarrow\quad
 \Big(\; \begin{ytableau}
1 & 3 & 4\\ 
2 
\end{ytableau} \;,\quad
 \begin{ytableau}
\none&\none & 6 \\
5 & 7 & 8 
\end{ytableau}\; \Big) \nonumber \\
&\hspace{30pt} \quad\rightsquigarrow\quad
 \Big(\; 
 \begin{ytableau}
1 & 3 & 4\\ 
2 
\end{ytableau} \;,\quad
 \begin{ytableau}
1 & 2 & 4\\ 
3 
\end{ytableau}\; \Big). \label{eq:young1to2}
\end{align}
  
Since standard Young tableaux of the shape $[n-k,k]$ 
correspond to a basis of the corresponding irrep of $\symgpn$, we are done by (a).
\end{proof}

\ssec{The Irrep \sswap}

Recall 
$\ham{\clique{n}}$ denotes the Hamiltonian the swap matrices associate with the clique $\clique{n}$ on $n$ vertices, 
$\hampol{\clique{n}}$ denotes the element of the  \sswap \
associated to the clique and that $\hamirrep{\clique{n}}{\irrpsym}$ denotes the representation of $\hampol{\clique{n}}$ in the $\irrpsym$ irrep of the symmetric group $\symgpn$.

\begin{lem}\label{lem:calc}
The function of $k$ in 
 \cref{eq:traceHKn}
 is strictly increasing for $0\leq k\leq \frac n2$.
\end{lem}

\begin{proof}
Simply observe that the function's derivative w.r.t.~$k$ is
\[
2(n+1) -4k \geq 2(n+1)-2n =2 > 0. \qedhere
\]
\end{proof}

\begin{prop}\label{prop:Cink}
Let $\hampol{\clique n}\in\salg_n$ denote the Hamiltonian polynomial for the $n$-clique. Then the
ideal  $\cI_{\nk}$  of polynomials which annihilate
$\rep{[\nk]}$, is generated by $\sidl$ 
together with $\hampol{\clique n}-\eta_{\nk}$.
\end{prop}

\begin{proof}
Let $\ialg{\lambda}_n$ denote the simple algebra of the  irrep $\rep\lambda$
of $\symgpn$ for $\lambda\vdash n$. It is isomorphic to $M_{\chara{\lambda}e}(\C) $.
By \Cref{cor:swap}, 
\beq\label{eq:isomialg}
\salg_n \cong \bigoplus_{k=0}^{\lfloor \frac n2\rfloor}
\ialg{{[\nk]}}_n.
\eeq
With this notation,
\[
\salg_n /\cI_{\nk} \cong 
\ialg{{[\nk]}}_n 
.
\]

By \Cref{lem:clique_ham_identity} and \Cref{lem:irrpsym_value}, $\hampol{\clique n}-\eta_{\nk}\in\cI_{\nk}$.
Since by \Cref{lem:calc}, the map $k\mapsto \eta_{\nk}$ is injective for $0\leq k\leq n/2$, the image under the isomorphism of
\cref{eq:isomialg} of
$\hampol{\clique n}-\eta_{\nk}\in\cI_{\nk}$ 
is zero in the $k$-th entry and a nonzero number in each of the others.
In particular,\looseness=-1
\[
\salg_n / \cI(\hampol{\clique n}-\eta_{\nk}) \cong 
\ialg{{[\nk]}}_n \cong \salg_n /\cI_{\nk}.
\]
Since $\cI(\hampol{\clique n}-\eta_{\nk}) \subseteq \cI_{\nk}$,
this concludes the proof.
\end{proof}

\ssec{Structure of Swap Variable Polynomials}
\label{ssec:sswap_further_properties}

Now we describe some further properties of the Swap Algebra. 
The main result of this subsection is a sharp upper bound of $\lceil n/2 \rceil$ on the maximum degree of elements in the swap algebra, along with simplified form for elements in this algebra. Later, we will observe that this upper bound gives a corresponding upper bound on the level at which the hierarchy of semidefinite programs constructed in \Cref{sec:swap_ncSoS} gives an exact solution to the Quantum Max Cut problem. We head in that direction via a lemma.

Define the \df{support graph $G_p$ of a polynomial} $p$  in the variables $s_{ij}$ to have vertices corresponding to integers which appear as indices $i$ or $j$ for some $s_{ij}$
in $p$, with $(i,j)$ an edge iff $s_{ij}$
appears in $p$.
A polynomial in $\sidl$
is called \df{degree reducing}
provided it has exactly one highest degree term.

\sssec{Degree reducing polynomials in 
\texorpdfstring{$\sidl$}{Iswap}}

\begin{lem} 
\label{lem:degred mon}
For $n \geq 4$ and $1 \leq i,j,k,\ell \leq n$ with no two indices being equal, there is  a degree reducing polynomial in $\sidl$ whose highest degree term
has the form
\ben[\rm(1)]
\item
$s_{jk}s_{ik}s_{\ell k}$.
 \quad a term with support  graph a 
  three edge star;

\item 
$s_{ij} s_{jk} s_{k \ell} $, $s_{jk}s_{ij}s_{k \ell}$, or $s_{jk}s_{k \ell}s_{ij}$, \quad a term with support  graph a 
three edge line.
\een 

  Observe that any of the three edge line monomials equals a star graph monomial modulo a single triangle pair substitution.
\end{lem}
Here a
\df{triangle pair relation} is a polynomial of the 
form $s_{ij} s_{jk}- s_{jk} s_{ki}$.
It lies in $\sidl$ and 
relates behavior of edge pairs in the triangle which is the 
support of the polynomial.

\begin{proof}
Straightforward matrix multiplication can be used to verify the following matrix identity:
\[
\begin{split}
2\sw_{jk}\sw_{ik}\sw_{\ell k}  & = 1 - \sw_{ij} - \sw_{i\ell} - \sw_{jk} - \sw_{k\ell} + \sw_{ij}\ \sw_{ik}\\
& \phantom{={}}  + \sw_{ij}\ \sw_{i\ell}+ \sw_{ij}\ \sw_{k\ell} + \sw_{ik}\ \sw_{i\ell}\\
& \phantom{={}} - \sw_{ik}\ \sw_{j\ell}  + 
 \sw_{i\ell}\ \sw_{jk} + \sw_{jk}\ \sw_{j\ell},
\end{split}
\]
which implies the same identity for the $s_{ij}$ by \Cref{thm:swap}.

The proof of the three-edge-line case and the final paragraph of the lemma amount to the identities\looseness=-1
\begin{align*}
    s_{ij}s_{jk}s_{k\ell} = s_{ij}s_{j\ell}s_{jk},\\
    s_{jk}s_{ij}s_{k\ell} = s_{ik}s_{jk}s_{\ell k},\\
    s_{jk}s_{k\ell}s_{ij} = s_{j\ell}s_{jk}s_{ij}
\end{align*}
so we can always reduce to the first case.
\end{proof}

\begin{lem}\label{lem:degred-disjoint-lines}
    For $n \geq 6$ and $1 \leq i,j,k,a,b,c \leq n$ with no two indices being equal, there is  a degree reducing polynomial in $\sidl$ whose highest degree term
has the form
$$s_{ij}s_{jk}s_{a b}s_{b c}, \qquad \text{
 a term whose support  graph is
  two disjoint two edge lines.}$$
\end{lem}

\begin{proof}
It is straightforward to verify the following identity for $2^6\times2^6$ matrices from which the lemma follows using \Cref{thm:swap} by reindexing  if needed:
\begin{multline*}
4 \sw_{12}    \sw_{23}    \sw_{45}    \sw_{56} =3 - 3 \sw_{12} - 3 \sw_{13} - 3 \sw_{23} - 3 \sw_{45}\\ - 3 \sw_{46} - 3 \sw_{56} + 2 \sw_{12}  \sw_{13} + 3 \sw_{12}  \sw_{45} + 
 3 \sw_{12}  \sw_{46}\\ + 3 \sw_{12}  \sw_{56} + 3 \sw_{13}  \sw_{45} + 3 \sw_{13}  \sw_{46} + 3 \sw_{13}  \sw_{56}\\ + \sw_{14}  \sw_{25} - \sw_{14}  \sw_{26} - 
 \sw_{14}  \sw_{35} + \sw_{14}  \sw_{36}\\ - \sw_{15}  \sw_{24} + \sw_{15}  \sw_{26} + \sw_{15}  \sw_{34} - \sw_{15}  \sw_{36}\\ + \sw_{16}  \sw_{24} - 
 \sw_{16}  \sw_{25} - \sw_{16}  \sw_{34} + \sw_{16}  \sw_{35}\\ + 3 \sw_{23}  \sw_{45} + 3 \sw_{23}  \sw_{46} + 3 \sw_{23}  \sw_{56} + \sw_{24}  \sw_{35}\\ - 
 \sw_{24}  \sw_{36} - \sw_{25}  \sw_{34} + \sw_{25}  \sw_{36} + \sw_{26}  \sw_{34}\\ - \sw_{26}  \sw_{35} + 2 \sw_{45}  \sw_{46} - 2 \sw_{12}  \sw_{13}  \sw_{45}\\ - 
 2 \sw_{12}  \sw_{13}  \sw_{46} - 2 \sw_{12}  \sw_{13}  \sw_{56}\\ - 2 \sw_{12}  \sw_{45}  \sw_{46} - 2 \sw_{13}  \sw_{45}  \sw_{46}\\ - 
 2 \sw_{14}  \sw_{25}  \sw_{36} + 2 \sw_{14}  \sw_{26}  \sw_{35}\\ + 2 \sw_{15}  \sw_{24}  \sw_{36} - 2 \sw_{15}  \sw_{26}  \sw_{34}\\ - 
 2 \sw_{16}  \sw_{24}  \sw_{35} + 2 \sw_{16}  \sw_{25}  \sw_{34}\\ - 2 \sw_{23}  \sw_{45}  \sw_{46}.
\end{multline*}\qedhere
\end{proof}

\sssec{Useful graph properties}

\begin{lem} \label{lem:subgraph-edge-relationship}
    Let $G$ be a connected $n$ vertex graph.
    \ben[\rm(1)]
    \item
    If $n \geq 4$, then $G$ contains a three edge line or a three edge star graph as a subgraph.
    \item 
    If $G$ has more than three edges, then $G$ contains a three edge line or a three edge star or a triangle as a subgraph.
\een 
\end{lem}

\begin{proof}
    Let $G$ be as above. There exists some spanning tree $T$ of $G$ as $G$ is connected. Given two points $u$ and $v$ in $T$, there is a unique path connecting them. The length of this path is the graph distance from $u$ to $v$. Now, pick some point in $T$ and call it $r$. We define $V_i$ to be the set of vertices in $T$ that are distance $i$ from $r$. We know that $V_0 = \{ r \}$. If $V_1$ contains three or more elements, then $G$ must contain a three edge star graph as a subgraph. If $V_1$ contains fewer than three elements, then as $n \geq 4,$ we must have an element in $V_2$. Thus, $G$ contains a three edge line.

    The proof of the second assertion is obvious.
\end{proof}

\begin{lem}
\label{lem:threeFourEdges}
  Suppose $G$ is an $n$ vertex graph with $|\E(G)| \geq \lceil \frac{n}{2} \rceil  +1 $.
  Then $G$
  must have 
  \ben[\rm(1)]
  \item
a single connected component with at least three edges, hence  $G$ contains a triangle, a three edge line, or a three edge star as a subgraph; 
 \ or
  \item
two connected components each 
containing a 2 edge line.
\een 
\end{lem}

  \proof 
   To see this, consider $G_1,\dots,G_k$, the connected components of $G$.
   The point is that when
$|\E(G)|$ is small the number of possible configurations is small:
   \ben[\rm(a)]
     \item
         \label{it:neven}
$|\E(G)|= \frac{n}{2} =  \lceil \frac{n}{2} \rceil$  with $n$ even implies  each $G_i$ has exactly one edge.
       \item
       \label{it:nodd}
$|\E(G)|=  \lceil \frac{n}{2} \rceil$ with $n$ odd 
implies  each $G_i$ except one, say $G_1$,
has exactly one edge and $G_1$ is a 2 edge line.
        
           \item
           \label{it:nevenPlus}
$|\E(G)|= \frac{n}{2} +1 =  \lceil \frac{n}{2} \rceil + 1$ with $n$ even implies
 each $G_i$ except two, say $G_1$ and $G_2$,  
have 1 edge and $G_1$ and $G_2$ each contain (actually equal)  2 edge lines.

\item 
\label{it:noddP}
$|\E(G)|= \frac{n}{2} +1 =  \lceil \frac{n}{2} \rceil + 1$ with $n$ odd implies
 each $G_i$ except one, say $G_1$,
has exactly one edge and $G_1$ has 3 edges, so is a three edge line, a triangle or a three edge star graph.
         \een 
Bigger $|\E(G)| $ just adds edges to some of these components,
thereby  justifying inequality in the theorem statement.
  \qed

\sssec{Monomials associated to their support graphs}

\begin{lem} \label{lem:monomial-exchange}
    Any monomial in the $n$-vertex swap variables of the form $p = q s_{ij}$ can be written in the form $p = s_{ij} q'$, for $\deg(q') \leq \deg(q)$.
\end{lem}

\begin{proof}
    For any $s_{k\ell}$ and $s_{ij}$,
    we have 
    $s_{k\ell}s_{ij} = s_{ij}s_{e}$ for some edge $e$. Indeed,
    we consider 5 cases: (1) if $\{i,j\}\cap\{ k,\ell\}=\varnothing$,
    then $s_e= s_{k \ell}$; (2) if
    $j=k$, then
    $s_e=s_{ik}$; (3) if $i=k, $ then $s_e=s_{j\ell}$;
   (4) if $i=\ell$, then $s_e=s_{kj}$;
       (5) if $j=\ell$, then $s_e=s_{ik}$.
    Thus, applying the above inductively, we get $q s_{ij} = s_{ij} q'$ for $\deg(q) \geq \deg(q')$.
\end{proof}

\begin{lem} \label{lem:square-deg-red}
Any monomial in the $n$-vertex swap variables of the form $p = s_{ij} q s_{ij}$ 
can be written as a monomial of degree at most $\deg(q)$.
In particular, 
for fixed ${i,j}$ any 
monomial in the $n$-vertex swap variables can be assumed to contain
at most a single $s_{ij}$.
\end{lem}

\begin{proof}
    Applying \Cref{lem:monomial-exchange}, we get $s_{ij} q s_{ij} = s_{ij}^2 q' = q'$ for $\deg(q) \geq \deg(q').$

    The second statement follows from the first by induction.
\end{proof}

\begin{lem} \label{lem:three-edge-line}
    A monomial in the $n$-vertex swap variables of the form 
    $p = q_0 \prod_{j = 1}^k \left(s_{e_j} q_j\right)$ with the $e_j$ being distinct edges 
    equals a monomial $\left(\prod_{j=1}^k s_{e_j}\right) q$ of the same degree or smaller.
\end{lem}

\begin{proof}
    Suppose $p = q_0 \prod_{j = 1}^k \left(s_{e_j} q_j\right)$ as above. Then one application of \Cref{lem:monomial-exchange} gives us $p = s_{e_1}q_0'q_1 \prod_{j = 2}^k \left(s_{e_j} q_j\right)$. Two more applications yield $p = s_{e_1}s_{e_2}q_0''q_1'q_2 \prod_{j = 3}^k \left(s_{e_j} q_j\right)$. Repeating this process inductively, we obtain $p = \left(\prod_{j=1}^ks_{e_k}\right) q_0^{(n)} q_1^{(n-1)} \cdots q_{n-1}' q_n$. Note that at each step, the degree did not increase, so taking $q = q_0^{(n)} q_1^{(n-1)} \cdots q_{n-1}' q_n$ we obtain the desired result.
\end{proof}

\sssec{Polynomials in swap variables}

After considerable preparation we now give and prove 
any polynomial $p$ in swaps can be reduced to a 
simple form $q$.
The proof is constructive
(and does not use  Gr\"obner bases).

\begin{thm}
\label{thm:swapDegree}
    Any  polynomial $p$ in the $n$-vertex swap variables,
    is  mod $\sidl$ equal to some polynomial $q$ with $\deg(q) \leq \lceil \frac{n}{2} \rceil$.

    Moreover, 
one can take $q$ to have  each of its terms a monomial of the form
 \begin{align}
        m= m_1 m_2...m_r\qquad mod \ \ \sidl
    \end{align}
    where $m_1, m_2, \dots, m_r $ are commuting monomials of  all of degree one except possibly one has degree two. Also each pair
    $m_i $, $m_j$ with $i \not =j$ are supported on vertex disjoint
    graphs.
\end{thm}

\begin{proof}
Let $\tilde p$ be a polynomial of minimum degree within $p + \sidl$.
  By \Cref{lem:square-deg-red} we may assume that in any term $\tau$ of $\tilde p$
  the degree of any particular $s_{ij}$ is at most $1$.
  Suppose that $\deg(\tau) \geq  \lceil \frac{n}{2}\rceil +1$. 
  Then $|\E(G_\tau)| 
  \geq \lceil \frac{n}{2}\rceil +1$, so
  by \Cref{lem:threeFourEdges} the graph 
   $G_\tau$ contains 
   \ben[\rm(1)]
   \item 
   three edges 
   ${e_1} , {e_2} , {e_3} $ which form  a 3 edge line, or 3 edge star, or triangle; \ \ or 
   \item 
   two vertex disjoint 2 edge lines.
   \een 
In the first case we
   can use \Cref{lem:three-edge-line} to obtain $\tau = s_{e_1} s_{e_2} s_{e_3} q'$ where $e_1,e_2,e_3$ are the three edges in the three edge subgraph described above. 
    By \Cref{lem:degred mon}
    we have a degree reducing relation for monomials supported on three edge lines, three edge star graphs, and triangles which combined with $\tau$ produce a polynomial equivalent to it mod $\sidl$ 
    but of degree less than $\deg(\tau)$. This
    contradicts $\tau$ having minimal degree.

    To prove the second case \Cref{lem:degred-disjoint-lines}
    works similarly to give a degree drop contradiction.
    This proves the first part of the theorem.
    
Proof of the second part of the theorem. 
Since $\deg(q) 
  \leq \lceil \frac{n}{2}\rceil$
  we see 
\Cref{lem:threeFourEdges} \Cref{it:neven}, \Cref{it:nodd}
and
\Cref{lem:three-edge-line} 
imply that each term $m$ of $q$ 
has the asserted form.

\end{proof}

We end this section with a proof that the upper bound given in the preceding theorem is tight when $n$ is even. 

\begin{thm}
For any even integer $n$, the monomial $s_{12}s_{34}...s_{n-1,n} \in \salg_n$ cannot be written as a sum of monomials of degree less than $n /2$.  

Thus, for any $n$, we require polynomials of degree at least $\lfloor n /2 \rfloor$ to express all elements in the $n$ qubit swap algebra $\salg_n$. 
\end{thm}

\begin{proof}
Consider the case where $n$ is even. We prove the result for swap matrices, from which the theorem is immediate by \Cref{thm:swap}. Consider the matrix $\sw_{12}\sw_{34}...\sw_{n-1,n}$. Expanding in the Pauli basis we see that it is supported on Pauli matrices of weight $n$ (i.e. it's expansion contains degree $n$ products of Pauli matrices). But any product of at most $n/2 - 1$ swaps acts on at most $n-2$ qubits, and so is supported on Pauli matrices of weight at most $n-2$. Since the Pauli matrices form a orthogonal basis for $M_2(\mathbb{C})^{\otimes n}$, this shows the matrix $\sw_{12}\sw_{34}...\sw_{n-1,n}$ cannot be equal to a sum of products of swap  matrices with degree at most $n/2 - 1$. 

The result for odd $n$ is immediate by considering the monomial $\sw_{12}\sw_{34}...\sw_{n-2,n-1}$.
\end{proof}

\def\mmp{{\cM}}

\sec{The Swap Algebra and the Non-Commutative Sum of Squares Hierarchy}
\label{sec:swap_ncSoS}

In this section we construct an ncSoS hierarchy for the \sswap. We begin by reviewing the general ncSoS technique in \Cref{ssec:ncSoS} and then show how it can be applied specifically to the swap algebra in \Cref{sec:swap_ncSoSd4}.

Previously an ncSoS hierarchy applied over the Pauli algebra, the \emph{Quantum Lasserre Hierarchy}, has been to used to upper (or lower) bound the maximum (or minimum) eigenvalue of local Hamiltonians \cite{brandao2013product, gharibian2019almost}. 
The Quantum Lasserre Hierarchy is general and capable of addressing any qubit Hamiltonian problem. Our hierarchy is distinct in that it is specific to the swap algebra. But in principle, an ncSoS hierarchy can be constructed for any suitably presented algebra. Hamiltonian problems studied in physics have algebraic structures associated with them. Thus, a message here is that ncSoS techniques can be adapted to the algebra of Hamiltonians and potentially give improved bounds to eigenvalue problems.

\ssec{A Monomial Order -- grlex}
\label{sec:monorder}
In computations with an algebra and their presentation via ideals we often  need to place an order on monomials.
In our situation the variables are
$\{s_{ij} \mid 1\leq i<j\leq n\}$
and we define a  
\df{graded lexicographic order (grlex)} $<$
on them as follows.

\ben[\rm(1)]
\item Order the alphabet as
$$s_{1 1} < s_{1 2} < \cdots < s_{1 n} <  s_{23}< 
s_{2 4} < \cdots < s_{2 n} < \cdots < s_{n n};$$

\item 
for any two monomials $a$ and $b$ in the $s_{i j}$,
take
$a<b$ if $\deg(a)< \deg(b)$;

\item 
  if $\deg(a)= \deg(b)$, then look at them
  as words in the $s_{i j}$ and sort them as you would in a dictionary according to alphabetical order.
\een 

\ssec{Non-Commutative Sum Of Squares Hierarchy}

\label{ssec:ncSoS}

Let $\cI$ be a $*$-ideal in $\free{\C}{s}$.
Suppose $h$ is a symmetric polynomial, by which we mean that under the  involution $\phantom{}^*$, we have $h^* = h$.
Define $\nu_d(h)$ to be the lower limit of upper bounds $\nu$ making\looseness=-1
\begin{align}
\label{eq:posss}
    \nu - h \in  \SOS_{2d}{} + \cI,
\end{align}
where $\SOS_{2d}$ denotes the set of all 
sums of squares of polynomials 
in the variables $s_{i j}$
each having degree $\leq d$.

Note that if $\nu - h \in \SOS_{2d}{} + \cI$ for any $d$ we must also have that $\pi(\nu - h) \geq 0$ for all representations $\pi : \free{\C}{s} / \cI \rightarrow \cB(\cH)$. 
Thus, we have that any such $\nu$, and in particular any $\nu_d(h)$, gives an upper bound on the max eigenvalue of $h$ under all representations of $\free{\C}{s} / \cI$.
For a fixed $\cI$ we call this process the \df{$d$th relaxation} and the least upper bound $\nu_d(h)$ we call the \df{$d$th relaxed value}. Key to analyzing and computing this bound is expressing 
$\SOS_{2d}$ and membership in $\cI$ succinctly.

Applying this formalism with $\cI=\sidl$
gives an upper bound 
\begin{align}
\nu_d(h) \geq \maxeig(h(\sw)),
\end{align}
where $h(\sw)$ denotes the matrix obtained by substituting each element $s_{ij}$ in the polynomial $h$ with the corresponding swap matrix $\sw_{ij}$\footnote{So, in particular, observe that we have $\hampol{G}(\sw) = \ham{G}$ for any graph $G$.}, and we will investigate this particular situation more in the following section. For now we return to the case of a general ideal $\cI$.

The Veroneses are column vectors, denoted $V_d(n)$,
which consist 
of degree $d$ monomials in the $n(n-1)/2$ variables $s_{ij},\ i<j $, ordered w.r.t.~grlex. 
This makes it possible to test membership in $\SOS_{2d}$ with the help of semidefinite programming (SDP).

\begin{lem}\label{lem:sosSDP}
Let $h=h^*\in\free{\C}s$ be of degree $\leq2d$. Then $h\in\SOS$ iff there is a positive semidefinite matrix $\Gamma $ such that
\beq\label{eq:sosSDP}
h=V_d(n)^* \Gamma  V_d(n).
\eeq
Finding such a $\Gamma $ can be done with an SDP.
\end{lem}

\begin{proof}
This is well-known and routine \cite{klepBook}. \cref{eq:sosSDP}
yields a system of linear equations on the entries of $\Gamma $, so finding a positive semidefinite $\Gamma $ satisfying these linear constraints amounts to a feasibility SDP.
\end{proof}

\begin{cor}\label{cor:sosSDP}
Let $h=h^*\in\free{\C}s$. Then $h\in\SOS_{2d}+\cI$ iff there is a positive semidefinite matrix $\Gamma $ such that
\beq\label{eq:sosSDPI}
h-V_d(n)^* \Gamma  V_d(n) \in \cI.
\eeq
Assuming a linear algebra basis for $\free{\C}s/\cI$ or a ``good'' generating set\footnote{i.e., a Gr\"obner basis, cf.~\Cref{sec:GB} below} for $\cI$  is known, finding such a $\Gamma $ can be done with an SDP.
\end{cor}

\begin{proof}
The first part of the statement follows as in \Cref{lem:sosSDP}.
Then, to translate \cref{eq:sosSDPI} into a linear system on the entries of $\Gamma $, we need to solve the ideal membership problem for $\cI$; this can be done using a linear algebraic basis for the space $\free{\C}s/\cI$ or via Gr\"obner bases. For a longer discussion of this issue see \Cref{sssec:Finding_Linear_Constraints}. 
\end{proof}

With this, \cref{eq:posss} can be expressed as the SDP
\beq\label{eq:ncSOS}
\begin{split}
\nu_d(h)=\inf & \; \nu \\
\text{s.t.} & \; \Gamma \succeq0\\
&\; \nu-h-V_d(n)^*\Gamma V_d(n)\in\cI.
\end{split}
\eeq

\begin{rmk}
\label{rem:Id} 
One can  approximate \cref{eq:ncSOS} by using truncated ideals.
Given generators $g_1,\ldots,g_e$ for the ideal $\cI$, form the
degree $2d$ truncation of $\cI$ as follows:
\[
\cI_{2d}:=\spn\{u g_k v \mid u,v\text{ words in }s_{ij}, \; k=1,\ldots,e,\; \deg(u g_k v)\leq2d \}.
\]
(Beware, frequently, $\cI_{2d}\subsetneq\cI\cap\free{\C}{s}_{2d}$!)
Membership in $\cI_{2d}$ can be expressed as a linear system on the coefficients in the linear combination. This yields the following approximation to \cref{eq:ncSOS}:
\beq\label{eq:weakncSOS}
\begin{split}
\bunderline{\nu}_d(h)=\inf & \; \nu \\
\text{s.t.} & \; \Gamma \succeq0\\
&\; \nu-h-V_d(n)^*\Gamma V_d(n)\in\cI_{2d}.
\end{split}
\eeq 
Since $\cI_{2d}\subsetneq\cI\cap\free{\C}{s}_{2d}$ in general,
$\bunderline{\nu}_{d}(h)$ 
of \cref{eq:weakncSOS}
can be strictly greater than
${\nu}_d(h)$ of \cref{eq:ncSOS}. 
\qed
\end{rmk}

Typically to solve a noncommutative sos SDP problem one applies standard duality and solves the associated dual SDP. We give its interpretation as an SDP involving pseudomoments.
Following a standard Lagrangian duality argument, the dual SDP to the one in \cref{eq:ncSOS} is 
\beq\label{eq:ncSOS'}
\begin{split}
\rotatebox[origin=c]{180}{$\nu$}_d(h) = \sup & \; L( h) \\
\text{s.t.} & \; L \in (\SOS_{2d}{}+\cI)^\vee\\
&\; L(1)=1.
\end{split}
\eeq
Here $(\SOS_{2d}{}+\cI)^\vee$ denotes the dual cone to the cone $ \SOS_{2d}{} + \cI$, 
\[
(\SOS_{2d}{}+\cI)^\vee = 
\big\{
L: \free{\C}{s}_{2d} \to \C \mid  L \text{ $*$-linear   
    with } L(\SOS_{2d})\subseteq\RR_{\geq0},  \ L(\cI\cap\free{\C}{s}_{2d} )=\{0\} \big\}.
\]
As above, by replacing $\cI$ in \cref{eq:ncSOS'} by its truncation $\cI_{2d}$, we obtain the approximation
\beq\label{eq:weakncSOS'}
\begin{split}
\rotatebox[origin=c]{180}{$\bunderline{\nu}$}_d(h) = \sup & \; L( h) \\
\text{s.t.} & \; L \in (\SOS_{2d}{}+\cI_{2d})^\vee\\
&\; L(1)=1.
\end{split}
\eeq

\begin{rmk}
Weak duality always holds for a primal-dual SDP pair, i.e., 
\beq\label{eq:weakduality}
\nu_d(h) \geq 
\rotatebox[origin=c]{180}{$\nu$}_d(h),
\eeq
and under natural mild conditions, strong duality holds, that is,
we have equality in \cref{eq:weakduality}.
This holds, for example, in the presence of so-called Slater points, a condition that is satisfied for $\cI=\sidl$, cf.~\Cref{prop:slater} below. 
\qed
\end{rmk}

\def\nuflip{{\rotatebox[origin=c]{180}{$\nu$}
}}

Consider 
the pseudomoment matrix pattern $\mmp$ with symbolic entries
$$\mmp_d ^ n:= V_d(n) V_d(n)^*.$$

\begin{lem}
For $L:\free
{\C}s_{2d}\to\C$ $*$-linear
, 
$L(\SOS_{2d})\subseteq\RR_{\geq0}$ iff
$(I\otimes L)(\mmp_d ^n )\succeq0$.
\end{lem}

\begin{proof}
For $p\in\SOS_{2d}$, assume $p=\sum_j p_j^*p_j$, and write each
$p_j=V_d(n)^*\vec{p}_j $, where $\vec p_j$ is the (column) vector of coefficients of $p_j$. Then
\[
L(p)=\sum_j L(p_j^*p_j) = \sum_j L(\vec p_j^{\;*} V_d(n)V_d(n)^* \vec p_j) = \sum_j  \vec p_j^{\;*} (I\otimes L)(\mmp_d(n)) \vec p_j,
\]
from which the conclusion follows.
\end{proof}

We call the entries of $\cM_d(L):=(I\otimes L)(\mmp_d ^n)$  the pseudomoments (of degree $\leq2d$) of $L$. 
We can now rewrite \cref{eq:ncSOS'} as an SDP as follows:
\beq\label{eq:ncSOS''}
\begin{split}
\rotatebox[origin=c]{180}{$\nu$}_d(h) = \sup{} & \;  L(h)  \\
\text{s.t.} & \; \cM_d(L)\succeq0\\
&\; \cM_d(L)_{1,1}=1 \\
&\; L(\cI\cap\free{\C}{s}_{2d} )=\{0\}.
\end{split}
\eeq
The objective function $L(h)$ can be equivalently presented as
\beq\label{eq:gramObj}
L(h)=\langle \cM_d(L),\Gamma_h \rangle,
\eeq
where $\Gamma_h$ is a (not necessarily positive semidefinite) Gram matrix for $h$, i.e., $h=V_d(n)^*\Gamma_hV_d(n)$. We note that the right-hand side of \cref{eq:gramObj} is independent of the Gram representation 
$\Gamma_h$ chosen.

\begin{rmk} \label{rmk:real_valued_mm}\label{Note:real_linear_functionals} 
In general the supremum above, e.g.~in \cref{eq:ncSOS'} is taken  all $*$-linear functionals $L: \free
{\C}s_{2d} \rightarrow \mathbb{C}$. However in the case where $h^* = h\in\free{\RR} s$ and $\cI^* = \cI$ is generated by polynomials in $\free{\RR} s$, we can simplify this supremum and only consider $*$-linear functionals $L : \free
{\RR}s_{2d} \rightarrow \mathbb{R} $. To see why, assume we are in this case and note that for any function $L$ achieving this supremum, we also have that the function $L'$ defined by $L'(p) = \frac{1}{2}\left( L(p) + \overline{L(p)}\right)$ also achieves this supremum. Since $L': \free
{\RR}s_{2d} \rightarrow \mathbb{R} $ this simplification is justified. \qed
\end{rmk}

\begin{ex}
\label{ex:las1}
Consider the case where $\cI =\{0\}$. In this case the constraint 
$L(\cI\cap\free{\C}{s}_{2d} )=\{0\}$ is automatically satisfied.
\\
Take $n=3$, $d=1$.
Then
$V_1(3)= ( 1, s_{12}, s_{13}, s_{23})^*$
and an arbitrary degree one polynomial $p\in\free{\C}s$ has the form
\begin{align} 
p =  p_0+ p_1 s_{12} + p_2 s_{13} + p_ 3 s_{23} =
    V_1(3)^* \vec p \quad \text{with} \quad \vec p: = (p_0, p_1, p_2, p_3)^T \ \text{and}
\\
 p^*p = \vec p^{\;*} \cM_1^3 \vec p
 \quad\text{ with }\quad 
\mmp_1^3 = 
\bem 1 & s_{12} &  s_{13} & s_{23} \\
s_{12} & s_{12} ^2 & s_{12} s_{13} & s_{12} s_{23}\\
s_{13} & s_{13} s_{12} & s_{13} ^2 & s_{13}  s_{23}\\
s_{23} & s_{23} s_{12} & s_{23}  s_{13} & s_{23} ^2
\eem
\end{align}
If $L:\free{\RR}s_{2}\to\mathbb{R}$ is $*$-linear, then
\[
\begin{split}
\cM_1(L) & =
\bem
L(1) & L(s_{12}) & L( s_{13}) & L(s_{23}) \\
L(s_{12}) & L(s_{12} ^2) & L(s_{12} s_{13}) & L(s_{12} s_{23})\\
L(s_{13}) & L(s_{13} s_{12}) & L(s_{13} ^2) & L(  s_{13} s_{23})\\
L(s_{23}) & L(s_{23} s_{12}) & L(s_{23}  s_{13}) & L(s_{23} ^2)
\eem
\\
& =
\bem L(1) & L(s_{12}) & L( s_{13}) & L(s_{23}) \\
L(s_{12}) & L(s_{12} ^2) & L(s_{12} s_{13}) & L(s_{12} s_{23})\\
L(s_{13}) & L(s_{12} s_{13} ) & L(s_{13} ^2) & L(  s_{13} s_{23})\\
L(s_{23}) & L( s_{12}s_{23}) & L( s_{13}s_{23} ) & L(s_{23} ^2)
\eem 
\end{split}
\]
is a self-adjoint matrix.

\bs 
Consider the Hamiltonian given by the polynomial
$h= s_{12} + s_{13} + s_{23}$.
The classical first relaxation~\cref{eq:ncSOS''} when the ideal $\cI$
is $\{0\}$ (noting that we can restrict to linear functionals $L : \free{\RR}s_{2d} \rightarrow \mathbb{R}$ by \Cref{Note:real_linear_functionals}) becomes:
\begin{align*}
\nuflip_1(h)= 
\max & \;  \ell_{12} + \ell_{13} + \ell_{23}\\
\text{s.t. } & \; \\[-.5cm]
& \bem 1 & \ell_{12} & \ell_{13} & \ell_{23} \\
\ell_{12} & \ell_{12,12} & \ell_{12,13} & \ell_{12,23}\\
\ell_{13} & \ell_{12,13} & \ell_{13,13} & \ell_{13,23}\\
\ell_{23} & \ell_{12,23} & \ell_{13,23} & \ell_{23,23}
\eem \succeq 0\\
 = \infty \phantom{=} &
\end{align*}
where $\ell_{12}, \ell_{13},  \ell_{23}, \ell_{12,12}, \ell_{12,13}, \ell_{12,23}, \ell_{13,23} \in \mathbb{R} $ are arbitrary, and should be thought of as giving the value of the linear functional $L$ applied to the variables $s_{ij}$ and monomials in them in the matrix $\cM_1(L)$ above. 

This gives an upper bound on the eigenvalues of 
the sum of three (arbitrary) self-adjoint matrices which is vacuous.
However, we soon see other ideals where the first relaxation bound has substance, for example
in \Cref{ex:sym_las}.

\end{ex}

\begin{ex}\label{ex:sym_las}

We repeat \Cref{ex:las1} but with $\cI$ changed. We now take $\cI = \cI^{S_n}$, which we define to be the ideal generated by the elements
\begin{align}
\label{eq:sym3}
     s_{ij}^2 - 1, \qquad 
s_{ij} s_{jk} - s_{ik} s_{ij},
\qquad
s_{ij}s_{kl}-s_{kl}s_{ij}
 \qquad i,j,k,l \text{ all distinct}
\end{align}
which we can think of as enforcing the defining relations 
\begin{align} \label{eq:sym_id}
s_{ij}^2 = 1, \qquad 
s_{ij} s_{jk} = s_{ik} s_{ij},
\qquad
s_{ij}s_{kl}=s_{kl}s_{ij}
 \qquad i,j,k,l \text{ all distinct}
\end{align}
of the symmetric group $\symgpn$. We will see in a moment how enforcing the condition $L(\cI\cap\free{\C}{s}_{2d} )=\{0\}$ leads to linear constraints on the pseudomoment matrix pattern.

We begin with the pseudomoment matrix pattern
\begin{align}\label{eq:M13}
\mmp_1^3 = 
\bem 1 & s_{12} &  s_{13} & s_{23} \\
s_{12} & s_{12} ^2 & s_{12} s_{13} & s_{12} s_{23}\\
s_{13} & s_{13} s_{12} & s_{13} ^2 & s_{13}  s_{23}\\
s_{23} & s_{23} s_{12} & s_{23}  s_{13} & s_{23} ^2
\eem    
\end{align}
obtained in \Cref{ex:las1}. Now viewing these elements as elements of the algebra $\free{\C}{s} / \cI^{S_n}$ (or, equivalently, applying the identities given in \Cref{eq:sym_id} and recalling that $s_{ij} = s_{ji}$ for any~$i,j$) we see this pseudomoment matrix pattern is equivalent to the pseudomoment matrix pattern  
\begin{align}
\mmp_1^3 = 
\bem 1 & s_{12} &  s_{13} & s_{23} \\
s_{12} & 1 & s_{12} s_{13} & s_{13} s_{12}\\
s_{13} & s_{13} s_{12} & 1 & s_{12}  s_{13}\\
s_{23} & s_{12} s_{13} & s_{13}  s_{12} & 1
\eem    
= 
\bem 1 & s_{12} &  s_{13} & s_{23} \\
s_{12} & 1 & s_{12} s_{13} & (s_{12} s_{13})^* \\
s_{13} & (s_{12} s_{13})^* & 1 & s_{12}  s_{13}\\
s_{23} & s_{12} s_{13} & (s_{12}  s_{13})^* & 1
\eem    
\end{align}

Writing $\cM^3_1$ in this way we see that the first relaxation is:
\beq
\begin{aligned}\label{eq:wasteSDP}
\nuflip_1(h)= 
\max & \;  \ell_{12} + \ell_{13} + \ell_{23}\\
\text{s.t. } & \; \\[-.5cm]
& \bem 1 & \ell_{12} & \ell_{13} & \ell_{23} \\
\ell_{12} & 1 & \ell_{12,13} & (\ell_{12,13})^*\\
\ell_{13} & (\ell_{12,13})^* & 1 & \ell_{12,13}\\
\ell_{23} & \ell_{12,13} & (\ell_{12,13})^* & 1
\eem \succeq 0.\\
\end{aligned}
\eeq
Where we can take $\ell_{12}, \ell_{13}, \ell_{23},\ell_{12,13}$ to be real valued by \Cref{rmk:real_valued_mm}. But then we also have $\ell_{12,13} = (\ell_{12,13})^* $ and the relaxation simplifies slightly further to
\begin{align*}
\nuflip_1(h)= 
\max & \;  \ell_{12} + \ell_{13} + \ell_{23}\\
\text{s.t. } & \; \\[-.5cm]
& \bem 1 & \ell_{12} & \ell_{13} & \ell_{23} \\
\ell_{12} & 1 & \ell_{12,13} & \ell_{12,13}\\
\ell_{13} & \ell_{12,13} & 1 & \ell_{12,13}\\
\ell_{23} & \ell_{12,13} & \ell_{12,13} & 1
\eem \succeq 0\\
 = 3. \phantom{=\ } &
\end{align*}
We refer the reader to \Cref{appd:Ex38} for the second relaxation.

\end{ex}

In the example above we simplified the pseudomoment matrix pattern in a somewhat ad-hoc manner. In that case it is not too difficult to verify that that simplification was complete, i.e. that there were no linear constraints amongst entries of the pseudomoment matrix pattern that were missed when translating to the SDP. But for larger moment matrices or more complicated ideals it is not always obvious when the pseudomoment matrix pattern is fully reduced. In the next section (\Cref{sssec:Finding_Linear_Constraints}) we will discuss some techniques that can be used to find all the linear constraints amongs entries of an SDP that are enforced by a given ideal. 

\subsubsection{Finding Linear Constrains in the Moment Matrix}

\label{sssec:Finding_Linear_Constraints}

We now illustrate some terminology and discuss possible approaches to finding all the linear constraints in the SDP computing the $d$th relaxed value $\rotatebox[origin=c]{180}{$\nu$}_d(h)$ of some element $h \in \free{\C}{s} / \cI$ when working with a general $*$-ideal $\cI$.
For concreteness, we will frequently refer back to the example in the previous section where we computed these linear constraints at level $d=1$ for the ideal $\cI^{S_3}$.

There are two general approaches that will be discussed in this paper.

\ben[\rm(1)]
\item 
\label{it:symRules}
{\it Equations become rules.}
Every element of $\cI$ can be thought of as enforcing equality between sums of monomials. A particularly nice case is when the equality is just between pairs of monomials -- as occurred in~\Cref{eq:sym_id}. 
Given two equivalent monomials, we  want to 
pick a suitable representative. 
We can use the grlex monomial order of \Cref{sec:monorder} to do this: 
the larger monomial in  grlex  order are replaced by the smaller.
This procedure 
applied to the equalities in \Cref{eq:sym_id} yields the following  ``replacement rules''
\begin{align}
\label{eq:S3GB}
s_{ij}^2 \to 1,                  \quad
s_{23} s_{13} \to s_{12} s_{23}, \quad
s_{13} s_{12} \to s_{12} s_{23}, \\
s_{23} s_{12} \to s_{12} s_{13}, \quad
s_{13} s_{23} \to s_{12} s_{13}.
\end{align}
 When we apply the rules to a polynomial $p$, for example,
 take
 $p= s_{23} s_{12}  s_{13}^2 + 
  s_{12}^2 
 $, then
 we get another polynomial 
 $\tilde p= s_{12} s_{13} + 1$,
 having the key property $p- \tilde p \in \cI^{S_n}$.
Applying these rules to the monomials in the initial moment matrix pattern of \Cref{ex:sym_las} clearly gives 
$\mmp_3^1$ of \cref{eq:M13} above.
 
 For general ideals and their defining equations (or generators)
 we can similarly associate a list grlex based of rules.
Then for some $p$, after applying rules,  $\tilde p$ may depend
 on the order in which the rules are applied and even worse 
 $p \in \cI$ might not reduce to $\tilde p =0$.
 Later, in \Cref{sec:GB}, 
 we show how this problem can be fixed
 by selecting rules corresponding to 
 a set of generators for $\cI$
 called a  Gr\"obner Basis (GB).
 Indeed the rules \Cref{eq:S3GB} do 
correspond to a GB for $S_3$ w.r.t.~the grlex monomial order. A longer discussion of Gr\"obner Basis can be found in \Cref{sec:GB}.

 \item {\it Linear algebra basis.} 
 \label{it:symLinBasis}
 An alternate approach is possible if a good understanding of the quotient algebra
 $\cA^{\cI^{S_3}}_3 := \free{\C}{s}/\cI^{S_3} $ is available.
 Namely, if a linear basis for the quotient space can be obtained (at least for images of polynomials up to a certain degree), then simple linear algebra allows us to form (lower) levels of the relaxation hierarchy.

For instance, a basis for the image of $\free{\C}{s}_{2} $ in
$\cA^{\cI^{S_3}}_3$
is
$$ 1,\ s_{12} , \ s_{13}, \
 s_{12}s_{13}, \ s_{13} s_{12},
 $$
 since all monomials in $\free{\C}{s}_{2} $
 are a linear combination of these modulo 
 the $\cI^{S_3}$,
 as can be derived from \Cref{eq:sym3}.
Thus
 producing a  moment matrix pattern for the ideal $S_3$ is done by using 
 the trivial linear expansion of monomials in terms of this basis,
 $$
s_{23} s_{13} = s_{12} s_{23}, \quad
s_{13} s_{12} = s_{12} s_{23}, \quad 
s_{23} s_{12} = s_{12} s_{13}, \quad
s_{13} s_{23} = s_{12} s_{13}.
 $$
 \een 

 The symmetric group $S_3$  is so simple that  the distinction between the rule approach and the linear basis approach is merely semantic. 
 However, for swaps the difference  has substance. 
 Indeed,
\Cref{sec:swap_ncSoSd4} takes the linear basis approach
while \Cref{sec:GB} takes the Gr\"obner Basis approach.

\ssec{Non-Commutative Sum Of Squares Hierarchy for the Swap algebra}
\label{sec:swap_ncSoSd4}

In this section we restrict our attention to the swap algebra. We begin with a brief discussion of convergence of Sum of Squares relaxations in this algebra. Then we turn our attention to implementations of the $1$st and $2$nd relaxations in the swap algebra using linear algebraic bases. We begin in \Cref{sssec:first_relaxation_swaps} with an example of the $1$st relaxation using a linear algebraic basis, then in \Cref{sssec:linear_basis_degree_2_swaps}, construct a linear algebraic basis for monomials of degree at most two in the swap algebra. In \Cref{sssec:2nd_swap_relax} we discuss an implementation of the $2$nd swap relaxation using a linear algebraic basis for swap algebra monomials of degree at most four which we construct in \Cref{ssec:swap34}. We end this section with a discussion of some numerical results obtained using the $2$nd swap relaxation.

\begin{prop}\label{prop:slater}
Strong duality holds between the primal-dual pair of SDPs in \cref{eq:ncSOS} and \cref{eq:ncSOS''}.
\end{prop}

\begin{proof}
It suffices to find a strictly feasible point $L$ for the dual SDP in \cref{eq:ncSOS''}. Equivalently, a strictly positive linear functional $\ell$ on the Swap algebra $\salg_n$, or equivalently, on $\swa_n$. But this is easy: simply take the trace.
\end{proof}

\begin{thm}\label{eq:lassconverges}
For $h=h^*\in\salg_n$,
the sequences $\nu_d(h)$ and $\rotatebox[origin=c]{180}{$\nu$}_d(h)$ 
defined in \cref{eq:ncSOS} and \cref{eq:ncSOS''}
converge monotonically to $\maxeig(h(\sw))$. More precisely,
for $d\geq \lceil \frac n2\rceil$,
\[
\nu_d(h)= \maxeig(h(\sw))=
\rotatebox[origin=c]{180}{$\nu$}_d(h).
\]
\end{thm}

\begin{proof}
The convergence statement follows from the Helton-McCullough Positivstellensatz \cite{HM04} since the algebra
$\salg_n$ is finite-dimensional and hence archimedean.
The convergence is finite by \Cref{thm:swapDegree}.
\end{proof}

\sssec{First relaxation for the Swap algebra}
\label{sssec:first_relaxation_swaps}
We now present in some detail the SDP arising from the first
noncommutative sum of squares relaxation for the Swap algebra. We proceed as in \Cref{ex:las1} and \Cref{ex:sym_las} and define
$V_1(n)=(1, s_{12}, s_{13},\ldots, s_{1n}, s_{23},\ldots,s_{n-1\,n})^*$ and consider
$\cM_1(L)$ for some $*$-linear functional $L:\free{\C}s_{2}\to\C$
which satisfies $L(\free{\RR}s_{2})\subseteq\RR$.

To encode the constraint $L(\sidl\cap \free{\C}s_2)=\{0\}$, we need to understand the linear space spanned by monomials of degree at most two in the swap algebra. We construct a basis $\cB_2$ for this space together with expansions for all the other products in terms of this basis to encode \cref{eq:ncSOS''}. This follows the approach outined in~\Cref{it:symLinBasis} of~\Cref{sssec:Finding_Linear_Constraints}.

\begin{ex}\label{ex:las1b}
 Swap Algebra, $n=3$.
A basis $\cB_2$ for
the linear space spanned by monomials of degree at most two in the swap algebra
is given by the entries of $V_1(3)$ together
with one element, e.g., $s_{12}s_{13}$. All the other quadratic terms in the $s_{ij}$ can be expressed with these as follows:
\[
\begin{split}
s_{ij}^2&=1\\
s_{12}s_{23} & = -1+s_{12}+s_{13}+s_{23}-s_{12}s_{13}\\
s_{13}s_{12} & = -1+s_{12}+s_{13}+s_{23}-s_{12}s_{13} \\
s_{13}s_{23} & = s_{12}s_{13}\\
s_{23}s_{12} & = s_{12}s_{13}\\
s_{23}s_{13} & = -1+s_{12}+s_{13}+s_{23}-s_{12}s_{13}
\end{split}
\]
With this 
a unital $*$-linear 
$L$ with $L(\sidl\cap \free{\C}s_2)=\{0\}$
is determined by 3 real numbers $\ell_{ij}=L(s_{ij})$
and a complex number $q=L(s_{12}s_{13})$.
Thus $\cM_1(L)$ 
simplifies into
\[
\begin{split}
\cM_1(L) & =
\bem L(1) & L(s_{12}) & L( s_{13}) & L(s_{23}) \\
L(s_{12}) & L(s_{12} ^2) & L(s_{12} s_{13}) & L(s_{12} s_{23})\\
L(s_{13}) & L(s_{12} s_{13} )^* & L(s_{13} ^2) & L(  s_{13} s_{23})\\
L(s_{23}) & L( s_{12}s_{23})^* & L( s_{13}s_{23} )^* & L(s_{23} ^2)
\eem
 \\
 &=
\bem 1 & \ell_{12} & \ell_{13} & \ell_{23} \\
\ell_{12} & 1 & q & -1+\ell_{12}+\ell_{13}+\ell_{23}-q \\
\ell_{13} & q^* & 1 & q\\
\ell_{23} & -1+\ell_{12}+\ell_{13}+\ell_{23}-q^* &q^* & 1
\eem.
\end{split}
\]
Conversely, each $\cM_1(L)$ of this form yields a unital $*$-linear $L$ 
with $L(\sidl\cap \free{\C}s_2)=\{0\}$.
This 
makes it straightforward to write down the SDP of \cref{eq:ncSOS''}
for the objective function $L(h)$ for $h=\ell_{12} + \ell_{13} + \ell_{23}$ in this case.
\begin{align*}
\max & \;  \ell_{12} + \ell_{13} + \ell_{23}\\
\text{s.t. } & \; \\[-.5cm]
&\bem 1 & \ell_{12} & \ell_{13} & \ell_{23} \\
\ell_{12} & 1 & q & -1+\ell_{12}+\ell_{13}+\ell_{23}-q \\
\ell_{13} & q & 1 & q\\
\ell_{23} & -1+\ell_{12}+\ell_{13}+\ell_{23}-q &q & 1
\eem \succeq 0\\
 \phantom{=\ } & = 3.
\end{align*}
Here $ \ell_{ij}\in\RR$, 
and as explained in \Cref{Note:real_linear_functionals}, we have without loss of generality assumed that $q=L(s_{12}s_{13})\in\RR$.
\end{ex}
The examples illustrate SDP relaxations  using
linear algebra constraints to capture $\sidl$ when $n=3$.
The rest of this section and related appendices 
give the machinery needed, 
namely $\cB_3, \cB_4$,  for any $n$ and 
for $2${nd} swap relaxations.

\sssec{Linear space spanned by the products of at most two swap matrices}

\label{sssec:linear_basis_degree_2_swaps}

\begin{prop}\label{prop:swap2}
A basis $\cB_2$ for
the linear space spanned by monomials of degree at most two in the swap algebra $\swa_n$
is given by 
\[
\begin{split}
 I  &  \\ 
 \sw_{ij} & \quad  i<j \\
 \sw_{ij}\sw_{ik} & \quad i<j<k\\
 \sw_{ij}\sw_{k\ell} & \quad  i<j,\; i<k<\ell. 
\end{split}
\]
\end{prop}

\begin{proof}
Let us first prove the spanning property ob $\cB_2$. 
Since two swap matrices with disjoint indices commute,
it suffices to consider products of two swap matrices, where one of the indices repeats. Letting $i<j<k$, there are six such products, namely
\begin{align*}
\begin{autobreak}
\sw_{ij}\sw_{ik}, \;
\sw_{ik}\sw_{ij}, \; 
\sw_{ij}\sw_{jk}, \;
\sw_{jk}\sw_{ij}, \;  
\sw_{ik}\sw_{jk},\;
\sw_{jk}\sw_{ik}.
\end{autobreak}
\end{align*}
The first of these has been included in $\cB_2$. The second product can be expressed with the first one and linear terms in $\sw_{pq}$ using \cref{eq:Btri}. Further, by a routine calculation we have\footnote{While the identities given in this section can be easily verified since they involve matrices of small to moderate size, we have in fact discovered many of them with the help of noncommutative Gr\"obner bases; cf.~\Cref{sec:GB} below.}
\[
\begin{split}
\sw_{ij}\sw_{jk} & = -1 +\sw_{ij}+\sw_{ij}+\sw_{jk}- \sw_{ij}\sw_{ik} \in \spn \cB_2 \\
\sw_{jk}\sw_{ij} & = \sw_{ij}\sw_{ik}\in \cB_2 \\
\sw_{ik}\sw_{jk} & = \sw_{ij}\sw_{ik}\in\cB_2\\
\sw_{jk}\sw_{ik} & =  -1 +\sw_{ij}+\sw_{ij}+\sw_{jk}- \sw_{ij}\sw_{ik}\in \spn\cB_2 .
\end{split}
\]

We now turn to the linear independence of $\cB_2$. Assume there are scalars $\alpha,\beta_{ij},\gamma_{ijk},\delta_{ijk\ell}$ satisfying
\beq\label{eq:linDep}
\alpha I + \sum_{i<j} \beta_{ij}  \sw_{ij} 
+ \sum_{i<j<k} \gamma_{ijk} \sw_{ij}\sw_{ik} +
\sum_{\substack{i<j\\[.5mm] i<k<\ell}} \delta_{ijk\ell} \sw_{ij}\sw_{k\ell} =0.
\eeq
Before expanding the left-hand side of   \cref{eq:linDep} in terms of the Pauli $\sigma_W$, note that for $i<j<k$,
\[
\begin{split}
4 \sw_{ij}\sw_{ik} & = 1 + \sigma_X^i  \sigma_X^j + \sigma_X^i  \sigma_X^k + \sigma_X^j  \sigma_X^k + \sigma_Y^i  \sigma_Y^j + \sigma_Y^i  \sigma_Y^k + \sigma_Y^j  \sigma_Y^k + 
 \sigma_Z^i  \sigma_Z^j + \sigma_Z^i  \sigma_Z^k + \sigma_Z^j  \sigma_Z^k \\ 
& \phantom{={}} + i \big( \sigma_X^i  \sigma_Y^j  \sigma_Z^k - 
  \sigma_X^i  \sigma_Z^j  \sigma_Y^k -  \sigma_Y^i  \sigma_X^j  \sigma_Z^k +  \sigma_Y^i  \sigma_Z^j  \sigma_X^k + 
  \sigma_Z^i  \sigma_X^j  \sigma_Y^k -  \sigma_Z^i  \sigma_Y^j  \sigma_X^k \big).
\end{split}
\]
Likewise, letting $i<j<k<\ell$, we have
\[
\begin{split}
4 \sw_{ij}\sw_{k\ell} & = 
1 + \sigma_X^i  \sigma_X^j + \sigma_X^k  \sigma_X^\ell + \sigma_Y^i  \sigma_Y^j + \sigma_Y^k  \sigma_Y^\ell + \sigma_Z^i  \sigma_Z^j + \sigma_Z^k  \sigma_Z^\ell \\
 & \phantom{={}}  + 
 \sigma_X^i  \sigma_X^j  \sigma_X^k  \sigma_X^\ell + \sigma_X^i  \sigma_X^j  \sigma_Y^k  \sigma_Y^\ell + \sigma_X^i  \sigma_X^j  \sigma_Z^k  \sigma_Z^\ell + 
 \sigma_Y^i  \sigma_Y^j  \sigma_X^k  \sigma_X^\ell \\
 & \phantom{={}}   + \sigma_Y^i  \sigma_Y^j  \sigma_Y^k  \sigma_Y^\ell + \sigma_Y^i  \sigma_Y^j  \sigma_Z^k  \sigma_Z^\ell + 
 \sigma_Z^i  \sigma_Z^j  \sigma_X^k  \sigma_X^\ell + \sigma_Z^i  \sigma_Z^j  \sigma_Y^k  \sigma_Y^\ell + \sigma_Z^i  \sigma_Z^j  \sigma_Z^k  \sigma_Z^\ell.
\end{split}
\]
If $i<k<j<\ell$, then
\[
\begin{split}
4 \sw_{ij}\sw_{k\ell} & = 
1 + \sigma_X^i  \sigma_X^j + \sigma_X^k  \sigma_X^\ell + \sigma_Y^i  \sigma_Y^j + \sigma_Y^k  \sigma_Y^\ell + \sigma_Z^i  \sigma_Z^j + \sigma_Z^k  \sigma_Z^\ell  \\
 & \phantom{={}}  + 
 \sigma_X^i  \sigma_X^k  \sigma_X^j  \sigma_X^\ell + \sigma_X^i  \sigma_Y^k  \sigma_X^j  \sigma_Y^\ell + \sigma_X^i  \sigma_Z^k  \sigma_X^j  \sigma_Z^\ell + 
 \sigma_Y^i  \sigma_X^k  \sigma_Y^j  \sigma_X^\ell \\
 & \phantom{={}}   + \sigma_Y^i  \sigma_Y^k  \sigma_Y^j  \sigma_Y^\ell + \sigma_Y^i  \sigma_Z^k  \sigma_Y^j  \sigma_Z^\ell + 
 \sigma_Z^i  \sigma_X^k  \sigma_Z^j  \sigma_X^\ell + \sigma_Z^i  \sigma_Y^k  \sigma_Z^j  \sigma_Y^\ell + \sigma_Z^i  \sigma_Z^k  \sigma_Z^j  \sigma_Z^\ell
\end{split}
\]
Finally, when $i<k<\ell<j$, then
\[
\begin{split}
4 \sw_{ij}\sw_{k\ell} & = 
1 + \sigma_X^i  \sigma_X^j + \sigma_X^k  \sigma_X^\ell + \sigma_Y^i  \sigma_Y^j + \sigma_Y^k  \sigma_Y^\ell + \sigma_Z^i  \sigma_Z^j + \sigma_Z^k  \sigma_Z^\ell \\
 & \phantom{={}}   + 
 \sigma_X^i  \sigma_X^k  \sigma_X^\ell  \sigma_X^j + \sigma_X^i  \sigma_Y^k  \sigma_Y^\ell  \sigma_X^j + \sigma_X^i  \sigma_Z^k  \sigma_Z^\ell  \sigma_X^j + 
 \sigma_Y^i  \sigma_X^k  \sigma_X^\ell  \sigma_Y^j  \\
 & \phantom{={}}  + \sigma_Y^i  \sigma_Y^k  \sigma_Y^\ell  \sigma_Y^j + \sigma_Y^i  \sigma_Z^k  \sigma_Z^\ell  \sigma_Y^j + 
 \sigma_Z^i  \sigma_X^k  \sigma_X^\ell  \sigma_Z^j + \sigma_Z^i  \sigma_Y^k  \sigma_Y^\ell  \sigma_Z^j + \sigma_Z^i  \sigma_Z^k  \sigma_Z^\ell  \sigma_Z^j
\end{split}
\]

We are now ready to expand \cref{eq:linDep}. 
We shall make heavy use of the Pauli basis as given in \cref{eq:pauliBasis}.
Consider the term 
$\sw_{ij}\sw_{k\ell}
$ next to $\delta_{ijk\ell}$. If $i<j<k<\ell$, then the expansion
will contain $\sigma_X^i  \sigma_X^j  \sigma_Y^k  \sigma_Y^\ell $,
a term that cannot appear anywhere else in \cref{eq:linDep}.
If $i<k<j<\ell$, then the expansion contains $ \sigma_X^i  \sigma_Y^k  \sigma_X^j  \sigma_Y^\ell $, again a term that cannot appear anywhere else in \cref{eq:linDep}. Finally, if $i<k<\ell<j$,
then the same conclusion holds for 
$\sigma_X^i  \sigma_Y^k  \sigma_Y^\ell  \sigma_X^j $.
Thus $\delta_{ijk\ell}=0$ for all $i,j,k,\ell$.

We next consider $\gamma_{ijk}\sw_{ij}\sw_{ik}$.
Here the expansion contains $\sigma_X^i  \sigma_Y^j  \sigma_Z^k$ that is again a term that cannot appear elsewhere in \cref{eq:linDep}. Thus all $\gamma_{ijk}=0$.  

Similar but easier reasoning will also imply $\beta_{ij}=0$ and finally $\alpha=0$ proving linear independence of $\cB_2$.
\end{proof}

\begin{rmk}
The set $\cB_2$ has 
\[
1+{n \choose 2}+{n \choose 3}+3 {n\choose 4} = \frac{1}{24} \left(3 n^4-14 n^3+33 n^2-22 n+24\right)
\]
elements. \qed
\end{rmk}

\sssec{Second relaxation for the Swap algebra}
\label{sssec:2nd_swap_relax}

We continue our discussion of relaxation by describing the second relaxation.
This requires sets of monomials $\cB_3, \cB_4$ (to be described later in \Cref{ssec:swap3,sssec:linswap4} which are degree 3 and 4 analogs of $\cB_2$.

In order to build the second relaxation, we replace $V_2(n)$ 
by the vector $\vec\cB_2$ consisting of the swap basis $\cB_2$ determined in \Cref{prop:swap2}.
The corresponding symbolic pseudomoment matrix $\vec\cB_2\vec\cB_2^*$ will
contain products of up to four swaps. To form the SDPs in \cref{eq:ncSOS} or \cref{eq:ncSOS''} we  need to understand the linear space spanned by all products of up to four Swap matrices,
a topic we describe in some detail below in \Cref{ssec:swap34}.

\begin{algorithm}[ht]
 \SetKwInOut{Input}{Input}\SetKwInOut{Output}{Output}
\Input{Graph $G=(V,E)$ on $n$ vertices}
\Output{Solution to the $2$nd relaxation of quantum max-cut for $\hampol G$.
Alternately, ``upper bound on the minimum eigenvalue of the QMC Hamiltonian''
}
\BlankLine

Form $\vec\cB_2$\;
Form the  symbolic pseudomoment matrix $\cM_2:=\vec\cB_2\vec\cB_2^*$\;
Express each entry in $\cM_2$ as a linear combination of elements of $\cB_4$ (see \Cref{sssec:linswap4}) to obtain $\cM_2'$\;
Replace each distinct term appearing in $\cM_2'$ with a new (scalar) variable; call the resulting matrix $\cM_2''$\;
Solve the SDP
    \beq
\label{eq:pauliSOS2}
\rotatebox[origin=c]{180}{$\nu$}_2(\hampol G) = \sup \{ \langle \cM_2'',\Gamma_G\rangle \mid 
\cM_2''\succeq0,\;
(\cM_2'')_{1,1}=1\},
 \eeq
where $\Gamma_G$ is any matrix satisfying $\hampol G=\vec\cB_2^* \Gamma_G\vec\cB_2$\;
\BlankLine
\KwRet{\rotatebox[origin=c]{180}{$\nu$}$_2(\hampol G)$}
\caption{$2$nd relaxation of the quantum max-cut}\label{algo:polySDPswap}
 \end{algorithm}

\sssec{Swap relaxation behaves well  in experiments }
We implemented \Cref{algo:polySDPswap} in Mathematica  (using the out-of-the-box semidefinite optimization module in Mathematica) on all graphs with $\leq8$ vertices:

\begin{prop}\label{prop:lass28}
For  $n\leq 8$ the 2nd relaxed value of an $n$ vertex Quantum Max Cut Hamilton with uniform edge weights is up to the tolerance of $10^{-7}$ equal to the max eigenvalue of that Hamiltonian.  
\end{prop}

It would be interesting to find the smallest graph on which the second relaxation is not exact.

\begin{rmk}
\Cref{prop:lass28} is surprising when compared to classical (commutative) max cut relaxation which performs much worse. 

For example, the second Lasserre relaxation is not exact
for the 5 cycle. Also, while we do not give comparative statistics, the first classical relaxation is worse than 
the one for swaps, e.g., even the triangle is classically not
first Lasserre exact. \qed
\end{rmk}

\subsubsection{Segue} 

Since, clearly, it is not feasible to find linear algebra bases for linear spaces spanned by products of $d>4$ swaps by hand, we turn to Gr\"obner bases in next.

\sec{Gr\"obner Bases}\label{sec:GB}

This section gives a few observations about Gr\"obner Bases (GB)  with an eye toward computation.\looseness=-1

\ssec{Our Gr\" obner Basis Set Up}

 Our presentation is aimed at readers who have a basic familiarity  with GBs; 
a standard reference in the commutative setting
is \cite{cox15book}, 
while \cite{moraNCGB,greenNCGB,phdNCGB} describe the appropriate nc analogs.
Noncommutative GBs have properties  similar  to those of  commutative GB with the dramatic exception that a noncommutative GB might not be finite. 
However,
we start with some conceptual motivation
to provide a little context
for readers who are unfamiliar with GBs.

\sssec{Motivation for why we need Gr\"obner generators vs
any old  set of generators for an ideal}

The reader is advised to 
review \Cref{ex:sym_las}
which foreshadowed some of the ideas here.

Now we describe precisely the two big 
drawbacks of doing calculations on an ideal $\cI$
using a set of generators for $\cI$ which is too small.
Let 
$B$ be a set of generators for 
$\cI$
a two sided ideal in the free algebra  $\free{\C}s$,  and let $\cA:=   \free{\C}s/\cI$.
A basic question is: how do we tell if two 
elements $a, \tilde a$ of $\cA$ are the same?
Equivalently, do two given 
nc polynomials $p, \tilde p$ have difference
$\delta:=p -\tilde p $ in $\cI$?

The simplest naïve approach is to 
use the graded lexicographic (grlex) monomial order 
introduced in \Cref{sec:monorder}. 
Then each polynomial $b \in B$
is a weighted sum of monomials; one of these monomials is bigger than the rest in the monomial order. Denote it by $\LT(b)$ and
 construct a replacement rule, denoted $r_b$ 
via 
$$
\LT(b)  \to \ b - \LT(b)
$$
Now apply the rules $r_B$ gotten from the 
polynomials in $B$
repeatedly to a polynomial $p$
until they no longer cause any changes\footnote{The fact that we are using grlex -- an ``admissible monomial order'' -- implies that the successive application of rules will stop.};
this leaves us with a 
polynomial decomposition
$$
p=\overline p^B
+ \ \delta.
$$
where
$ \delta \in \cI
$.

All of this does not tell us if $p$
actually is in the ideal $\cI$ and the remainder $\overline p^B$ might even depend on the order in which the rules are applied. This is extremely unsatisfying.
Next, 
Gr\"obner Bases to the rescue:
 $B$ is called a \df{Gr\"obner Bases} if the remainder $\overline p^B$ of a polynomial $p$  is $0$ iff $p\in\cI$. 
Equivalently,
  $\overline p^B$ can be looked at as the remainder 
  of the division of $p$ by $B$.
  Hence, $\overline p^B$ 
  is uniquely determined, independent of how the rules  are applied, and
  is called \df{the canonical form of $p$ modulo $\cI$}
  (w.r.t. the grlex order).
So far we only have established semantics to the effect that GBs 
solve our main difficulties, but the next theorem
has impact.

\begin{thm}
    [\cite{moraNCGB}]
    A finitely generated ideal $\cI$ in $\free{\C}s$ 
    has a Gr\"oebner Basis $GB_\cI$ w.r.t.~grlex,
     which may be infinite.
\end{thm}

\begin{proof}[Comments on proof]
Loosely speaking, the noncommutative Buchberger criterion \cite[Theorem 5.1]{moraNCGB}
states that $B$ is a GB iff each $S$-polynomial
built off $B$
can be expressed in terms of the elements of $B$
with control on the grlex order of multipliers.
Thus
algorithms for building  GBs work by producing 
$S$-polynomials for pairs
$a_i,a_j$ of (not necessarily distinct)  polynomials which are generators  of $\cI$.
Here, $S$-polynomials are ones  of the form
\beq\label{eq:Spoly}
S_{i,j}(w_i,w_i',w_j,w_j')=
\frac1{\LC(a_i)}w_ia_iw_i'-
\frac1{\LC(a_j)}w_ja_jw_j'
\eeq
for words $w_i,w_i',w_j,w_j'$ in $x$ satisfying
\beq\label{eq:obstr}
w_i\LT(a_i)w_i'=
w_j\LT(a_j)w_j'.
\eeq
Here $\LC(a)$ denotes the coefficient of the leading term of $a$.

At each step 
in construction of a GB, one considers a collection of polynomials $\hat B$.
If the ``remainder'' of an $S$-polynomial
after division by $\hat B$ is nonzero, one adds it
to $\hat B$ and repeats the process.
Ultimately (maybe in an infinite number of steps)  $\hat B$
grows to $B$, a GB.
\end{proof}
 
\begin{rmk}
    \ben[\rm(1)]
    \item In 
this exposition we used the grlex monomial order. However, many (but not all) monomial orders would work in the same way with the same properties. A linear order on monomials is a \df{monomial order} if \cref{it:monOrder} holds, and is called 
admissible if \cref{it:admiss} holds as well:
\begin{enumerate}[\rm(a)]
\item\label{it:monOrder}
$m_1<m_2$ implies $m m_1m'<mm_2m'$ for all monomials $m,m',m_1,m_2$;
\item\label{it:admiss}
the given ordering is a well-ordering on the set of monomials, i.e., every descending chain of monomials becomes eventually
stationary.
\end{enumerate}
We refer the reader to \cite{moraNCGB,greenNCGB} for details.

\item
An equivalent characterization of $B$ being a GB is:
the  set
of leading terms $\LT(B)$ of elements of $B$ generates
the same monomial ideal $\LT(\cI)$ as all leading terms of $\cI$.

\item
While we do not go into this, 
given an admissible monomial order,
to 
every ideal and
we can associate a 
``reduced'' GB \cite{phdNCGB}, and these are unique. Loosely speaking, 
a reduced GB is one in which no polynomial is redundant and none of the monomials appearing in a polynomial of the GB are redundant. 

The standard GB packages output  a reduced GBs as the default.
    \item 
Any maximal set of linearly independent remainders among $\{\overline p^B\mid p\in\free{\C}s\}$ for a GB $B$ forms a 
basis for the quotient algebra $\free{\C}s/\cI$.

\item 
GB algorithms are very time and memory consuming both in the commutative and the non-commutative case.
For the complexity analysis of the state-of-the-art $F_5$ algorithm for commutative GBs see \cite{GBcomplexity}.
    \een
\qed 
\end{rmk}

\ssec{Gr\"obner Bases for
the Paulis}

The purpose of this subsection is to demonstrate Gr\"obner bases in action in a toy example, namely for the Paulis, where it leads to the expected result. However, 
the choice  of the monomial order matters as we illustrate
in \Cref{ex:complicated}.

Consider the free algebra
$\free{\C}{x_1,\ldots,x_n,y_1,\ldots,y_n,z_1,\ldots,z_n}$ and 
its quotient by the ideal $\cIp$ generated by
\beq\label{eq:pauliGB}
 \begin{gathered}
x_j^2=y_j^2=z_j^2=1,\quad x_jy_j=i z_j, \quad y_jx_j=-i z_j,\\
[u_j,w_k]=0 \quad \text{for $j\neq k$ and }u,w\in\{x,y,z\}
\end{gathered}
\eeq
for $j,k= 1,\dots, n$.
The quotient algebra
$\free{\C}{x,y,z}/\cIp$ is isomorphic to the algebra generated by the Paulis $\{\sigma_W^j\mid 1\leq j\leq n,\; W\in\{X,Y,Z,I\}\}$,
that is, to $M_{2^n}(\C)$.

\begin{ex}\label{ex:pauliGB4}
With the aid of a computer algebra system we computed the GB in the case $n=4$. 
With respect to the graded lex order with $x_1<y_1<z_1<x_2<\cdots<z_4$
it is given by the following 90 polynomials, all of which come from \cref{eq:pauliGB}:
\begin{align*}
\begin{autobreak}
\phantom{=}
-1 + x_ 1^2,\;  -1 + x_ 2^2,\;  -1 + x_ 3^2,\;  -1 + x_ 4^2,\;  
 -1 + y_ 1^2,\;  -1 + y_ 2^2,\;  -1 + y_ 3^2,\;  -1 + y_ 4^2,\;  
 -1 + z_ 1^2,\;  -1 + z_ 2^2,\;  -1 + z_ 3^2,\;  -1 + z_ 4^2,\;  
 -i z_ 1 + x_ 1 y_ 1,\;  -i z_ 2 + x_ 2 y_ 2,\;  
 -i z_ 3 + x_ 3 y_ 3,\;  -i z_ 4 + x_ 4 y_ 4,\;  i z_ 1 + y_ 1 x_ 1,\;  
 i z_ 2 + y_ 2 x_ 2,\;  i z_ 3 + y_ 3 x_ 3,\;  i z_ 4 + y_ 4 x_ 4,\;  
 x_ 1 x_ 2 - x_ 2 x_ 1,\;  x_ 1 y_ 2 - y_ 2 x_ 1,\;  x_ 1 z_ 2 - z_ 2 x_ 1,\;  
 -x_ 2 y_ 1 + y_ 1 x_ 2,\;  y_ 1 y_ 2 - y_ 2 y_ 1,\;  y_ 1 z_ 2 - z_ 2 y_ 1,\;  
 -x_ 2 z_ 1 + z_ 1 x_ 2,\;  -y_ 2 z_ 1 + z_ 1 y_ 2,\;  z_ 1 z_ 2 - z_ 2 z_ 1,\;  
 x_ 1 x_ 3 - x_ 3 x_ 1,\;  x_ 1 y_ 3 - y_ 3 x_ 1,\;  x_ 1 z_ 3 - z_ 3 x_ 1,\;  
 -x_ 3 y_ 1 + y_ 1 x_ 3,\;  y_ 1 y_ 3 - y_ 3 y_ 1,\;  y_ 1 z_ 3 - z_ 3 y_ 1,\;  
 -x_ 3 z_ 1 + z_ 1 x_ 3,\;  -y_ 3 z_ 1 + z_ 1 y_ 3,\;  z_ 1 z_ 3 - z_ 3 z_ 1,\;  
 x_ 1 x_ 4 - x_ 4 x_ 1,\;  x_ 1 y_ 4 - y_ 4 x_ 1,\;  x_ 1 z_ 4 - z_ 4 x_ 1,\;  
 -x_ 4 y_ 1 + y_ 1 x_ 4,\;  y_ 1 y_ 4 - y_ 4 y_ 1,\;  y_ 1 z_ 4 - z_ 4 y_ 1,\;  
 -x_ 4 z_ 1 + z_ 1 x_ 4,\;  -y_ 4 z_ 1 + z_ 1 y_ 4,\;  z_ 1 z_ 4 - z_ 4 z_ 1,\;  
 -x_ 1 x_ 2 + x_ 2 x_ 1,\;  x_ 2 y_ 1 - y_ 1 x_ 2,\;  x_ 2 z_ 1 - z_ 1 x_ 2,\;  
 -x_ 1 y_ 2 + y_ 2 x_ 1,\;  -y_ 1 y_ 2 + y_ 2 y_ 1,\;  y_ 2 z_ 1 - z_ 1 y_ 2,\;  
 -x_ 1 z_ 2 + z_ 2 x_ 1,\;  -y_ 1 z_ 2 + z_ 2 y_ 1,\;  
 -z_ 1 z_ 2 + z_ 2 z_ 1,\;  x_ 2 x_ 3 - x_ 3 x_ 2,\;  x_ 2 y_ 3 - y_ 3 x_ 2,\;  
 x_ 2 z_ 3 - z_ 3 x_ 2,\;  -x_ 3 y_ 2 + y_ 2 x_ 3,\;  y_ 2 y_ 3 - y_ 3 y_ 2,\;  
 y_ 2 z_ 3 - z_ 3 y_ 2,\;  -x_ 3 z_ 2 + z_ 2 x_ 3,\;  -y_ 3 z_ 2 + z_ 2 y_ 3,\;  
 z_ 2 z_ 3 - z_ 3 z_ 2,\;  x_ 2 x_ 4 - x_ 4 x_ 2,\;  x_ 2 y_ 4 - y_ 4 x_ 2,\;  
 x_ 2 z_ 4 - z_ 4 x_ 2,\;  -x_ 4 y_ 2 + y_ 2 x_ 4,\;  y_ 2 y_ 4 - y_ 4 y_ 2,\;  
 y_ 2 z_ 4 - z_ 4 y_ 2,\;  -x_ 4 z_ 2 + z_ 2 x_ 4,\;  -y_ 4 z_ 2 + z_ 2 y_ 4,\;  
 z_ 2 z_ 4 - z_ 4 z_ 2,\;  -x_ 1 x_ 3 + x_ 3 x_ 1,\;  x_ 3 y_ 1 - y_ 1 x_ 3,\;  
 x_ 3 z_ 1 - z_ 1 x_ 3,\;  -x_ 1 y_ 3 + y_ 3 x_ 1,\;  -y_ 1 y_ 3 + y_ 3 y_ 1,\;  
 y_ 3 z_ 1 - z_ 1 y_ 3,\;  -x_ 1 z_ 3 + z_ 3 x_ 1,\;  -y_ 1 z_ 3 + z_ 3 y_ 1,\;  
 -z_ 1 z_ 3 + z_ 3 z_ 1,\;  -x_ 2 x_ 3 + x_ 3 x_ 2,\;  x_ 3 y_ 2 - y_ 2 x_ 3,\;  
 x_ 3 z_ 2 - z_ 2 x_ 3,\;  -x_ 2 y_ 3 + y_ 3 x_ 2,\;  -y_ 2 y_ 3 + y_ 3 y_ 2,\;  
 y_ 3 z_ 2 - z_ 2 y_ 3,\;  -x_ 2 z_ 3 + z_ 3 x_ 2,\;  -y_ 2 z_ 3 + z_ 3 y_ 2,\;  
 -z_ 2 z_ 3 + z_ 3 z_ 2,\;  x_ 3 x_ 4 - x_ 4 x_ 3,\;  x_ 3 y_ 4 - y_ 4 x_ 3,\;  
 x_ 3 z_ 4 - z_ 4 x_ 3,\;  -x_ 4 y_ 3 + y_ 3 x_ 4,\;  y_ 3 y_ 4 - y_ 4 y_ 3,\;  
 y_ 3 z_ 4 - z_ 4 y_ 3,\;  -x_ 4 z_ 3 + z_ 3 x_ 4,\;  -y_ 4 z_ 3 + z_ 3 y_ 4,\;  
 z_ 3 z_ 4 - z_ 4 z_ 3,\;  -x_ 1 x_ 4 + x_ 4 x_ 1,\;  x_ 4 y_ 1 - y_ 1 x_ 4,\;  
 x_ 4 z_ 1 - z_ 1 x_ 4,\;  -x_ 1 y_ 4 + y_ 4 x_ 1,\;  -y_ 1 y_ 4 + y_ 4 y_ 1,\;  
 y_ 4 z_ 1 - z_ 1 y_ 4,\;  -x_ 1 z_ 4 + z_ 4 x_ 1,\;  -y_ 1 z_ 4 + z_ 4 y_ 1,\;  
 -z_ 1 z_ 4 + z_ 4 z_ 1,\;  -x_ 2 x_ 4 + x_ 4 x_ 2,\;  x_ 4 y_ 2 - y_ 2 x_ 4,\;  
 x_ 4 z_ 2 - z_ 2 x_ 4,\;  -x_ 2 y_ 4 + y_ 4 x_ 2,\;  -y_ 2 y_ 4 + y_ 4 y_ 2,\;  
 y_ 4 z_ 2 - z_ 2 y_ 4,\;  -x_ 2 z_ 4 + z_ 4 x_ 2,\;  -y_ 2 z_ 4 + z_ 4 y_ 2,\;  
 -z_ 2 z_ 4 + z_ 4 z_ 2,\;  -x_ 3 x_ 4 + x_ 4 x_ 3,\;  x_ 4 y_ 3 - y_ 3 x_ 4,\;  
 x_ 4 z_ 3 - z_ 3 x_ 4,\;  -x_ 3 y_ 4 + y_ 4 x_ 3,\;  -y_ 3 y_ 4 + y_ 4 y_ 3,\;  
 y_ 4 z_ 3 - z_ 3 y_ 4,\;  -x_ 3 z_ 4 + z_ 4 x_ 3,\;  -y_ 3 z_ 4 + z_ 4 y_ 3,\;  
 -z_ 3 z_ 4 + z_ 4 z_ 3
\end{autobreak}
\end{align*}
 Thus the ``natural generators" for $\cIp$
are themselves a GB.
\end{ex}

\begin{prop}\label{prop:pauliGB}
With respect to the graded lex order under which 
\[x_1<y_1<z_1<x_2<\cdots<z_n,\]
polynomials arising from
\cref{eq:pauliGB} 
together with
\beq\label{eq:addPauliGB}
x_jz_j+iy_j, \; z_jx_j-iy_j,\;
y_jz_j-ix_j, \; z_jy_j+ix_j, \quad 1\leq j\leq n
\eeq
form a Gr\"obner basis for $\cIp$. It has $\frac 92n(n+1)$ elements.
\end{prop}

\begin{proof}
Firstly, note that the polynomials in \cref{eq:addPauliGB} belong to $\cIp$. For instance,
\[
\begin{split}
x_jz_j+iy_j&=
-i y_j(z_j^2-1)-y_j(x_jy_j-iz_j)z_j+(x_jy_j-iz_j)y_jz_j\\ 
&\phantom{=}{}+
(y_jx_j+iz_j)y_jy_j-x_j(y_j^2-1)z_j.
\end{split}
\]

We now apply the noncommutative Buchberger algorithm (see, e.g., \cite{moraNCGB}), to check that the constructed set $\cB$ of polynomials is a GB. Firstly, self-obstructions. 
Given one of the polynomials $p\in\cB$ it involves at most two indices $i,j$. Hence all its nontrivial self-obstructions can also involve only these two indices. But modulo the GB these all reduce to 0 (by \Cref{ex:pauliGB4}). Likewise, given polynomials $p_1,p_2\in\cB$, they involve at most four distinct indices. These same indices are the only ones that can appear in a  nontrivial 
 S-polynomial constructed from $p_1,p_2$. But, again by \Cref{ex:pauliGB4}, these must reduce to $0$ since $p_1,p_2$ appear in the GB associated to the chosen four indices.

 To count the number of elements in this GB, note there
 are $3n$ polynomials of the form $w_j^2=1$ for $w\in\{x,y,z\}$ and
 $1\leq j\leq n$. There are $6n$ polynomials of the form $w_ju_j\pm iv_j$ with $\{w,u,v\}=\{x,y,z\}$.
 Finally, there are $3^2\cdot {n\choose 2}$ commutators in $\cB$. Adding these numbers we get the desired count.
\end{proof}

\begin{ex}\label{ex:complicated}
Changing the monomial order to grlex under which
\[x_1<x_2<\cdots<x_n<y_1<\cdots<z_n,\]
the obtained GB is much more complicated. For instance, with $n=2$ the GB contains the degree three polynomial
$
y_2  z_1  z_2 - i x_2  z_1,
$
with $n=3$ the GB contains the degree four polynomial
$
y_3  z_1  z_2 z_3 - i x_3  z_1 z_2,
$
etc.
\end{ex}

\sssec{The \texorpdfstring{$d$th}{dth} relaxation of the quantum max-cut
can be computed in polynomial time}

\Cref{prop:pauliGB} makes it possible to compute the 
$d$th relaxation of the quantum max-cut in polynomial time.

\begin{algorithm}[H]
 \SetKwInOut{Input}{Input}\SetKwInOut{Output}{Output}
\Input{Graph $G=(V,E)$ on $n$ vertices, level of relaxation $d$}
\Output{Solution to the $d$th relaxation of quantum max-cut for $\hampol G$}
\BlankLine

Find the Veronese vector $V_d(n)$ 
ontaining all monomials of degree $\leq$ in the swap variables,
and express them in terms of symbolic Paulis $x_j,y_j,z_j$\;
Replace each entry of $V_d(n)$ by its remainder upon division with the GB of $\cIp$\;
Compute a basis for the span of all the entries of $V_d(n)$\;
Form the vector $\cV_d(n)$ with the basis
as its entries\; 
Find $\cM_d(n):=\cV_d(n)\cV_d(n)^*$\;
Replace each entry of $\cM_d(n)$ by its remainder upon division with the GB of $\cIp$\;
Replace each distinct term appearing in $\cM_d(n)$ with a new (scalar) variable; call the resulting matrix $\cM_d(L)$\;
Solve the SDP
    \beq
\label{eq:pauliSOS}
\rotatebox[origin=c]{180}{$\nu$}_d(\hampol G) = \sup \{ \langle \cM_d(L),\Gamma_G\rangle \mid 
\cM_d(L)\succeq0,\;
\cM_d(L)_{1,1}=1\},
 \eeq
where $\Gamma_G$ is any matrix satisfying $\hampol G=\cV_k(n)^* \Gamma_G\cV_k(n)$\;
\BlankLine
\KwRet{\rotatebox[origin=c]{180}{$\nu$}$_d(\hampol G)$}
\caption{$d$th relaxation of the quantum max-cut supported by Paulis}\label{algo:polySDP}
 \end{algorithm}

\begin{thm}\label{thm:polySDP}
\Cref{algo:polySDP} computes the value of the $d$th relaxation of the quantum max-cut in polynomial time.
\end{thm}

\begin{proof}
Each step takes polynomial time. The only argument required is to establish that the constructed SDP of \cref{eq:pauliSOS} can be solved in polynomial time using standard interior point solvers. 
But this follows from SDP complexity theory (see, e.g., \cite[Section 1.9]{deKlerk02}); we only need the existence of Slater points (which we established in the proof of \Cref{prop:slater})
together with bounds on the feasible region of the SDP.
Since our variables $s_{ij}$ are involutions, the diagonal of $\cM_d(L)$ in \cref{eq:pauliSOS} is constantly $1$ yielding a 
bound of $2$ on each positive semidefinite $\cM_d(L)$.
\end{proof}

\begin{rmk}\label{rmk:Pro21}
As pointed to us by Claudio Procesi,
an alternative method for providing a 
polynomial time SDP hierarchy for the quantum max-cut
can be based on the notion of good permutations \cite{Procesi2021}.
For $d\in\N$, a permutation $\sigma\in S_n$ is \emph{$d$-bad} if
there are $1\leq i_1 < i_2 < \cdots < i_d \leq n$ with
$\sigma(i_1) >\sigma(i_2) > \cdots > \sigma(i_d) $. 
Otherwise it is \emph{$d$-good}. Since each permutation is a product of transpositions, this notion can also be applied to certain elements of the swap algebra, e.g., to products of the $s_{ij}$.
Procesi \cite[Theorem 8]{Procesi2021} shows that $3$-good permutations form a basis of the swap algebra. Further, the proof of his theorem gives a recursive method for expressing an element of the swap algebra as a linear combination w.r.t.~this basis.

Hence providing an ordering for the set of $3$-good permutations one can build increasing size Veronese vectors to be used in the ncSoS hierarchy. (Note that the transposition $(i\ j)$  with $j-i>1$ is not $3$-good, but an induction on $j-i$ together with \Cref{eq:Brels} of \Cref{defn:sswap1} can be used to express $s_{ij}$ in terms of $3$-good permutations.)
It would be interesting to investigate this further and compare 
the performance of this method with the one presented here. \qed
\end{rmk}

\ssec{Gr\"obner Bases for
the Swap Algebra}

Ideally one would want to simplify \Cref{algo:polySDP} to avoid 
expanding everything in terms of Paulis. This is indeed possible if one computes the Gr\"obner basis for the Swap algebra.
Unfortunately, the GB for the Swap algebra appears more involved that the one for the Paulis, and we were unable to identify its form in general.
However, for small $n$ it can be explicitly computed.

Consider the lex order on pairs $(i,j)$, i.e.,
$(i,j)<(k,\ell)$ if $i<k$ or $i=k$ and $j<\ell$, and order
swaps $s_{ij}$ w.r.t.~lex order on the indices.
We used Magma \cite{magma} 
to bootstrap the computation of the GBs. Namely, we computed GB for some small $n$, say $n=4$. Then 
replace indices $(1,2,3,4)$ with any increasing 4-subset of $(1,2,3,4,5)$ to obtain a generating set for $\sidl_5$. On this  generating set we next run the GB algorithm to obtain a GB. Then rinse and repeat. The GB for $n=4$ is given in \Cref{app:GB34}.

\begin{prop}\label{prop:swapGB3}
The GB for $\sidl_3$ is given by the following eight polynomials:
\beq\label{eq:swapGB3}
\begin{gathered}
-1 + s_{12}^2,\;  1 - s_{12} - s_{13} - s_{23} + s_{12}   s_{13} + 
 s_{12}   s_{23},\\  1 - s_{12} - s_{13} - s_{23} + s_{12}   s_{13} + 
 s_{13}   s_{12},\;  -1 + s_{13}^2,\\  -s_{12}   s_{13} + s_{13}   s_{23},\;  -s_{12}   s_{13} + 
 s_{23}   s_{12},\\  1 - s_{12} - s_{13} - s_{23} + s_{12}   s_{13} + s_{23}   s_{13},\;  -1 + s_{23}^2
\end{gathered}
\eeq
\end{prop}

\begin{proof}
Verifying this is a straightforward, though tedious calculation. Alternately, this can be obtained with the help of a computer algebra system such as Magma or NCAlgebra under Mathematica.
\end{proof}

\ssec{\texorpdfstring{$d$th}{dth} Swap Relaxation}

Finally, we have all the ingredients needed to give the algorithm for computing the $d$th relaxation based purely in terms of the swaps.
It generalizes \Cref{algo:polySDPswap} from $d=2$ to arbitrary $d$.

\begin{algorithm}[ht]
 \SetKwInOut{Input}{Input}\SetKwInOut{Output}{Output}
\Input{Graph $G=(V,E)$ on $n$ vertices, level of relaxation $d$,\\ GB$_n$=Gr\"obner basis for $\sidl_n$}
\Output{Solution to the $d$th relaxation of quantum max-cut for $\hampol G$}
\BlankLine

Find $V_d(n)$\;
Replace each entry of $V_d(n)$ by its remainder upon division with GB$_n$\;
Compute a basis for the span of all the entries of $V_d(n)$\;
Form the vector $\cV_d(n)$ with the basis
as its entries\; 
Find $\cM_d(n):=\cV_d(n)\cV_d(n)^*$\;
Replace each entry of $\cM_d(n)$ by its remainder upon division with GB$_n$\;
Replace each distinct term appearing in $\cM_d(n)$ with a new (scalar) variable; call the resulting matrix $\cM_d(L)$\;
Solve the SDP
    \beq
\label{eq:pauliSOSswap}
\rotatebox[origin=c]{180}{$\nu$}_d(\hampol G) = \sup \{ \langle \cM_d(L),\Gamma_G\rangle \mid 
\cM_d(L)\succeq0,\;
\cM_d(L)_{1,1}=1\},
 \eeq
where $\Gamma_G$ is any matrix satisfying $\hampol G=\cV_d(n)^* \Gamma_G\cV_d(n)$\;
\BlankLine
\KwRet{\rotatebox[origin=c]{180}{$\nu$}$_d(\hampol G)$}
\caption{$d$th relaxation of the quantum max-cut}\label{algo:polySDPpure}
 \end{algorithm}

This algorithm, like \Cref{algo:polySDP} runs in finite time with similar justifications. However, this algorithm requies as input a Gr\"obner basis for $\sidl_n$. Finding such a Gr\"obner basis for finite $d$ and arbitrary $n$ may not be possible in polynomial time.

\ssec{Irrep Swap Relaxations}
\label{sec:irrep_Sos}

As seen in \Cref{cor:spec_decomp},
$\displaystyle  \maxeig(\ham{G}) = \max_{k=0}^{\lfloor \frac n2\rfloor}\left(
\maxeig(\hamirrep{G}{[\nk]})\right)$. 
In this section we explain how to form SDP relaxations to find $\maxeig(\hamirrep{G}{[\nk]})$
for a given graph $G$ on $n$ vertices and $k\leq n/2$.

From the proof of \Cref{lem:irrpsym_value} (in particular \cref{eq:eta_hat_value}) 
we have that the following constraint is satisfied inside the $[\nk]$ irrep
\beq\label{eq:lineq}
\sum_{(i,j) \in \E(K_n)}\rho_{[n-k,k]}((i \ j ))  =\hamclinth{[\nk]} I
= \left( \binom{n}{2} + k^2 - k(n+1) \right) I.
\eeq
Enforcing this constraint is equivalent to enforcing that 
\begin{align}
\hampol{\clique n} = \eta_{\nk} \qquad \Longleftrightarrow \qquad \hampol{\clique n} - \eta_{\nk} = 0.
\end{align}
and so, by \Cref{prop:Cink} the algebra formed by the $s_{ij}$ variables satisfying this extra relation is isomorphic to the relevant Irrep \sswap. Stated differently, \Cref{prop:Cink} shows that all constraints in the $[n-k,k]$ irrep can be derived from \Cref{eq:lineq} above along with the relations given in \Cref{defn:sswap1}.

We can force the swap variables to satisfy the same constraint by requiring that one of them, namely $s_{n-1\,n}$, (the last one in our chosen monomial order) can be written in terms of the others:
\begin{align} 
\label{eq:jucys}
s_{n-1 \; n} = - \sum_{i<j \leq  n, \; i \not = n-1} s_{ij} + \hamclinth{[\nk]}.
\end{align}

We can now adapt \Cref{algo:polySDPpure} to find a relaxation for 
$\maxeig(\hamirrep{G}{[\nk]})$.

\sssec{Using the Gr\"obner basis for swaps}

\begin{algorithm}[H]
 \SetKwInOut{Input}{Input}\SetKwInOut{Output}{Output}
\Input{Graph $G=(V,E)$ on $n$ vertices, $k\leq n/2$, level of relaxation $d$\\ GB$_n$=Gr\"obner basis for $\sidl_n$}
\Output{Solution to the $d$th relaxation for $\maxeig(\hamirrep{G}{[\nk]})$}
\BlankLine

Find $V_d(n)$\;
\While{$V_d(n)$ has not stabilized}{
  Replace each entry of $V_d(n)$ by its remainder upon division with GB$_n$\;
  In each of the entries replace $s_{n-1\, n}$ with the expression derived from \cref{eq:jucys} }
Compute a basis for the span of all the entries of $V_d(n)$\;
Form the vector $\cV_d(n)$ with the basis
as its entries\; 
Find $\cM_d(n):=\cV_d(n)\cV_d(n)^*$\;
\While{$\cM_d(n)$ has not stabilized}{
  Replace each entry of $\cM_d(n)$ by its remainder upon division with GB$_n$\;
  In each of the entries replace $s_{n-1\, n}$ with the expression derived from \cref{eq:jucys} }
Replace each distinct term appearing in $\cM_d(n)$ with a new (scalar) variable; call the resulting matrix $\cM_d(L)$\;
Solve the SDP
    \beq
\label{eq:pauliSOSswap2}
\rotatebox[origin=c]{180}{$\nu$}_d(\hamirrep{G}{[\nk]}) = \sup \{ \langle \cM_d(L),\Gamma_G\rangle \mid 
\cM_d(L)\succeq0,\;
\cM_d(L)_{1,1}=1\},
 \eeq
where $\Gamma_G$ is any matrix satisfying $\hampol G'=\cV_d(n)^* \Gamma_G\cV_d(n)$; here $\hampol G'$ is $\hampol G$, where $s_{n-1\,n}$ was replaced as before\;
\BlankLine
\KwRet{\rotatebox[origin=c]{180}{$\nu$}$_d(\hamirrep{G}{[\nk]})$}
\caption{$d$th relaxation for $\maxeig(\hamirrep{G}{[\nk]})$}\label{algo:polySDPpuren-k}
 \end{algorithm}

We now discuss the runtime of \Cref{{algo:polySDPpuren-k}}.
Replacing $s_{n-1\, n}$ with the expression derived from \cref{eq:lineq}
produces monomials in the $s_{ij}$ that are all smaller than the original one 
in the monomial order. 
We also note that there are ${n \choose 2} = n(n-1)/2$ distinct monomials of degree one in the swap algebra, and hence at most $(n(n-1)/2)^d$ monomials of degree at most $d$. Thus, the two while loops in \Cref{algo:polySDPpuren-k} terminate after at most   $(n(n-1)/2)^d$ many steps. For finite $d$ \Cref{algo:polySDPpuren-k} then runs in polynomial time, provided a Gr\"obner basis is supplied as input to the algorithm.

\sssec{Using the Gr\"obner basis for irreps}

To avoid having to deal with the two loops in \Cref{algo:polySDPpuren-k}, we can find a GB for the ideal $\cI_{\nk}$.
Recall from \Cref{prop:Cink} that $\cI_{\nk}$ is generated by
$\sidl$ 
together with $\hampol{\clique n}-\eta_{\nk}$.
So we can either find the Gr\"obner basis GB$_n$ for $\sidl$,
then add the  `clique polynomial',
$\hampol{\clique n}-\eta_{\nk}$,
and run the GB algorithm again. Or, one takes the generating set of polynomials 
in \Cref{rmk:s_ij_abuse_of_notation} for $\sidl$, adds the 
clique 
polynomial, and runs a GB. Once the Gr\"obner basis GB$_{\nk}$ for $\cI_{\nk}$
is available, one simple adapts \Cref{algo:polySDPpure} 
by replacing GB$_n$
 by GB$_{\nk}$.
Thus we obtain \Cref{algo:polySDPpurepuren-k}.

\begin{algorithm}[H]
 \SetKwInOut{Input}{Input}\SetKwInOut{Output}{Output}
\Input{Graph $G=(V,E)$ on $n$ vertices, $k\leq n/2$, level of relaxation $d$\\ GB$_\nk$=Gr\"obner basis for $\cI_{\nk}$}
\Output{Solution to the $d$th relaxation for $\maxeig(\hamirrep{G}{[\nk]})$}
\BlankLine

Find $V_d(n)$\;
Replace each entry of $V_d(n)$ by its remainder upon division with GB$_\nk$\;
Compute a basis for the span of all the entries of $V_d(n)$\;
Form the vector $\cV_d(n)$ with the basis
as its entries\; 
Find $\cM_d(n):=\cV_d(n)\cV_d(n)^*$\;
Replace each entry of $\cM_d(n)$ by its remainder upon division with GB$_\nk$\;
Replace each distinct term appearing in $\cM_d(n)$ with a new (scalar) variable; call the resulting matrix $\cM_d(L)$\;
Solve the SDP
    \beq
\label{eq:pauliSOSswap3}
\rotatebox[origin=c]{180}{$\nu$}_d(\hampol G) = \sup \{ \langle \cM_d(L),\Gamma_G\rangle \mid 
\cM_d(L)\succeq0,\;
\cM_d(L)_{1,1}=1\},
 \eeq
where $\Gamma_G$ is any matrix satisfying $\hampol G \equiv \cV_d(n)^* \Gamma_G\cV_d(n)$ $\mod$ GB$_\nk$\;
\BlankLine
\KwRet{\rotatebox[origin=c]{180}{$\nu$}$_d(\hamirrep{G}{[\nk]})$}
\caption{$d$th relaxation for $\maxeig(\hamirrep{G}{[\nk]})$}\label{algo:polySDPpurepuren-k}
 \end{algorithm}

 As in the case of \Cref{algo:polySDPpuren-k,algo:polySDPpure} this algorithm has polynomial runtime provided a Gr\"obner basis is provided as input for the Algorithm. The Gr\"obner bases GB$_{4-k,k}$ for $k \in \{1,2\}$ are given in \Cref{app:GB34}.

\sec{Finding Eigenvalues via Clique Decompositions
of Graphs}
\label{sec:eig_via_clique_decomp}

The QMC problem is dramatically easier to solve on smaller graphs. So, it would be ideal 
if we could cheaply compute the solution to QMC on a large graph $G$ from the solutions to QMC on a number of smaller graphs. 
There is an obvious way this can be implemented if the graph $G$ is not connected. 
In this case, we can solve QMC on each of the connected components of $G$ and add up the solutions to obtain the solution for $G$.

Less obvious is that a similar decomposition is possible when the complement of a graph $G$ is disconnected. 
Suppose, for instance, that an $n$ vertex graph $G$ has a complement $G^c$ with two connected components $G_1^c$ and $G_2^c$. In this case, the QMC Hamiltonian for $G$ can be written as
$$
\ham{G} = \ham{\clique{n}} - \ham{G_1^c} - \ham{G_2^c},
$$
where $\clique{n}$ denotes an $n$-vertex clique. 
\Cref{lem:clique_ham_identity} gives that $\ham{\completion{G}}$ on the $[\nk]$ irrep 
is equal to $\hamclint{\nk} I$, and since $G^c_1$ and $G^c_2$ are disjoint we obtain
 $$\spec ( \hamirrep{G}{[n-k,k]}) = \{\hamclint{[\nk]}\} - \spec(\hamirrep{G_1^c}{[\nk]})  - \spec(\hamirrep{G_2^c}{[\nk]}),$$
where the addition and subtraction is the Minkowski sum of sets.
Combining this observation with the Young branching rule for irreps,
we can solve the QMC problem on $G$ by first solving the problem on the smaller graphs $G_1^c$ and $G^c_2$.

In this section we fill in details above and develop  this type of graph decomposition
``to its fullest." As a result, we obtain algorithms 
for approximating eigenvalues and even computing them in exact arithmetic in some special cases.
We begin by doing a simple example in more detail.

\ssec{The Star Graph}
\label{subsec:Star_Graph}

As a first example, we consider the $n$-vertex star graph, which we denote $\starn$. 
Eigenvalues of the corresponding QMC Hamiltonian $\ham{\starn}$ have been computed previously \cite{lieb1962ordering}. Here we re-derive these results using the methods of this section.

\begin{lem}
    For any $n$ and $k \leq  n / 2 $ the matrix $\hamirrep{\starn}{[\nk]}$ has two eigenvalues, whose values are given by
    \beq
    \begin{split}
    e_1 = 2(n - k + 1), \qquad
    e_2 = 2k.
    \end{split}
    \eeq
   If $n$ is even and $k=n/2$, then $\hamirrep{\starn}{[n/2,n/2]}$ is $(n+2)$ times the identity matrix.
\end{lem}

\begin{proof}
 Consider first the case $k<n/2$.
 Label the vertices of the star graph so that the $n$-th vertex corresponds to the center of the star. 
 Then $\starn^c= \clique{n-1} $ which yields \ 
 $ \starn = \clique{n} - \clique{n-1} $, so
    \begin{align}
    \hampol{\starn} = \hampol{\clique{n}} - \hampol{\clique{n-1}}.
    \end{align}
    Apply the $[\nk]$ irrep to get
    \begin{align}\label{eq:Hstarnk}
    \hamirrep{\starn}{[\nk]} = \hamirrep{\clique{n}}{[\nk]} - \hamirrep{\clique{n-1}}{[\nk]}.
    \end{align}
    Now we consider the matrices $\hamirrep{\clique{n}}{[\nk]}$ and $\hamirrep{\clique{n-1}}{[\nk]}$. 
    
    \Cref{lem:clique_ham_identity,lem:irrpsym_value} immediately give 
    \begin{align}\label{eq:Hcliquenk}
        \hamirrep{\clique{n}}{[\nk]} = \hamclint{[\nk]} I = \left(2 k (n + 1) - 2 k^2\right) I.
    \end{align}
    However, these results do not immediately say anything about the matrix $\hamirrep{\clique{n-1}}{[\nk]}$.
    To deal with this element we invoke Young's branching rule, which states that the restriction of an irrep 
    $\lambda$ of $\symgpn$ to $S_{n-1}$ corresponds to a direct sum over all irreps which can be obtained by removing a single cell from the partition corresponding to $\lambda$. 
    In the case considered here, that means that $\hamirrep{\clique{n-1}}{[\nk]}$ can be decomposed into a direct sum of the matrices $\hamirrep{\clique{n-1}}{[n-k - 1, k]}$ and $\hamirrep{\clique{n-1}}{[n-k, k- 1]}$. The matrices $\hamirrep{\clique{n}}{[\nk]}$ and $\hamirrep{\clique{n-1}}{[\nk]}$ also commute (since $\hamirrep{\clique{n}}{[\nk]}$ is just a multiple of the identity), hence are simultaneously diagonalizable. Then the possible eigenvalues of $\hamirrep{\starn}{[\nk]}$ are given by a difference of the single eigenvalue of $\hamirrep{\clique{n}}{[\nk]}$ and the two possible eigenvalues of $\hamirrep{\clique{n-1}}{[\nk]}$. This gives 
    \begin{align*}
        e_1 &= \hamclint{[\nk]} - \hamclint{[n-k, k-1]} \\
        &= \left(2 k (n + 1) - 2 k^2\right) - 
        \left(2 (k-1) n - 2 (k-1)^2\right) \\
        &= 2k + 2n - 2k^2 + 2(k-1)^2\\
        &= 2(n - k + 1)
    \end{align*}
    and 
    \begin{align*}
        e_2 &= \hamclint{[\nk]} - \hamclint{[n-k-1, k]} \\
        &= \left(2 k (n + 1) - 2 k^2\right) - 
        \left(2 k n - 2 k^2\right) \\
        &= 2k.
    \end{align*}

    Now assume $n$ is even, and let $k=n/2$. Applying \cref{eq:Hstarnk} we have
    \beq\label{eq:Hstarnn1}
\hamirrep{\starn}{[n/2,n/2]} = \hamirrep{\clique{n}}{[n/2,n/2]} - \hamirrep{\clique{n-1}}{[n/2,n/2]},
    \eeq
    and by \cref{eq:Hcliquenk},
    \beq\label{eq:Hstarnn2}
\hamirrep{\clique{n}}{[n/2,n/2]} = \hamclint{[n/2,n/2]} I = \frac12n(n+2) I.
    \eeq
The difference to the first case occurs when applying the branching rule to compute $\hamirrep{\clique{n-1}}{[n/2,n/2]} $.
Since there is only one valid way to remove a box from a rectangular $2\times n/2$ Young tableaux, we obtain
\beq\label{eq:Hstarnn3}
\hamirrep{\clique{n-1}}{[n/2,n/2]} = \hamirrep{\clique{n-1}}{[n/2,n/2-1]} = \hamclint{[n/2,n/2-1]}I= \frac12(n^2-4)I.
\eeq
Combining \cref{eq:Hstarnn3} with \cref{eq:Hstarnn2} in
\cref{eq:Hstarnn1} yields
\[
\hamirrep{\starn}{[n/2,n/2]} = (n+2)I.\qedhere
\]
\end{proof}

\ssec{Graph Clique Decomposition and Simpler Hamiltonians}

In the previous example we showed how to compute the eigenvalues of the star graph Hamiltonian by writing it as a difference of cliques and then using Young's Branching Rule. In that example, the cliques considered differed in size by one vertex. But the same techniques also apply generally to any graph whose corresponding Hamiltonian can be decomposed as an arbitrary signed sum of clique Hamiltonians. In this section, we provide an algorithm which computes such a sum for a given graph $G$ whenever such a decomposition is possible. Then we show how to speed up the computation of some QMC Hamiltonian's eigenvalues using this decomposition. 
While this algorithm is straightforward, we do not know of an occurrence of it elsewhere in the literature, and so for completeness prove correctness of the algorithm and runtime guarantees in this paper. We also note that a similar, but distinct, decomposition of a graphs is considered in ~\cite{buchanan2021subgraph,pouzet2022boolean}. These papers investigate ways in which graphs can be decomposed into sums of cliques over $\mathbb{F}_2$ (i.e. decompositions where existence of an edge in the original graph corresponds to the parity of the number of cliques containing that edge). These decomposition can be applied more generally than the decomposition considered in this paper.

\sssec{The Tree Clique Decomposition}

\newcommand{\tcd}[1]{\mathcal{T}(#1)}

To keep track of how the QMC Hamiltonian associated with a graph $G$ decomposes into a sum of signed cliques we introduce an object which we call the \df{Tree Clique Decomposition} of $G$. This is introduced formally in the next definition.  In that definition and throughout the remainder of the paper we assume that all trees discussed are rooted (or equivalently directed) trees.

\begin{defn}[Tree Clique Decomposition]
\label{defn:tree_clique_decomp}
For any connected graph $G$, the tree clique decomposition of $G$, denoted $\tcd{G}$ consists of an $m$-vertex tree $T$ and set of $n$ connected graphs $\{G(v_1), ..., G(v_m)\}$, where each graph $G(v_i)$ is associated with a vertex $v_i$ of $T$ and the following properties hold:
\begin{enumerate}[\rm(1)]
    \item For the root vertex $v_1$ of $T$, we have $G(v_1) = G$.
    \item For any vertex $v_i$ of $T$ which is not a leaf vertex, let $c_1, ..., c_k$ be its children. Then we have 
    \begin{align}
        G(v_i)^c = \bigcup_{ j \in \{1,2,...,k\}} G(c_j),
    \end{align}
    where we understand the union of graphs to be their disjoint union.  \label{prop:tree_clique_children}
    \item For any leaf vertex $v_j$ of $T$ we have that $G(v_j)^c$ is connected or $G(v_j)^c$ is totally disconnected. 
\end{enumerate}
\end{defn}

Given a graph $G$ it is reasonably straightforward to compute the tree clique decomposition of $G$ by iteratively taking complements of graphs and then breaking them into their connected components. We make this process formal in \Cref{alg:tree-clique-decomposition}. 

\begin{algorithm}[H]
    \SetKwData{active}{activeTLeaves}
    \SetKwData{iter}{i}
    \SetKwInOut{Input}{Input}\SetKwInOut{Output}{Output}

    \Input{A graph $G$ to be decomposed.}
    \Output{A tree clique decomposition $\tcd{G} = \{T, \{G(v_1),..., G(v_m)\}\}$}
    \BlankLine
    Initialize $T$ as the singleton graph with vertex $v_1$\;
    \active := $\{v_1\}$\;
    \iter := 0 \;
    $G(v_1) := G$\;
    \While{\active is non-empty}{
        \iter := \iter + 1 \;
        remove a vertex from \active, call it $v_{\iter}$ \;
        \If{$G(v_{\iter})^c$ is disconnected and $G(v_{\iter})^c$ is not totally disconnected}{
            \ForEach{connected component $H$ of $G(v_{\iter})^c$}{
                Add a vertex $c$ as a child to vertex $v_{\iter}$ in the tree $T$\;
                $G(c) := H$\;
                Add $c$ to \active\;
            }
        }
    }
    \KwRet{$\{T, \{G(v_1), G(v_2), ..., G(v_m)\}\}$}
\caption{Tree Clique Decomposition}
\label{alg:tree-clique-decomposition}
\end{algorithm}

Now we prove some simple properties of the tree clique decomposition, then prove correctness and bound the runtime of \Cref{alg:tree-clique-decomposition}.

\begin{thm} \label{thm:tree_clique_decomp_properties}
    The tree clique decomposition $\tcd{G}$ of an $n$-vertex graph $G$ is unique (up to permutations amongst siblings in $T$). Additionally: 
    \begin{enumerate}[\rm (1)]
        \item The associated tree~$T$ has depth at most~$n$. \label{item:tree_clique_depth}
        \item \label{item:tree_clique_width} For any $i \leq n$ the total number of vertices contained in graphs indexed by vertices at depth $i$ in $T$ is at most $n$, that is 
        \begin{align}
            \sum_{w : \text{depth}(w) = i} \abs{\V(G(w))} \leq n. 
        \end{align}
    
    \end{enumerate}
\end{thm}

\begin{proof}
    We first show uniqueness of the tree clique decomposition. Assume not. Then there is a graph $G$ with two tree clique decompositions $\tcd{G} = \{T, \{G(v_1), G(v_2), ..., G(v_m)\}\}$ and $\tcd{G}' = \{T', \{G(w_1)', G(w_2)', ..., G(w_{m'})'\}\}$. Because both are tree clique decompositions we must have
    \begin{align}
    G(v_1) = G(w_1)' = G
    \end{align}
    where $w_1$ and $v_1$ are the root vertices of $T$ and $T'$ respectively. Now, let $v^*$ and $w^*$ be minimal vertices (in terms of depth) at which the child structure of the clique decompositions $\tcd{G}$ and $\tcd{G}'$ differ, meaning that either $v^*$ and $w^*$ have different numbers of children or that sets of graphs associated with these children are not identical. Either way, letting $C(v^*)$ be the children of $v^*$ and $C'(w^*)$ be the children of $w^*$, we must have 
    \begin{align}
        \bigcup_{c \in C(v^*)} G(c) \neq \bigcup_{c' \in C'(w^*)} G(c').
    \end{align}
    But because this was a minimal occurrence with respect to depth, we must also have 
    \begin{align}
        G(v^*) = G(w^*).
    \end{align}
    But then finally, because $\tcd{G}$ and $\tcd{G}'$ are both tree clique decompositions we have 
    \begin{align}
        G(v^*)^c = \bigcup_{c \in C(v^*)} G(c) \neq \bigcup_{c' \in C'(w^*)} G(c') = G(w^*)^c
    \end{align}
    and this contradiction proves the result.

    Next, we prove \Cref{item:tree_clique_depth}, that is we show the tree $T$ appearing in the tree clique decomposition $\tcd{G} = \{T, \{G(v_1), ..., G(v_m)\}\}$ of an $n$-vertex graph $G$ has depth at most $n$. By definition of the tree clique decomposition, for any non-leaf vertex $w \in T$ with child $c$ we must have that $G(w)^c$ is disconnected with $G(c)$ corresponding to a connected component of $G(w)^c$. But then $G(c)$ is necessarily a graph on fewer vertices than $G(w)$. It follows that the number of vertices in a graph $G(v)$ is a strictly decreasing function of the depth of of the vertex $v$ in the tree $T$. Then, since the root vertex $v_1$ of $T$ corresponds to the $n$ vertex graph $(v_1) = G$, the bound on the depth of $T$ follows. 

    To prove \Cref{item:tree_clique_width} we first observe that, for any vertex $w \in T$ with children $c_1,...,c_k$ we have 
    \begin{align}
        \sum_i \abs{\V(G(c_i))} = \abs{\V(G(w))},
    \end{align}
    since $\bigcup_i G(c_i) = G(w)^c$. Then we also have that the total number of vertices involved in graphs indexed at depth $i$ of the tree $T$ is a non-increasing function of $i$ and, in particular, letting $v_1$ be the root vertex of $T$ we have
    \begin{align}
        \sum_{w: \text{depth}(w) = i} \abs{\V(G(w))} \leq \abs{\V(G(v_1))} = n \; \forall i,
    \end{align}
    as desired.
\end{proof}

\begin{cor} 
        \Cref{alg:tree-clique-decomposition} computes the tree clique decomposition of the input graph $G$ in time $O(n^3)$. 
\end{cor}

\begin{proof}

    By inspection, the tree $T$ and associated graphs $G(v_1),...,G(v_m)$ returned by the algorithm satisfy all the properties of the tree clique decomposition outlined in \Cref{defn:tree_clique_decomp} so correctness of \Cref{alg:tree-clique-decomposition} is immediate. To bound the runtime of the algorithm first note that an $n$-vertex graph can be decomposed into its connected components in time $O(n^2)$ (using depth first or breadth first search). This is the dominant step in the inner loop of the algorithm, so the algorithm takes time at most $O(\abs{\V(G(w))}^2)$ to complete the inner loop that removes vertex $w$ from \texttt{activeTleaves}. Next, note that every vertex in $T$ appears in \texttt{activeTleaves} at most once. Then, letting $d$ denote the depth of $T$, the total runtime of the algorithm is bounded by: 
    \begin{align}
        O\left( \sum_{w \in \V(T)} \abs{\V(G(w))}^2 \right) &= O\left( \sum_{i = 1}^{d} \sum_{w: \text{depth}(w) = i} \abs{\V(G(w))}^2 \right) \\
        &\leq  O\left( \sum_{i = 1}^{d} n^2 \right) \leq O(n^3),
    \end{align}
    where the first inequality follows from \Cref{item:tree_clique_width} of \Cref{thm:tree_clique_decomp_properties} and the second inequality follows from \Cref{item:tree_clique_depth}. 
\end{proof}

\sssec{Clique decompositions and Hamiltonians}

Finally we show how the tree graph decomposition can be used to write the QMC Hamiltonian associated with a graph $G$ as a signed sum of clique Hamiltonians and smaller ``residual'' graph Hamiltonians. 

\begin{thm}
\label{thm:cliqueHam}
Let $G$ be a graph and $\tcd{G} = \{T, \{G(v_1), ..., G(v_n)\}\}$ be the tree-clique decomposition of $G$. Also, for any graph $G$, let $\completion{G}$ denote the complete graph on the vertices of $G$. Then the following claims hold:
\begin{enumerate}[\rm(1)]
    \item For any vertex $v \in T$ with children $c_1, ..., c_k$ we have 
    \begin{align}
        \ham{G(v)} = \ham{\completion{G(v)}} - \sum_{j \in \{1,...,k\}} \ham{G(c_j)}
    \end{align}
    \label{item:tree_clique_local_ham}
    \item Let $L$ denote the set of leaf vertices in $T$, and $R$ be all non-leaf vertices. Also, for any vertex $v \in T$, let $d(v)$ denote the depth of vertex $v$ in the the tree, with the root vertex having depth $d(v_1) = 0$.
    Then we have 
    \begin{align}
        \ham{G} = \sum_{r \in R} (-1)^{d(r)} \ham{\completion{G(r)}} + \sum_{l \in L}  (-1)^{d(l)} \ham{G(l)}
    \end{align}
    \label{item:tree_clique_global_ham}
\end{enumerate}
    
\end{thm}

\begin{proof}
Part \labelcref{item:tree_clique_local_ham} of the theorem follows from \Cref{prop:tree_clique_children} in \Cref{defn:tree_clique_decomp} which gives 
\begin{align}
    G(v)^c = \bigcup_{j \in \{1,...,k\}} G(c_j)
\end{align}
and hence
\begin{align}
    \ham{G(v)} &= \ham{\completion{G(v)}} - \ham{G(v)^c} = \ham{\completion{G(v)}} -  \sum_{j \in \{1,...,k\}} \ham{G(c_j)}.
\end{align}
Part \labelcref{item:tree_clique_global_ham} then follows from repeated application of \labelcref{item:tree_clique_local_ham}, beginning with the root vertex of~$T$. 
\end{proof}

\sssec{Example of clique decomposition, \texorpdfstring{\Cref{alg:tree-clique-decomposition}}{Algorithm 6}}

Before giving an explicit example of \Cref{alg:tree-clique-decomposition} we note any complete $k$-partite graph has a particularly simple tree 
clique decomposition which \Cref{alg:tree-clique-decomposition}
finds in one step.
This is because a complete $k$-partite graph is by definition the complement of the disjoint union of $k$ different complete graphs;
these are the output of the algorithm.

Now we give an explicit illustration of \Cref{alg:tree-clique-decomposition} on a graph which is not a complete $k$-partite graph.

\begin{figure}[H]
    \centering
    \begin{tikzpicture}[scale=2]
\node[draw, style={shape=circle, thick}] (7) at (0,0) {7};
\node[draw, style={shape=circle, thick}] (8) at (1,1) {8};
\node[draw, style={shape=circle, thick}] (9) at (1,-1) {9};
\node[draw, style={shape=circle, thick}] (10) at (2,0) {10};
\node[draw, style={shape=circle, thick}] (3) at (3,1.5) {3};
\node[draw, style={shape=circle, thick}] (2) at (4,1.5) {2};
\node[draw, style={shape=circle, thick}] (1) at (4,.5) {1};
\node[draw, style={shape=circle, thick}] (6) at (4,-1.5) {6};
\node[draw, style={shape=circle, thick}] (5) at (3,-1.5) {5};
\node[draw, style={shape=circle, thick}] (4) at (4,-.5) {4};
\draw (1) -- (2);
\draw (1) -- (3);
\draw (1) -- (4);
\draw (1) -- (10);
\draw (2) -- (3);  
\draw (2) -- (10);
\draw (3) -- (10);
\draw (4) -- (5);
\draw (4) -- (6);
\draw (4) -- (10);
\draw (5) -- (6);
\draw (5) -- (10);
\draw (6) -- (10);
\draw (7) -- (8);
\draw (7) -- (9);
\draw (7) -- (10);
\draw (8) -- (9);
\draw (8) -- (10);
\draw (9) -- (10);
\end{tikzpicture}
    \caption{This is the input graph to the algorithm. 
    The outputtree $T$ is in \Cref{fig:alg-tree-output-1}}
    and the graphs $G(v_j)$ are in
     \Cref{fig:alg-output-1}.
    \label{fig:alg-input-1}
\end{figure}
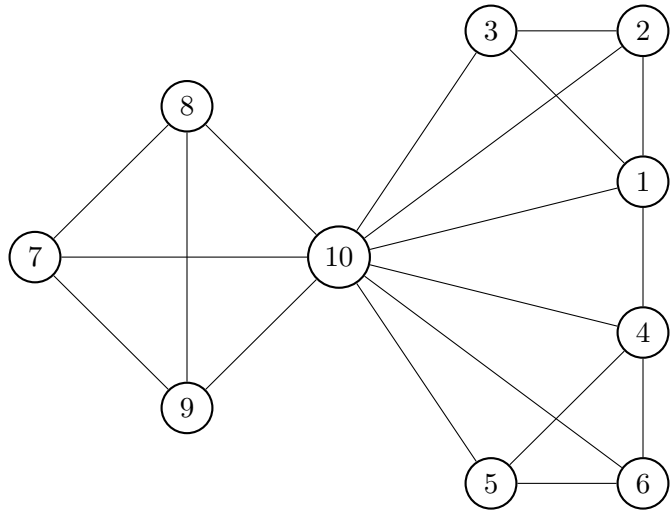

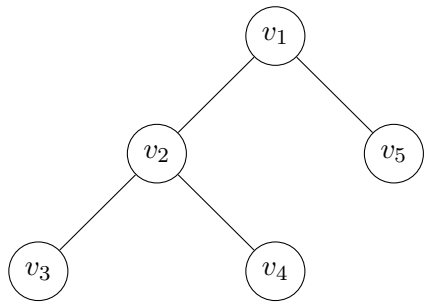
\begin{figure}[H]
    \centering
      \begin{tikzpicture}[yscale=2]
    \node[draw, shape=circle] (root) {$v_1$};
    \node[draw, shape=circle] (left) [below left=of root] {$v_2$};
    \node[draw, shape=circle] (right) [below right=of root] {$v_5$};
    \node[draw, shape=circle] (left2) [below left=of left] {$v_3$};
    \node[draw, shape=circle] (right2) [below right=of left] {$v_4$};
    
    \draw (root) -- (left) -- (left2);
    \draw (root) -- (right);
    \draw (left) -- (right2);
  \end{tikzpicture}
      \label{fig:alg-tree-output-1}
    \caption{This is the output $T$ from \Cref{alg:tree-clique-decomposition} run on the graph in \Cref{fig:alg-input-1}.}
\end{figure}

\begin{figure}[H]
    \centering
    \begin{tikzpicture}[xscale=.7,every node/.style={scale=0.6}]
\node[draw, style={shape=circle, thick}] (7) at (0,0) {7};
\node[draw, style={shape=circle, thick}] (8) at (1,1) {8};
\node[draw, style={shape=circle, thick}] (9) at (1,-1) {9};
\node[draw, style={shape=circle, thick}] (10) at (2,0) {10};
\node[draw, style={shape=circle, thick}] (3) at (3,1.5) {3};
\node[draw, style={shape=circle, thick}] (2) at (4,1.5) {2};
\node[draw, style={shape=circle, thick}] (1) at (4,.5) {1};
\node[draw, style={shape=circle, thick}] (6) at (4,-1.5) {6};
\node[draw, style={shape=circle, thick}] (5) at (3,-1.5) {5};
\node[draw, style={shape=circle, thick}] (4) at (4,-.5) {4};
\draw (1) -- (2);
\draw (1) -- (3);
\draw (1) -- (4);
\draw (1) -- (10);
\draw (2) -- (3);  
\draw (2) -- (10);
\draw (3) -- (10);
\draw (4) -- (5);
\draw (4) -- (6);
\draw (4) -- (10);
\draw (5) -- (6);
\draw (5) -- (10);
\draw (6) -- (10);
\draw (7) -- (8);
\draw (7) -- (9);
\draw (7) -- (10);
\draw (8) -- (9);
\draw (8) -- (10);
\draw (9) -- (10);
\node at (2,-2.5) {$G(v_1)$};
\end{tikzpicture}
\hfill
\begin{tikzpicture}[xscale=.7,every node/.style={scale=0.6}]
\node[draw, style={shape=circle, thick}] (7) at (0,0) {7};
\node[draw, style={shape=circle, thick}] (8) at (1,1) {8};
\node[draw, style={shape=circle, thick}] (9) at (1,-1) {9};
\node[draw, style={shape=circle, thick}] (3) at (3,1.5) {3};
\node[draw, style={shape=circle, thick}] (2) at (4,1.5) {2};
\node[draw, style={shape=circle, thick}] (1) at (5,.5) {1};
\node[draw, style={shape=circle, thick}] (6) at (4,-1.5) {6};
\node[draw, style={shape=circle, thick}] (5) at (3,-1.5) {5};
\node[draw, style={shape=circle, thick}] (4) at (5,-.5) {4};
\draw (3) -- (4);
\draw (3) -- (5);
\draw (3) -- (6);
\draw (2) -- (4);
\draw (2) -- (5);
\draw (2) -- (6);
\draw (1) -- (4);
\draw (1) -- (5);
\draw (1) -- (6);
\draw (7) -- (1);
\draw (7) -- (2);
\draw (7) -- (3);
\draw (7) -- (4);
\draw (7) -- (5);
\draw (7) -- (6);
\draw (8) -- (1);
\draw (8) -- (2);
\draw (8) -- (3);
\draw (8) -- (4);
\draw (8) -- (5);
\draw (8) -- (6);
\draw (9) -- (1);
\draw (9) -- (2);
\draw (9) -- (3);
\draw (9) -- (4);
\draw (9) -- (5);
\draw (9) -- (6);
\node at (3,-2.5) {$G(v_2)$};
\end{tikzpicture}
\hfill
\begin{tikzpicture}[xscale=.7,every node/.style={scale=0.6}]
\node[draw, style={shape=circle, thick}] (3) at (3,1.5) {3};
\node[draw, style={shape=circle, thick}] (2) at (4,1.5) {2};
\node[draw, style={shape=circle, thick}] (1) at (4,.5) {1};
\node[draw, style={shape=circle, thick}] (6) at (4,-1.5) {6};
\node[draw, style={shape=circle, thick}] (5) at (3,-1.5) {5};
\node[draw, style={shape=circle, thick}] (4) at (4,-.5) {4};
\draw (1) -- (2);
\draw (1) -- (3);
\draw (1) -- (4);
\draw (2) -- (3);  
\draw (4) -- (5);
\draw (4) -- (6);
\draw (5) -- (6);
\node at (3.5,-2.5) {$G(v_3)$};
\end{tikzpicture}
\hfill
\begin{tikzpicture}[xscale=.7,every node/.style={scale=0.6}]
\node[draw, style={shape=circle, thick},color=white] (1) at (0,-1.5) {7};
\node[draw, style={shape=circle, thick},color=white] (2) at (0,1.5) {7};
\node[draw, style={shape=circle, thick}] (7) at (0,0) {7};
\node[draw, style={shape=circle, thick}] (8) at (1,1) {8};
\node[draw, style={shape=circle, thick}] (9) at (1,-1) {9};
\draw (7) -- (8);
\draw (7) -- (9);
\draw (8) -- (9);
\node at (.5,-2.5) {$G(v_4)$};
\end{tikzpicture}
\hfill
\begin{tikzpicture}[xscale=.6,every node/.style={scale=0.6}]
\node[draw, style={shape=circle, thick},color=white] (1) at (0,-1.5) {7};
\node[draw, style={shape=circle, thick},color=white] (2) at (0,1.5) {7};
\node[draw, style={shape=circle, thick}] (10) at (0,0) {10};
\node at (0,-2.5) {$G(v_5)$};
\end{tikzpicture}
    \caption{This is the list of graphs which are output.
   The leaf graph, the graphs on which the algorithm terminates, correspond to $v_3$, $v_4$, and $v_5$. 
   The first of these
   is a clique, the second does not decompose further
   and the last  is an isolated point.}
    \label{fig:alg-output-1}
\end{figure}
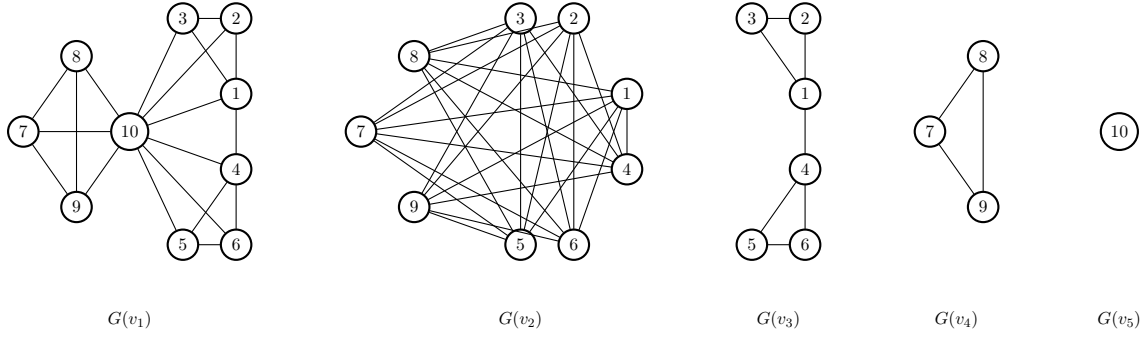

\ssec{Computing Eigenvalues from Clique Decomposition}
\label{sec:eigCliqueDecomp}

\newcommand{\comp}[1]{{#1}^c}
\newcommand{\rest}[2]{\mathrm{Rstrct}\left(#1,#2\right)}

Now we show how we can use the tree clique decomposition to speed up the computation of the eigenvalues of some QMC graph Hamiltonians. 

A key part of this argument will be a use of Young's branching rule to characterize the irreps that can arise when considering restriction of an $S_n$ irrep to $S_m$ for some $m < n$.  We introduce notation for this situation and prove its correctness in the next lemma. 

\def\cJ{{\mathcal J}}

\begin{lem} \label{lem:iterated_Young_branching}

Given any two row partition $[\nk]$ with $n$ boxes and positive integer $m < n$ let $\rest{m}{[\nk]}$ be the set of all 
two row Young diagrams with $m$ boxes total, at most $n-k$ boxes in the first row, and at most $k$ boxes in the second. Stated in notation:
\begin{align}
\label{eq:Rstrct}
\rest{m}{[\nk]} = \{ \irrp{m-j,j} \}_{j \in \cJ }
\end{align}
with  $ \cJ:=\{j :  m + k - n \leq j \leq min(k,m/2) \}$.

Then the restriction of an irrep $[n-k,k]$ of $S_n$ to $S_m$ (i.e., the action of that irrep on just the elements of $S_m$) is isomorphic to a direct sum over all irreps in 
$\rest{m}{[n-k,k]}$ where each irrep occurs with some non-zero multiplicity.

\end{lem}

\begin{proof}
A single application of Young's Branching rule tells us that the restriction of the irrep $\irrpsym$ to $S_{n-1}$ is isomorphic to a direct sum over the irreps that can be obtained by removing one box from $\irrpsym$ while leaving a valid Young diagram. 
Applying the branching rule again to each of these irreps gives a set of irreps whose direct sum is isomorphic to the restriction of $\irrpsym$ to $S_{n-2}$. Continuing this process inductively, we see the restriction of $\irrpsym$ to $S_m$ is isomorphic to a direct sum over all the irreps which can be obtained by removing $n-m$ boxes from $\irrpsym$ while leaving a valid Young diagram (with each irrep occurring with some non-zero multiplicity). Since we can only remove boxes in this process (and not add any) this is equivalent to a direct sum over all valid irreps with $m$ boxes total, at most $n-k$ boxes on the first row, and at most $k$ boxes on the second. But this is exactly the set of irreps in $\rest{m}{[\nk]}$, and we are done.
\end{proof}

\begin{lem} \label{lem:ham-max-eig}
    Let $G$ and $R$ be vertex-disjoint graphs on $n$ vertices and assume we know the spectra $\spec(\hamirrep{G}{n-k,k})$ and 
    $\spec(\hamirrep{R}{n-k,k})$, then the following hold:
    \begin{enumerate}[\rm(1)]
        \item \label{it:max-min-complement}
        The spectrum of $\hamirrep{\clique{n}}{n-k,k} - \hamirrep{G}{n-k,k}$ is $\{ \hamclint{\nk} - \alpha : \alpha \in \spec(\hamirrep{G}{\nk}) \}$;

        \item \label{it:max-min-disjoint-union}
        The spectrum of $\hamirrep{G \bigcup R}{n-k,k} = \hamirrep{G}{n-k,k} + \hamirrep{R}{n-k,k}$ for $G$ and $R$ disjoint is the
        Minkowski sum 
        $$\{ \alpha + \beta : \alpha \in \spec(\hamirrep{G}{\nk}), \; \beta \in \spec(\hamirrep{R}{\nk}) \}.$$
    \end{enumerate}
\end{lem}

\begin{proof}
    \Cref{it:max-min-complement} is an immediate corollary of \Cref{lem:clique_ham_identity}, but remains useful to state outright.

    To show \Cref{it:max-min-disjoint-union}, we begin with a toy example. 
    Suppose that $A^0$ is an $n$ by $n$ matrix with $\alpha \in \spec(A^0)$ and $B^0$ is an $m$ by $m$ matrix with $\beta \in \spec(B^0)$. 
    Hence, $A^0 u = \alpha u$ and $B^0 v = \beta v$. 
    Then, if $A = A^0 \otimes I_m$ and $B = I_n \otimes B^0$, we have that
    \begin{align}
        A (u \otimes v) = (A^0 u) \otimes (I_m v) = \alpha (u\otimes v) \label{eq:tensor-minkowski} \\
        B (u \otimes v) = (I_n u)\otimes (B^0 v) = \beta (u\otimes v).
    \end{align}
    Thus $u\otimes v$ is an eigenvector of $A + B$ corresponding to the eigenvalue $\alpha + \beta$. Thus, $\spec(A) + \spec(B) \subseteq \spec(A+B).$ If the spectra are viewed as multi-sets with multiplicity, there are $mn$ many element in each multi-set. Thus $\spec(A) + \spec(B) = \spec(A + B)$.

    Going back to the definition of the swap matrices \Cref{def:swapmatrices}, it is clear that if $G$ and $R$ are vertex-disjoint, then up to a canonical shuffle which (effectively) relabels the vertices we can reduce to the case of the toy example \cref{eq:tensor-minkowski} to show that the spectrum of $\hamirrep{G}{n-k,k} + \hamirrep{R}{n-k,k}$ equals the Minkowski sum $\spec(\hamirrep{G}{\nk}) + \spec(\hamirrep{R}{\nk})$.
\end{proof}

Next, we consider a connected graph $G$ and show we can characterize the eigenvalues of the irrep Hamiltonian $\hamirrep{G}{\lambda}$ of this graph in terms of the eigenvalues of irrep Hamiltonians of the connected graphs appearing in the complement of $G$. 

\begin{lem} \label{lem:irrham_spec_decomp}
Let $G$ be a connected graph on $n$ vertices, and let $\comp{G}$ denote its complement. Further, let $\comp{G}_1 , ..., \comp{G}_L$ denote the connected components of the graph $\comp{G}$ and let $\nu_1, \ldots , \nu_L$ be the number of vertices in each of these connected components. 

Then, for any two row irrep $\irrpsym$ of $S_n$ we have 
\begin{align}
\spec(\hamirrep{G}{\irrpsym}) = \left\{\{\hamclint{\irrpsym}\} - \sum_{j = 1}^L \left\{ \bigcup_{\altirrpsym \in \rest{\nu_j}{\irrpsym}} \spec(\hamirrep{\comp{G}_j}{\altirrpsym}) \right\} \right\}
\end{align}
where we understand addition and subtraction between sets to be Minkowski addition, so 
\begin{align}
    \{A\} \pm \{B\} = \{a \pm b \}_{a  \in A,\ b  \in B}.
\end{align}
\end{lem}

\begin{proof}
First note that we can write 
\begin{align}
    \hamirrep{G}{\lambda} = \hamirrep{\completion{G}}{\lambda} - \sum_{j = 1}^L \hamirrep{\comp{G}_{j}}{\lambda} \label{eq:irrepham_comp_decomp}
\end{align}
by definition of the graph complement. Additionally \Cref{lem:clique_ham_identity} gives that 
\begin{align}
\hamirrep{\completion{G}}{\irrpsym} = \hamclint{\irrpsym} I.
\end{align}
Combining this with \cref{eq:irrepham_comp_decomp} above immediately gives
\begin{align}
    \spec(\hamirrep{G}{\lambda}) = \left\{\{\hamclint{\irrpsym} \} - \spec\left(\sum_{j = 1}^L \hamirrep{\comp{G}_{j}}{\irrpsym}\right)  \right\}
\end{align}
where subtraction between sets again denotes Minkowski subtraction. Then, to prove the claim, all that remains is to show 
\begin{align}
\spec\left(\sum_{j = 1}^L \hamirrep{\comp{G}_{j}}{\irrpsym}\right)  =
\sum_{j = 1}^L \left\{ \bigcup_{\altirrpsym \in \rest{\nu_j}{\irrpsym}} \spec(\hamirrep{\comp{G}_j}{\altirrpsym}) \right\}.
\end{align}
We do this in two steps. First, note that because the graphs $\comp{G}_{1}, ..., \comp{G}_{L}$ are disjoint, we have 
\begin{align}
\spec\left(\sum_{j = 1}^L \hamirrep{\comp{G}_{j}}{\irrpsym}\right) = \sum_{j = 1}^L \left\{ \spec\left( \hamirrep{\comp{G}_{j}}{\irrpsym}\right) \right\}.
\label{eq:indep_graph_spec}
\end{align}
by \Cref{lem:ham-max-eig}.

Next, consider the irrep Hamiltonian 
$
\hamirrep{\comp{G}_{j}}{\irrpsym}
$
corresponding to a single connected component of $\comp{G}$. Recall that $G$ has $n$ vertices and $\comp{G}_{j}$ has $\nu_j$ vertices, and note that $\lambda$ specifies a two row irrep of $S_n$. For concreteness say this is the $[\nk]$ irrep. Then $\hamirrep{\comp{G}_{j}}{\irrpsym}$ corresponds to a representation of an element of $S_{\nu_j}$ inside the $\lambda$ irrep of $S_n$, i.e., the \textit{restriction} of a $S_n$ irrep to an element of $S_{\nu_j}$. 
But, by \Cref{lem:iterated_Young_branching}, this restriction is isomorphic to a direct sum over all irreps in $\rest{\nu_j}{\irrpsym}$ (each occurring with some nonzero multiplicity).
Then we conclude 
\begin{align}
    \spec\left( \hamirrep{\comp{G}_{j}}{\irrpsym}\right) = \bigcup_{\altirrpsym \in \rest{\nu_j}{\irrpsym}} \spec(\hamirrep{\comp{G}_j}{\altirrpsym}). \label{eq:connec_comp_spec}
\end{align}
Combining \cref{eq:connec_comp_spec,eq:indep_graph_spec} proves the result, and we are done. 
\end{proof}

The dimension of an irrep Hamiltonian's $\hamirrep{G}{\irrpsym}$ can scale exponentially with the number of vertices in the graph $G$. Thus, in general we should not expect to be able to keep track of the full spectra of $\hamirrep{G}{\irrpsym}$. However a straightforward corollary of \Cref{lem:irrham_spec_decomp} is that we can also keep track of the $r$ maximum and minimum values of a irrep Hamiltonian $\hamirrep{G}{\irrpsym}$ provided we know the $r$ maximum and minimum eigenvalues of the irrep Hamiltonians of the graphs appearing in the complement of $G$. 

\begin{cor} \label{cor:max_eig_from_complement}
Define $G$, $\comp{G}$, $\comp{G}_1 , ..., \comp{G}_L$ and $\rest{\nu}{[\nk]}$ as in \Cref{lem:irrham_spec_decomp}. Also, for any integer $r$ and finite set $S \subset R$, let $\rmax{r}(S)$ denote the $r$ largest elements of $S$ and $\rmin{r}(S)$ denote the $r$ smallest elements of $S$. Then we have 
\begin{align}
\rmax{r}\left(\spec(\hamirrep{G}{\irrpsym})\right) = \rmax{r}\left(\left\{\{\hamclint{\irrpsym}\} - \sum_{j = 1}^L \left\{ \bigcup_{\altirrpsym \in \rest{\nu_j}{\irrpsym}} \rmin{r}\left(\spec(\hamirrep{\comp{G}_j}{\altirrpsym})\right) \right\} \right\}\right).
\end{align}
In particular: 
\begin{align}
    \maxeig(\hamirrep{G}{\irrpsym}) &= \max \left(\left\{\{\hamclint{\irrpsym}\} - \sum_{j = 1}^L \left\{ \bigcup_{\altirrpsym \in \rest{\nu_j}{\irrpsym}} \mineig(\hamirrep{\comp{G}_j}{\altirrpsym}) \right\} \right\}\right) \\
    &= \hamclint{\irrpsym} - \sum_{j = 1}^L \left( \min_{\altirrpsym \in \rest{\nu_j}{\irrpsym}} \left( \mineig(\hamirrep{\comp{G}_j}{\altirrpsym}) \right)\right).
\end{align}
\end{cor}

\Cref{cor:max_eig_from_complement} shows that we can compute the maximum (resp. minimum) eigenvalues of the irrep Hamiltonian $\hamirrep{G}{\irrpsym}$ of a connected graph $G$ provided we know the minimum (resp. maximum) eigenvalues of the irrep Hamiltonian's of each connected component in the complement of $G$. 
We make this observation formally in \Cref{alg:inductive_eigenvalue_computation}. 

Before stating this algorithm, we remind the reader of the use the phrase ``all two row irrep Hamiltonians of $G$'' introduced in \Cref{sec:QMC_and_irreps}, which refers to the set of irrep Hamiltonians 
\begin{align}
   \{ \hamirrep{G}{[\nk]} : 1 \leq k \leq \lfloor n / 2 \rfloor \}. 
\end{align}
This convention is particularly important in the following algorithms, where the graphs $\tilde{G}$ (and hence the set of corresponding irreps) considered vary throughout the algorithm. 

 \begin{algorithm}[H]
 \SetKwInOut{Input}{Input}\SetKwInOut{Output}{Output}
\Input{
\begin{minipage}[t]{12cm}

\strut An $n$-vertex connected graph $G$ whose complement $G^c$ has connected components $\cG^c = \{G^c_1, ... G^c_k\}$.\\[-3mm]

\strut Min and max eigenvalues of all two row irrep Hamiltonians of $\tilde{G}$ for each $\tilde{G} \in \cG^c$. \\[-3mm]

\end{minipage}
}
\Output{Min and max eigenvalues all two row irrep Hamiltonians of $G$.}
\BlankLine

\ForEach{two row irrep $\lambda$ of $S_n$}{
$\maxeig(\hamirrep{G}{\irrpsym}) := \hamclint{\irrpsym} - \sum_{\tilde{G} \in \cG^c} \left( \min_{\altirrpsym \in \rest{\abs{\V(\tilde{G})}}{\irrpsym}} \left( \mineig(\hamirrep{\tilde{G}}{\altirrpsym}) \right)\right)$\; 
$\mineig(\hamirrep{G}{\irrpsym}) := \hamclint{\irrpsym} - \sum_{\tilde{G} \in \cG^c} \left( \max_{\altirrpsym \in \rest{\abs{\V(\tilde{G})}}{\irrpsym}} \left( \maxeig(\hamirrep{\tilde{G}}{\altirrpsym}) \right)\right)$\;
}
\caption{Inductive Step in an Eigenvalue Computation}
\label{alg:inductive_eigenvalue_computation}
 \end{algorithm}

Now by applying \Cref{alg:inductive_eigenvalue_computation} repeatedly we can compute the maximum and minimum eigenvalues of an irrep Hamiltonian $\hamirrep{G}{\irrpsym}$ (and hence the maximum and minimum eigenvalues of the QMC Hamiltonian $\ham{G}$) provided we know the maximum and minimum eigenvalues of all the irrep Hamiltonians of all graphs $G$ corresponding to the leaves in the tree clique decomposition of $G$. We make this process formal in \Cref{alg:tree_clique_eigenvalue_computation}.

\begin{algorithm}[H]
    \SetKwInOut{Input}{Input}\SetKwInOut{Output}{Output}
    \SetKw{Assert}{Assert}
    \Input{
    \begin{minipage}[t]{12cm}
    An $n$-vertex graph $G$ along with its tree clique decomposition $\tcd{G} = \{T, \{G(v_1), ..., G(v_m)\}\}$.  
    
    Min and max eigenvalues of all two row irrep Hamiltonians of $G(l)$ for every leaf vertex $l$ of $T$. 
    \end{minipage}
    }
    \Output{Min and Max eigenvalues of all two row irrep Hamiltonians of $G$.}
    \BlankLine

    d := depth($T$)\;
    \ForEach{$i$ in $(d-1, d - 2, ..., 1)$} {
        \ForEach{non-leaf vertex $w \in T$ at depth $i$}{
            \Assert that we know the max and min eigenvalues of all two row irrep Hamiltonians of $G(c)$ for each child $c$ of $w$\;
            Use \Cref{alg:inductive_eigenvalue_computation} to compute the max and min eigenvalues of all two row irrep Hamiltonians of $G(w)$ from the irrep Hamiltonians of all children $G(c)$\;
        }
    }
    \caption{Tree Clique Eigenvalue Computation}    \label{alg:tree_clique_eigenvalue_computation}
\end{algorithm}

Next we prove correctness and bounding the runtime of \Cref{alg:inductive_eigenvalue_computation,alg:tree_clique_eigenvalue_computation}, then giving some remarks concerning generalizations and consequences of these algorithms. 

 \begin{thm}
\Cref{alg:inductive_eigenvalue_computation,alg:tree_clique_eigenvalue_computation} both correctly compute the irrep eigenvalues of their input graph $G$. \Cref{alg:inductive_eigenvalue_computation} runs in time $O(n^2)$ and \Cref{alg:tree_clique_eigenvalue_computation} runs in time $O(n^3)$. 
\end{thm}

\begin{proof}
Correctness of \Cref{alg:inductive_eigenvalue_computation} follows immediately from \Cref{cor:max_eig_from_complement}. To bound its runtime, first note that there are at most $m/2 \in O(m)$ different two row irreps corresponding to any $m$-vertex graph $\tilde{G}$. Additionally, we have 
$
    \bigcup_{\tilde{G} \in \cG^c} = G^c
$
by construction and $\abs{\V(G^c)} = n$ by assumption. So the inner loop of \Cref{alg:inductive_eigenvalue_computation} involves maximizing/minimizing over partitions of at most $n$ elements and then summing over the results of that maximization/minimization, all of which can be done in time $O(n)$. The graph $G$ has at most $n$ vertices, hence at most $O(n) $ two row irreps, and so the outer loop of \Cref{alg:inductive_eigenvalue_computation} is repeated at most $O(n)$ times. The runtime of $O(n^2)$ follows. 

Correctness of \Cref{alg:tree_clique_eigenvalue_computation} follows from the observation that the \textbf{assert} statement in the inner loop of the algorithm is always satisfied, since we know the max and min eigenvalues of the irrep Hamiltonians for graphs corresponding to leaf vertices by assumption, and all other graph irrep Hamiltonians are computed from the ``bottom up'' by construction. 
To bound the runtime of \Cref{alg:tree_clique_eigenvalue_computation} note that, by \Cref{item:tree_clique_width} of \Cref{thm:tree_clique_decomp_properties} and the bound on the runtime of \Cref{alg:inductive_eigenvalue_computation} given above, for any $i \in \{d-1,...,1\}$, \Cref{alg:tree_clique_eigenvalue_computation} requires time at most $O(n^2)$ to compute the max and min irrep eigenvalues of all graphs $G(w)$ corresponding to vertices $w$ at depth $i$ in the tree clique decomposition of $G$. Then \Cref{item:tree_clique_depth} of \Cref{thm:tree_clique_decomp_properties} gives that $d \leq n$ and the upper bound of $O(n^3)$ follows. 
\end{proof}

\begin{rmk} 
A straightforward generalization of \Cref{alg:inductive_eigenvalue_computation,alg:tree_clique_eigenvalue_computation} using \Cref{cor:max_eig_from_complement} gives us an algorithm computing the $r$ highest and lowest eigenvalues of all irrep Hamiltonians of $G$ given the $r$ highest and lowest eigenvalues of all irrep Hamiltonians of all graphs indexed by leaves in the clique tree decomposition of $G$. The runtime of this algorithm remains polynomial in $r$ and $n$. \qed
\end{rmk}

\begin{rmk} 
    If all the graphs indexed by leaves in the clique tree decomposition of $G$ are cliques, then we can compute all eigenvalues of all irrep Hamiltonians of those graphs in exact arithmetic by \Cref{lem:clique_ham_identity,lem:irrpsym_value}. In this case all the eigenvalues of $G$ will be obtained by taking sums and differences of clique eigenvalues $\hamclint{\irrpsym}$ for different values of $\irrpsym$ and, in particular, all eigenvalues will be integer. 

    We give an example of a graph $G$ admitting such a decomposition in the next section. \qed
    \end{rmk}

\begin{rmk} 
    If we apply the same procedure as described in \Cref{alg:tree_clique_eigenvalue_computation} but take as input approximate eigenvalues of the irrep Hamiltonians of the graphs indexed by leaves in the clique tree decomposition of $G$ we obtain approximate eigenvalues of $G$. It is straightforward to show that the error in this approximation is at worst the sum of the approximation errors for the eigenvalues of each input graph. Additionally, if the algorithm is given upper/lower bounds on the minimum/maximum eigenvalues of all irrep Hamiltonians of graphs indexed by leaves in the tree clique decomposition of $G$, it produces upper/lower bounds on the maximum/minimum eigenvalues of $G$.

    One way to obtain these bounds is by running the ncSoS algorithm inside all two row irreps of the graphs indexed by leaves in the tree clique decomposition of $G$. Details of this use of ncSoS is given in \Cref{sec:irrep_Sos}. \qed
\end{rmk}

\subsection{Example of Clique Decomposition, its Hamiltonians and their Eigenvalues}
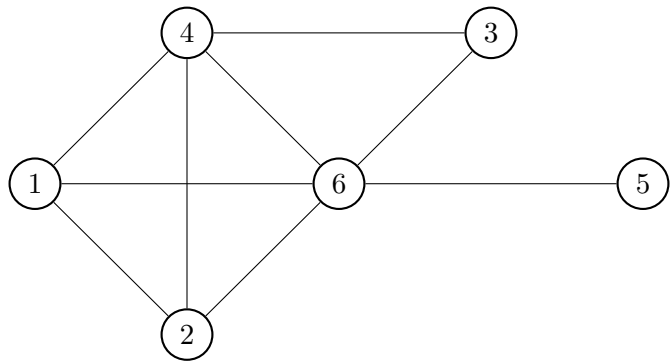
\begin{figure}[ht]
    \centering
    \begin{tikzpicture}[scale=2]
\node[draw, style={shape=circle, thick}] (1) at (0,0) {1};
\node[draw, style={shape=circle, thick}] (2) at (1,-1) {2};
\node[draw, style={shape=circle, thick}] (4) at (1,1) {4};
\node[draw, style={shape=circle, thick}] (6) at (2,0) {6};
\node[draw, style={shape=circle, thick}] (3) at (3,1) {3};
\node[draw, style={shape=circle, thick}] (5) at (4,0) {5};
\draw (1) -- (2);
\draw (1) -- (6);
\draw (1) -- (4);
\draw (2) -- (4);  
\draw (2) -- (6);
\draw (3) -- (4);
\draw (3) -- (6);
\draw (4) -- (6);
\draw (5) -- (6);
\end{tikzpicture}
    \caption{The graph $G$ to be analyzed in this example.}
    \label{fig:ex6vOutGs}
\end{figure}

In this subsection we apply the algorithms discussed previously in this section to analyze the graph $G$ shown in \Cref{fig:ex6vOutGs}.

\sssec{The tree clique decomposition of \texorpdfstring{$G$}{G} and the Hamiltonian decomposition}

\begin{figure}[ht]
    \centering
    \begin{tikzpicture}[yscale=1.5]
    \node[shape=circle,draw=black] (v1) at (0,0) {$v_1$};
    \node[shape=circle,draw=black] (v2) at (1,-1) {$v_2$};
    \node[shape=circle,draw=black] (v9) at (-1,-1) {$v_9$};
    \node[shape=circle,draw=black] (v3) at (0,-2) {$v_3$};
    \node[shape=circle,draw=black] (v8) at (2,-2) {$v_8$};
    \node[shape=circle,draw=black] (v4) at (-1,-3) {$v_4$} ;
    \node[shape=circle,draw=black] (v7) at (1,-3) {$v_7$} ;
    \node[shape=circle,draw=black] (v5) at (-2,-4) {$v_5$} ;
    \node[shape=circle,draw=black] (v6) at (0,-4) {$v_6$} ;
\draw (v9)--(v1)--(v2)--(v3)--(v4)--(v5);
\draw (v2)--(v8);
\draw (v3)--(v7);
\draw (v4)--(v6);
\end{tikzpicture}
    \caption{The output tree, $T$}
\end{figure}
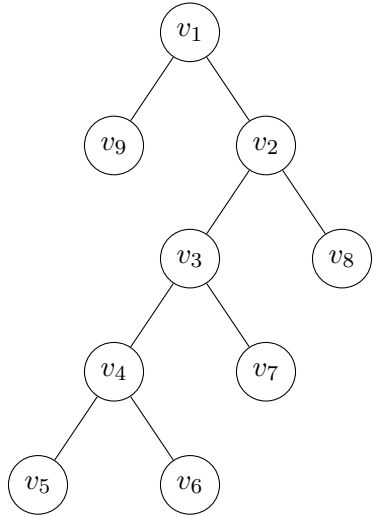

\begin{figure}[H]
    \centering
     \begin{tikzpicture}[xscale=.7,every node/.style={scale=0.6}]
\node[draw, style={shape=circle, thick}] (1) at (0,0) {1};
\node[draw, style={shape=circle, thick}] (2) at (1,-1) {2};
\node[draw, style={shape=circle, thick}] (4) at (1,1) {4};
\node[draw, style={shape=circle, thick}] (6) at (2,0) {6};
\node[draw, style={shape=circle, thick}] (3) at (3,1) {3};
\node[draw, style={shape=circle, thick}] (5) at (4,0) {5};
\draw (1) -- (2);
\draw (1) -- (6);
\draw (1) -- (4);
\draw (2) -- (4);  
\draw (2) -- (6);
\draw (3) -- (4);
\draw (3) -- (6);
\draw (4) -- (6);
\draw (5) -- (6);
\node at (2,-2) {$G(v_1)$};
     \end{tikzpicture}
     \hfill
     \begin{tikzpicture}[xscale=.7,every node/.style={scale=0.6}]
\node[draw, style={shape=circle, thick}] (1) at (0,0) {1};
\node[draw, style={shape=circle, thick}] (2) at (1,-1) {2};
\node[draw, style={shape=circle, thick}] (4) at (1,1) {4};
\node[draw, style={shape=circle, thick}] (3) at (3,1) {3};
\node[draw, style={shape=circle, thick}] (5) at (4,0) {5};
\draw (1) -- (3);
\draw (1) -- (5);
\draw (2) -- (5);  
\draw (2) -- (3);
\draw (3) -- (5);
\draw (4) -- (5);
\node at (2,-2) {$G(v_2)$};
\end{tikzpicture}
\hfill
     \begin{tikzpicture}[xscale=.7,every node/.style={scale=0.6}]
\node[draw, style={shape=circle, thick}] (1) at (0,0) {1};
\node[draw, style={shape=circle, thick}] (2) at (1,-1) {2};
\node[draw, style={shape=circle, thick}] (4) at (1,1) {4};
\node[draw, style={shape=circle, thick}] (3) at (3,1) {3};
\draw (1) -- (2);
\draw (1) -- (4);
\draw (2) -- (4);  
\draw (4) -- (3);
\node at (2,-2) {$G(v_3)$};
\end{tikzpicture}
\hfill
     \begin{tikzpicture}[xscale=.7,every node/.style={scale=0.6}]
\node[draw, style={shape=circle, thick}] (1) at (0,0) {1};
\node[draw, style={shape=circle, thick}] (2) at (1,-1) {2};
\node[draw, style={shape=circle, thick}] (3) at (3,1) {3};
\draw (1) -- (3) -- (2);
\node at (2,-2) {$G(v_4)$};
\end{tikzpicture}

\vspace{5mm}

     \begin{tikzpicture}[xscale=.7,every node/.style={scale=0.6}]
\node[draw, style={shape=circle, thick}] (1) at (0,0) {1};
\node[draw, style={shape=circle, thick}] (2) at (1,-1) {2};
\draw (1) -- (2) -- (2);
\node at (.5,-2) {$G(v_5)$};
\end{tikzpicture}
\hfill
     \begin{tikzpicture}[xscale=.7,every node/.style={scale=0.6}]
\node[draw, style={shape=circle, thick}] (3) at (3,-1) {3};
\node at (3,-2) {$G(v_6)$};
\end{tikzpicture}
\hfill
     \begin{tikzpicture}[xscale=.7,every node/.style={scale=0.6}]
\node[draw, style={shape=circle, thick}] (4) at (1,-1) {4};
\node at (1,-2) {$G(v_7)$};
\end{tikzpicture}
\hfill
     \begin{tikzpicture}[xscale=.7,every node/.style={scale=0.6}]
\node[draw, style={shape=circle, thick}] (5) at (4,-1) {5};
\node at (4,-2) {$G(v_8)$};
\end{tikzpicture}
\hfill
     \begin{tikzpicture}[xscale=.7,every node/.style={scale=0.6}]
\node[draw, style={shape=circle, thick}] (6) at (2,-1) {6};
\node at (2,-2) {$G(v_9)$};
\end{tikzpicture}
    \caption{The output graphs. The second row of graphs corresponds to the leaves $L$ of $T$, the first row to $R$.
    Note  $G(v_1)^c$ is the union of 
    its children, the two depth 1 graphs 
    $G(v_9)$ and     $G(v_2)$. 
    Likewise
    $G(v_2)^c$ is the union of two depth 2 graphs and 
     $G(v_3)^c$ is the union of two depth 3 graphs and
      $G(v_4)^c$ is the union of two depth 4 graphs.
      This can be read off from the tree.
    }
\end{figure}
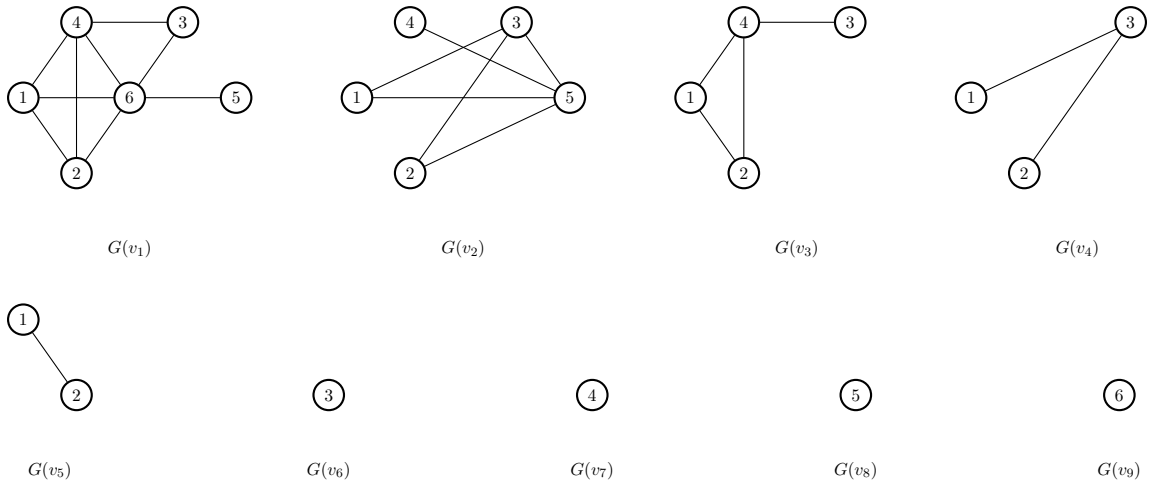

To write down the Hamiltonian clique decomposition for $G$,
first see that the leaves of tree $T$ are 
$L:=\{v_9, v_5, v_6, v_7, v_8 \}$ and the rest are 
$R:=\{v_1, v_2, v_3 , v_4 \}$.
The corresponding depths $d(v_j)$ are read off from $T$ 
and listed in \Cref{tab:Ex6vertGraph}

\begin{table}[ht]
    \centering
    \begin{tabular}{|c|c|c|c|c|c|c|c|}
        \hline
                     & $G(v_1)$ & $G(v_2)$ &  $G(v_3)$ &  $G(v_4)$ &  $G(v_5)$ \\
                     \hline
        Depth in $T$        & 0    &      1     &    2   & 3 &   4 \\
        SignHam  &       & $+$   &      -     &    $+$ & -       \\
        Dim $K(G(v))$         & 6  &      5   &  4    &    3   & 2   \\
        \hline 
        \end{tabular}
    \caption{Clique decomposition of graph $G$, ignoring single vertex graphs}
    \label{tab:Ex6vertGraph}
\end{table}
\noindent
and force the signs of the decomposing Hamiltonians
in the table.
By \Cref{thm:cliqueHam}
\Cref{item:tree_clique_global_ham}
the Hamiltonian $H_G$ has the decomposition:
\begin{align}
    \ham{G} = + \ham{\completion{G(v_1)}} - \ham{\completion{G(v_2)}} + \ham{\completion{G(v_3)}} -\ham{\completion{G(v_4)}} \\
    + \ham{G(v_5)} + \ham{G(v_6)} -\ham{G(v_7)} + \ham{G(v_8)} -\ham{G(v_9)}
\end{align}

Next use that $G(v_j)$ for $j=6,7,8,9$
are singletons, to see the corresponding Hamiltonians are 0;
also $\completion{G(v_5)} = {G(v_5)}$. Thus
\begin{align}
    \ham{G} = + \ham{\completion{G(v_1)}} - \ham{\completion{G(v_2)}} + \ham{\completion{G(v_3)}} -\ham{\completion{G(v_4)}} 
    + \ham{\completion{G(v_5)}},
\end{align}
a sum/difference of clique Hamiltonians.

    We mention that for this special example, one sees from \Cref{fig:ex6vOutGs}
the containments $V(G(v_j)) \supseteq V(G(v_{j+1})$ for $j=1,2,3,4$,
so
\begin{align}
     \completion{G(v_1)} \supseteq \completion{G(v_2)} \supseteq \completion{G(v_3)} \supseteq  \completion{G(v_4)} 
    \supseteq  \completion{G(v_5)} =  \completion{G(v_5)} .
\end{align}

\sssec{Eigenvalues}

Now we see how one can use 
\Cref{alg:tree_clique_eigenvalue_computation} to
compute max/min eigenvalues of $H_G$;
we restrict to one case
$
\maxeig(\hamirrep{G}{[3,3]})
$, since
the min eigenvalues and other irreps behave similarly.
Firstly, we list the data 
needed for eigenvalues (and also above).

\begin{table}[ht]
    \centering
    \begin{tabular}{|c|c|c|c|c|c|c|c|}
        \hline
             & $\clique{6}$  & $\clique{5}$ & $\clique{4}$ & $\clique{3}$  &  $\clique{2}$  \\
        \hline 
Sign & $+$ , max & $-$ , min  & $+$ ,  max &  $-$  , min  & $+$ ,  max 
        \\
        \hline 
&    $\hamclint{[3, 3]} =$  \ 24   & $\hamclint{[3, 2]}=$ 16   &  $\hamclint{[2, 2]}=$ 12 & $\hamclint{[2, 1]}=$ \ 6  &  $\hamclint{[2,0]}=$  0  \\   &      x  &   x   &   $\hamclint{[3, 1]}=$  8 & $\hamclint{[2, 1]}=$ 6 & $ \hamclint{[1, 1]}=$  4  \\
  &      x  &   x   &   x  & $\hamclint{[3, 0]}=$ 0 & $\hamclint{[2,0]}=$ 0  \\
             \hline 
    \end{tabular}
    \caption{$G$  and $\irrp{3,3}$. Data for computing $\maxeig(\hamirrep{G}{[3,3]})$.
    The number next to $[m,k]$ is $\hamclint{m,k}$.
    Entries to the right are needed in light of those to the left because of Young's Branching Rule.}
\end{table}

We shall apply \Cref{alg:inductive_eigenvalue_computation} repeatedly 
to compute $\maxeig(\hamirrep{G}{[3,3]})$.
\begin{align}
\maxeig(\hamirrep{G}{[3,3]}) & =
\hamclint{[3,3]} -  \left( \min_{\altirrpsym \in \rest{5} {[3,3]} }
\left( \mineig(\hamirrep{G(v_2)}{\altirrpsym} \right) \right). \\
& =
\hamclint{[3,3]} -   
 \mineig(\hamirrep{G(v_2)}{[3,2]})   \qquad and \\
\mineig(\hamirrep{G(v_2)}{[3,2]} ) & = \hamclint{[3,2]} - 
 \max_{\altirrpsym \in \rest{4} {[3,2]} }
\left( \maxeig(\hamirrep{G(v_3)}{\altirrpsym}  \right) \\
& =
 \hamclint{[3,2]} - \max
\left( \maxeig(\hamirrep{G(v_3)}{[3,1]}, \maxeig( \hamirrep{G(v_3)}{[2,2]} )   \right) 
 \end{align}
Combine these to get
\begin{align}
\maxeig(\hamirrep{G}{[3,3]}) =  \hamclint{[3,3]} -
 \hamclint{[3,2]} + \max
\left( \maxeig(\hamirrep{G(v_3)}{[3,1]}, \maxeig( \hamirrep{G(v_3)}{[2,2]} )   \right) 
\label{eq:eig33to22}
 \end{align}
Next 
 \begin{align}
     \maxeig(\hamirrep{G(v_3)}{[3,1]})& =\hamclint{3,1} - \min
     \left( \mineig(\hamirrep{G(v_4)}{[3,0]}, \mineig( \hamirrep{G(v_4)}{[2,1]} )  \right)
     \\
     \maxeig( \hamirrep{G(v_3)}{[2,2]} )& =\hamclint{2,2}- 
     \mineig(\hamirrep{G(v_4)}{[2,1]} ) 
 \end{align}
Before combining again
we compute the $\maxeig$ ingredients of 
these formulas:
\begin{align*}
     \mineig(\hamirrep{G(v_4)}{[3,0]}) & =
     \hamclint{3,0} -  \maxeig(\hamirrep{G(v_5)}{[2,0]})=
         \hamclint{3,0} -  \hamclint{[2,0]}
     \\
     \mineig( \hamirrep{G(v_4)}{[2,1]} )  & =
       \hamclint{2, 1} 
       - \max \left(  \maxeig(\hamirrep{G(v_5)}{[2,0]}, 
           \maxeig(\hamirrep{G(v_5)}{[1,1]}  \right)
       =  \hamclint{2, 1} 
       - \max \left(  \hamclint{[2,0]},   \hamclint{[1,1]}  \right)
           \end{align*} 
The last step in each line uses that $G(v_5)$ is a clique.
Now we combine all these, working backward to get
 \begin{align*}
     \maxeig(\hamirrep{G(v_3)}{[3,1]})& =\hamclint{3,1} - \min
     \left( 
           \hamclint{3,0} -  \hamclint{[2,0]} \ , \
       ( \hamclint{2, 1} 
       - \max \left(  \hamclint{[2,0]},   \hamclint{[1,1]}  \right) \right)
     \\
     \maxeig( \hamirrep{G(v_3)}{[2,2]} )& =\hamclint{2,2}- 
         ( \hamclint{2, 1} 
       - \max \left(  \hamclint{[2,0]},   \hamclint{[1,1]}  \right) )
 \end{align*}
and finally substituting these into \cref{eq:eig33to22}
gives
\begin{align*}
& \maxeig(\hamirrep{G}{[3,3]})  =  \hamclint{[3,3]} -
 \hamclint{[3,2]} \\
& + \max \left( 
   \hamclint{3,1} - \min
     \left( 
        \hamclint{3,0} -  \hamclint{[2,0]},
        \hamclint{2, 1} 
       - \max \left(  \hamclint{[2,0]},   \hamclint{[1,1]}  \right)\right) \ , \   
     \hamclint{2,2}- 
         \hamclint{2, 1} 
       + \max \left(  \hamclint{[2,0]},   \hamclint{[1,1]}  \right) \ \right).
 \end{align*}
 Now we substitute numbers in for the $\hamclint{m,k}$ and get 
\begin{align*}
 \maxeig(\hamirrep{G}{[3,3]})  & =
24 - 16 

+ \max \left( 
   8 - \min
     \left(  0 -  0, \   6  - \max \left(  0,  4  \right)\right) \ , \     12 -    6     + \max \left(  0,   4  \right) \ \right)
    =18,
 \end{align*}
which by our theory is 
$\maxeig(\hamirrep{G}{[3,3]})$,
as can be independently checked. We note that this ``working backwards'' step we just worked through is essentially the algorithm for computing max eigenvalues outlined in \Cref{alg:tree_clique_eigenvalue_computation}.

\makeatletter
\newcommand{\mycustomfontsize}{%
  \@setfontsize\mycustomfontsize{10.5pt}{12.5pt}
}
\makeatother

{
\addcontentsline{toc}{section}{\bibname}
\bibliographystyle{alphaurl}
\mycustomfontsize
\bibliography{ref}
}

\newpage 


\appendix

\section{Identities for the Proof of \texorpdfstring{\Cref{lem:swap3}}{Lemma B.1}}
\label{app:swap3}

{\tiny

}

\section{Linear space spanned by the products of at most three or four swap matrices}
\label{ssec:swap34}

In this appendix we construct linear algebraic bases $\cB_3$ and $\cB_4$ which are analogues of the basis $\cB_2$ constructed in \Cref{sssec:linear_basis_degree_2_swaps}. These bases are used in \Cref{algo:polySDPswap}, our degree 2 SDP relaxation in the swap matrices. 

\subsection{Linear space spanned by the products of at most three swap matrices}\label{ssec:swap3}

\begin{lem}\label{lem:swap3}
A basis $\cB_3$ for the linear span of all products of at most three swap matrices in the case $n=5$ is $\cB_2$ together with
\begin{multline*}
\sw_{12}\sw_{13}\sw_{45},\quad
\sw_{12}\sw_{14}\sw_{35},\quad
\sw_{12}\sw_{15}\sw_{34},\\
\sw_{13}\sw_{14}\sw_{25},\quad
\sw_{13}\sw_{15}\sw_{24},\quad
\sw_{14}\sw_{15}\sw_{23},\;
\end{multline*}
Alternately, 
$\sw_{14}\sw_{15}\sw_{23}$
could be replaced by
$\sw_{12}\sw_{34}\sw_{35}$  in $\cB_3$.
\end{lem}

\begin{proof}
The add-on statement follows from the equality
\[
\begin{split}
2 \sw_{14} \sw_{15} \sw_{23} & = \sw_{12} - \sw_{14} - \sw_{23} + \sw_{24} - \sw_{25} + \sw_{35} \\
& \phantom{={}} - \sw_{12} \sw_{34} - \sw_{12} \sw_{35} - 
 \sw_{12} \sw_{45} 
\\
& \phantom{={}} 
 + \sw_{13} \sw_{14} - \sw_{13} \sw_{15} - \sw_{13} \sw_{24} \\
& \phantom{={}} + \sw_{13} \sw_{25} + 
 \sw_{14} \sw_{15} + \sw_{14} \sw_{23} \\
& \phantom{={}} + \sw_{14} \sw_{25} + \sw_{15} \sw_{23} - \sw_{15} \sw_{24} \\
& \phantom{={}} + 
 \sw_{23} \sw_{45} - \sw_{24} \sw_{35} + \sw_{25} \sw_{34} \\
& \phantom{={}} - \sw_{34} \sw_{35} + 
 2 \sw_{12} \sw_{34} \sw_{35}\\
& \phantom{={}}  - 2 \sw_{13} \sw_{14} \sw_{25} + 2 \sw_{13} \sw_{15} \sw_{24}.
\end{split}
\]

For the spanning property 
it suffices to consider products of three swap matrices of the form $b \sw_{pq}$ for some quadratic $b\in\cB_2$, leaving us with $25\cdot 10-6=244$ cases to treat. The identities needed are given in \Cref{app:swap3}.

Finally, to establish linear independence,  assume 
there is a linear combination of elements of
$\cB_3$ that vanishes. Now consider the six cubic terms expanded in terms of the Paulis.
Then $\sigma_X^1 \sigma_Z^2\sigma_Y^3\sigma_X^4\sigma_X^5$
only appears in $\sw_{12}\sw_{13}\sw_{45}$, whence the coefficient next to it has to be $0$. Next,
$\sigma_X^1 \sigma_Y^2\sigma_X^3\sigma_Z^4\sigma_X^5$
only appears in $\sw_{12}\sw_{14}\sw_{35}$, so the latter also cannot appear in a linear dependence relation.
Similarly, 
$\sigma_X^1 \sigma_Y^2\sigma_X^3\sigma_X^4\sigma_Z^5$
eliminates $\sw_{12}\sw_{15}\sw_{34}$.
Next, 
$\sigma_X^1 \sigma_Y^2\sigma_Y^3\sigma_Z^4\sigma_Y^5$
appears in 
$\sw_{13}\sw_{14}\sw_{25}$, but not it
$\sw_{13}\sw_{15}\sw_{24}$ or $
\sw_{14}\sw_{15}\sw_{23}$,
thus also eliminating $\sw_{13}\sw_{14}\sw_{25}$.
Next, 
$\sigma_X^1 \sigma_Y^2\sigma_Y^3\sigma_Y^4\sigma_Z^5$
appears in 
$\sw_{13}\sw_{15}\sw_{24}$ but not in
$\sw_{14}\sw_{15}\sw_{23}$, so 
$\sw_{13}\sw_{15}\sw_{24}$ also cannot appear in a linear dependence relation.
Since $\sw_{13}\sw_{15}\sw_{24}$ contains degree five terms in the Paulis which cannot appear in an element of $\cB_2$, we conclude that 
$\cB_3$ is linearly independent.
\end{proof}

\begin{prop}\label{prop:swap3}
A basis $\cB_3$ for
the linear space spanned by monomials of degree at most three in the swap algebra $\swa_n$
consists of
$\cB_2$ and two types of cubics,
\begin{enumerate}[\rm(a)]
 \item[\rm (Type 6)]
 $\sw_{ij}\sw_{k\ell}\sw_{pq}$, \quad $i<j,\; k<\ell,\; p<q,\; (i,j)<(k,\ell)<(p,q)$\footnote{Pairs are compared w.r.t.~the lex ordering.}; 
  \item[\rm (Types 5)]
$\sw_{ij}\sw_{ik}\sw_{pq}$,\;
$\sw_{ij}\sw_{ip}\sw_{kq}$,\; 
$\sw_{ij}\sw_{iq}\sw_{kp}$,\\
$\sw_{ik}\sw_{ip}\sw_{jq}$,\; 
$\sw_{ik}\sw_{iq}\sw_{jp}$,\;
$\sw_{ip}\sw_{iq}\sw_{jk}$,\\
\mbox{}\hfill $i<j<k<p<q$.
\end{enumerate}
\end{prop}

\begin{proof}
The spanning property follows from \Cref{lem:swap3}, so it suffices to establish linear independence of $\cB_3$.

We shall mimic the proof of \Cref{prop:swap2}. 
Assume there is a linear dependence among elements of $\cB_3$.
Consider the cubic terms in the swap matrices in this dependence.
Firstly, let us look at the (Type 6) terms.
The expansion of $\sw_{ij}\sw_{k\ell}\sw_{pq}$ 
in terms of the Paulis will yield a term
$\sigma_X^i\sigma_X^j\sigma_Y^k\sigma_Y^\ell\sigma_Z^p\sigma_Z^q$
that only occurs in one of the (Type 6) elements.
Thus the linear dependence cannot contain any (Type 6) elements.

Next, consider (Type 5) elements.
Since given $i<j<k<p<q$, not each of the six corresponding (Type 5) cubics has unique degree 5 terms in terms of the Paulis
(cf.~proof of \Cref{lem:swap3}), 
some care is needed.
As in \Cref{lem:swap3}, for each $i<j<k<p<q$, uniqueness of
degree five terms eliminates the first four in each list of associated (Type 5) cubics, namely
$\sw_{ij}\sw_{ik}\sw_{pq}$,\;
$\sw_{ij}\sw_{ip}\sw_{kq}$,\; $\sw_{ij}\sw_{iq}\sw_{kp}$,\;
$\sw_{ik}\sw_{ip}\sw_{jq}$. Once these are eliminated, we can as in \Cref{lem:swap3} also eliminate the remaining two (Type 5)s for each $i<j<k<p<q$. We are thus left with only terms from $\cB_2$ which are linearly independent by \Cref{prop:swap2}.
\end{proof}

\begin{rmk}
The cardinality of $\cB_3$ is
\begin{multline*}
\#\cB_3=
\#\cB_2 + \frac1{3!} {n\choose 2}{n-2\choose 2}{n-4\choose 2} +  6 {n\choose 5}\\ 
=\frac{1}{240} \left(5 n^6-63 n^5+335 n^4-845 n^3+1100 n^2-532 n+240\right).
\end{multline*} \qed
\end{rmk}

\subsection{Linear space spanned by the products of at most four swap matrices}\label{sssec:linswap4}

While we shall not bore the reader with the full details,
the situation does change a bit in degree four products of the swap matrices. Namely, products of four swap matrices on disjoint indices cease to be linearly independent over smaller products, as shown by the following identity:

{\scriptsize
\begin{align*}
 & \hspace{-5mm} \sw_{18} \sw_{27} \sw_{36} \sw_{45} = \\
& \phantom{+{}}{} \sw_{18} \sw_{27} \sw_{34} \sw_{56} + \sw_{18} \sw_{25} \sw_{36} \sw_{47} - \sw_{18} \sw_{25} \sw_{34} \sw_{67}
\\ & - \sw_{18} \sw_{23} \sw_{47} \sw_{56} +   
   \sw_{18} \sw_{23} \sw_{45} \sw_{67} + \sw_{16} \sw_{27} \sw_{38} \sw_{45}\\ 
& - \sw_{16} \sw_{27} \sw_{34} \sw_{58} - \sw_{16} \sw_{25} \sw_{37} \sw_{48} +      \sw_{16} \sw_{25} \sw_{34} \sw_{78}\\
& - \sw_{16} \sw_{24} \sw_{38} \sw_{57} + \sw_{16} \sw_{24} \sw_{37} \sw_{58} + \sw_{16} \sw_{23} \sw_{48} \sw_{57}\\
& -      \sw_{16} \sw_{23} \sw_{45} \sw_{78} - \sw_{15} \sw_{26} \sw_{38} \sw_{47} + \sw_{15} \sw_{26} \sw_{37} \sw_{48}\\
& + \sw_{15} \sw_{24} \sw_{38} \sw_{67} -      \sw_{15} \sw_{24} \sw_{37} \sw_{68} - \sw_{15} \sw_{23} \sw_{48} \sw_{67}\\
& + \sw_{15} \sw_{23} \sw_{47} \sw_{68} - \sw_{14} \sw_{27} \sw_{38} \sw_{56} + \sw_{14} \sw_{27} \sw_{36} \sw_{58}\\
& + \sw_{14} \sw_{26} \sw_{38} \sw_{57} - \sw_{14} \sw_{26} \sw_{37} \sw_{58} + \sw_{14} \sw_{25} \sw_{37} \sw_{68}\\
& -      \sw_{14} \sw_{25} \sw_{36} \sw_{78} - \sw_{14} \sw_{23} \sw_{57} \sw_{68} + \sw_{14} \sw_{23} \sw_{56} \sw_{78} \\
& - \sw_{13} \sw_{26} \sw_{48} \sw_{57} +      \sw_{13} \sw_{26} \sw_{47} \sw_{58} + \sw_{13} \sw_{25} \sw_{48} \sw_{67} \\
& - \sw_{13} \sw_{25} \sw_{47} \sw_{68} - \sw_{13} \sw_{24} \sw_{58} \sw_{67} +      \sw_{13} \sw_{24} \sw_{57} \sw_{68} \\
& + \sw_{12} \sw_{38} \sw_{47} \sw_{56} - \sw_{12} \sw_{38} \sw_{45} \sw_{67} - \sw_{12} \sw_{36} \sw_{47} \sw_{58}\\
& +      \sw_{12} \sw_{36} \sw_{45} \sw_{78} + \sw_{12} \sw_{34} \sw_{58} \sw_{67} - \sw_{12} \sw_{34} \sw_{56} \sw_{78} \\
& - \frac{1}{2} \sw_{38} \sw_{47} \sw_{56} +      \frac{1}{2} \sw_{38} \sw_{45} \sw_{67} + \frac{1}{2} \sw_{36} \sw_{47} \sw_{58}
 - \frac{1}{2} \sw_{36} \sw_{45} \sw_{78} \\
& - \frac{1}{2} \sw_{34} \sw_{58} \sw_{67} +      \frac{1}{2} \sw_{34} \sw_{56} \sw_{78} + \frac{1}{2} \sw_{27} \sw_{38} \sw_{56} - \frac{1}{2} \sw_{27} \sw_{38} \sw_{45}\\
& - \frac{1}{2} \sw_{27} \sw_{36} \sw_{58} +      \frac{1}{2} \sw_{27} \sw_{36} \sw_{45} + \frac{1}{2} \sw_{27} \sw_{34} \sw_{58} - \frac{1}{2} \sw_{27} \sw_{34} \sw_{56}\\
& + \frac{1}{2} \sw_{26} \sw_{48} \sw_{57} -      \frac{1}{2} \sw_{26} \sw_{47} \sw_{58} - \frac{1}{2} \sw_{26} \sw_{38} \sw_{57} + \frac{1}{2} \sw_{26} \sw_{38} \sw_{47}\\
& + \frac{1}{2} \sw_{26} \sw_{37} \sw_{58} -      \frac{1}{2} \sw_{26} \sw_{37} \sw_{48} - \frac{1}{2} \sw_{25} \sw_{48} \sw_{67} + \frac{1}{2} \sw_{25} \sw_{47} \sw_{68}\\
& - \frac{1}{2} \sw_{25} \sw_{37} \sw_{68} +      \frac{1}{2} \sw_{25} \sw_{37} \sw_{48} + \frac{1}{2} \sw_{25} \sw_{36} \sw_{78} - \frac{1}{2} \sw_{25} \sw_{36} \sw_{47}\\
& - \frac{1}{2} \sw_{25} \sw_{34} \sw_{78} +      \frac{1}{2} \sw_{25} \sw_{34} \sw_{67} + \frac{1}{2} \sw_{24} \sw_{58} \sw_{67} - \frac{1}{2} \sw_{24} \sw_{57} \sw_{68}\\
& - \frac{1}{2} \sw_{24} \sw_{38} \sw_{67} + \frac{1}{2} \sw_{24} \sw_{38} \sw_{57} + \frac{1}{2} \sw_{24} \sw_{37} \sw_{68} - \frac{1}{2} \sw_{24} \sw_{37} \sw_{58}\\
& + \frac{1}{2} \sw_{23} \sw_{57} \sw_{68} -      \frac{1}{2} \sw_{23} \sw_{56} \sw_{78} + \frac{1}{2} \sw_{23} \sw_{48} \sw_{67} - \frac{1}{2} \sw_{23} \sw_{48} \sw_{57}\\
& - \frac{1}{2} \sw_{23} \sw_{47} \sw_{68} +      \frac{1}{2} \sw_{23} \sw_{47} \sw_{56} + \frac{1}{2} \sw_{23} \sw_{45} \sw_{78} - \frac{1}{2} \sw_{23} \sw_{45} \sw_{67}\\
&  + \frac{1}{2} \sw_{18} \sw_{47} \sw_{56} -      \frac{1}{2} \sw_{18} \sw_{45} \sw_{67} - \frac{1}{2} \sw_{18} \sw_{36} \sw_{47} + \frac{1}{2} \sw_{18} \sw_{36} \sw_{45}\\
& + \frac{1}{2} \sw_{18} \sw_{34} \sw_{67} -      \frac{1}{2} \sw_{18} \sw_{34} \sw_{56} - \frac{1}{2} \sw_{18} \sw_{27} \sw_{56} + \frac{1}{2} \sw_{18} \sw_{27} \sw_{45}\\
& + \frac{1}{2} \sw_{18} \sw_{27} \sw_{36} - \frac{1}{2} \sw_{18} \sw_{27} \sw_{34} + \frac{1}{2} \sw_{18} \sw_{25} \sw_{67} - \frac{1}{2} \sw_{18} \sw_{25} \sw_{47}\\
& - \frac{1}{2} \sw_{18} \sw_{25} \sw_{36} +      \frac{1}{2} \sw_{18} \sw_{25} \sw_{34} - \frac{1}{2} \sw_{18} \sw_{23} \sw_{67} + \frac{1}{2} \sw_{18} \sw_{23} \sw_{56}\\
& + \frac{1}{2} \sw_{18} \sw_{23} \sw_{47} -      \frac{1}{2} \sw_{18} \sw_{23} \sw_{45} - \frac{1}{2} \sw_{16} \sw_{48} \sw_{57} + \frac{1}{2} \sw_{16} \sw_{45} \sw_{78}\\
& + \frac{1}{2} \sw_{16} \sw_{38} \sw_{57} -      \frac{1}{2} \sw_{16} \sw_{38} \sw_{45} - \frac{1}{2} \sw_{16} \sw_{37} \sw_{58} + \frac{1}{2} \sw_{16} \sw_{37} \sw_{48}\\
& - \frac{1}{2} \sw_{16} \sw_{34} \sw_{78} +      \frac{1}{2} \sw_{16} \sw_{34} \sw_{58} + \frac{1}{2} \sw_{16} \sw_{27} \sw_{58} - \frac{1}{2} \sw_{16} \sw_{27} \sw_{45}\\
& - \frac{1}{2} \sw_{16} \sw_{27} \sw_{38} + \frac{1}{2} \sw_{16} \sw_{27} \sw_{34} - \frac{1}{2} \sw_{16} \sw_{25} \sw_{78} + \frac{1}{2} \sw_{16} \sw_{25} \sw_{48}\\
& + \frac{1}{2} \sw_{16} \sw_{25} \sw_{37} -      \frac{1}{2} \sw_{16} \sw_{25} \sw_{34} - \frac{1}{2} \sw_{16} \sw_{24} \sw_{58} + \frac{1}{2} \sw_{16} \sw_{24} \sw_{57} \\
& + \frac{1}{2} \sw_{16} \sw_{24} \sw_{38} -      \frac{1}{2} \sw_{16} \sw_{24} \sw_{37} + \frac{1}{2} \sw_{16} \sw_{23} \sw_{78} - \frac{1}{2} \sw_{16} \sw_{23} \sw_{57}\\
& - \frac{1}{2} \sw_{16} \sw_{23} \sw_{48} +      \frac{1}{2} \sw_{16} \sw_{23} \sw_{45} + \frac{1}{2} \sw_{15} \sw_{48} \sw_{67} - \frac{1}{2} \sw_{15} \sw_{47} \sw_{68}\\
& - \frac{1}{2} \sw_{15} \sw_{38} \sw_{67} +      \frac{1}{2} \sw_{15} \sw_{38} \sw_{47} + \frac{1}{2} \sw_{15} \sw_{37} \sw_{68} - \frac{1}{2} \sw_{15} \sw_{37} \sw_{48}\\
& - \frac{1}{2} \sw_{15} \sw_{26} \sw_{48} + \frac{1}{2} \sw_{15} \sw_{26} \sw_{47} + \frac{1}{2} \sw_{15} \sw_{26} \sw_{38} - \frac{1}{2} \sw_{15} \sw_{26} \sw_{37} \\
& + \frac{1}{2} \sw_{15} \sw_{24} \sw_{68} -      \frac{1}{2} \sw_{15} \sw_{24} \sw_{67} - \frac{1}{2} \sw_{15} \sw_{24} \sw_{38} + \frac{1}{2} \sw_{15} \sw_{24} \sw_{37}\\
& - \frac{1}{2} \sw_{15} \sw_{23} \sw_{68} +      \frac{1}{2} \sw_{15} \sw_{23} \sw_{67} + \frac{1}{2} \sw_{15} \sw_{23} \sw_{48} - \frac{1}{2} \sw_{15} \sw_{23} \sw_{47}\\
& + \frac{1}{2} \sw_{14} \sw_{57} \sw_{68} -      \frac{1}{2} \sw_{14} \sw_{56} \sw_{78} - \frac{1}{2} \sw_{14} \sw_{38} \sw_{57} + \frac{1}{2} \sw_{14} \sw_{38} \sw_{56}\\
& - \frac{1}{2} \sw_{14} \sw_{37} \sw_{68} +      \frac{1}{2} \sw_{14} \sw_{37} \sw_{58} + \frac{1}{2} \sw_{14} \sw_{36} \sw_{78} - \frac{1}{2} \sw_{14} \sw_{36} \sw_{58}\\
& - \frac{1}{2} \sw_{14} \sw_{27} \sw_{58} + \frac{1}{2} \sw_{14} \sw_{27} \sw_{56} + \frac{1}{2} \sw_{14} \sw_{27} \sw_{38} - \frac{1}{2} \sw_{14} \sw_{27} \sw_{36} \\
& + \frac{1}{2} \sw_{14} \sw_{26} \sw_{58} -      \frac{1}{2} \sw_{14} \sw_{26} \sw_{57} - \frac{1}{2} \sw_{14} \sw_{26} \sw_{38} + \frac{1}{2} \sw_{14} \sw_{26} \sw_{37}\\
& + \frac{1}{2} \sw_{14} \sw_{25} \sw_{78} -      \frac{1}{2} \sw_{14} \sw_{25} \sw_{68} - \frac{1}{2} \sw_{14} \sw_{25} \sw_{37} + \frac{1}{2} \sw_{14} \sw_{25} \sw_{36}\\
& - \frac{1}{2} \sw_{14} \sw_{23} \sw_{78} +      \frac{1}{2} \sw_{14} \sw_{23} \sw_{68} + \frac{1}{2} \sw_{14} \sw_{23} \sw_{57} - \frac{1}{2} \sw_{14} \sw_{23} \sw_{56}\\
& + \frac{1}{2} \sw_{13} \sw_{58} \sw_{67} -      \frac{1}{2} \sw_{13} \sw_{57} \sw_{68} - \frac{1}{2} \sw_{13} \sw_{48} \sw_{67} + \frac{1}{2} \sw_{13} \sw_{48} \sw_{57}\\
& + \frac{1}{2} \sw_{13} \sw_{47} \sw_{68} - \frac{1}{2} \sw_{13} \sw_{47} \sw_{58} - \frac{1}{2} \sw_{13} \sw_{26} \sw_{58} + \frac{1}{2} \sw_{13} \sw_{26} \sw_{57}\\
& + \frac{1}{2} \sw_{13} \sw_{26} \sw_{48} -      \frac{1}{2} \sw_{13} \sw_{26} \sw_{47} + \frac{1}{2} \sw_{13} \sw_{25} \sw_{68} - \frac{1}{2} \sw_{13} \sw_{25} \sw_{67} \\
& - \frac{1}{2} \sw_{13} \sw_{25} \sw_{48} +      \frac{1}{2} \sw_{13} \sw_{25} \sw_{47} - \frac{1}{2} \sw_{13} \sw_{24} \sw_{68} + \frac{1}{2} \sw_{13} \sw_{24} \sw_{67} \\
& + \frac{1}{2} \sw_{13} \sw_{24} \sw_{58} -      \frac{1}{2} \sw_{13} \sw_{24} \sw_{57} - \frac{1}{2} \sw_{12} \sw_{58} \sw_{67} + \frac{1}{2} \sw_{12} \sw_{56} \sw_{78}\\
& + \frac{1}{2} \sw_{12} \sw_{47} \sw_{58} -      \frac{1}{2} \sw_{12} \sw_{47} \sw_{56} - \frac{1}{2} \sw_{12} \sw_{45} \sw_{78} + \frac{1}{2} \sw_{12} \sw_{45} \sw_{67}\\
& + \frac{1}{2} \sw_{12} \sw_{38} \sw_{67} - \frac{1}{2} \sw_{12} \sw_{38} \sw_{56} - \frac{1}{2} \sw_{12} \sw_{38} \sw_{47} + \frac{1}{2} \sw_{12} \sw_{38} \sw_{45} \\
& - \frac{1}{2} \sw_{12} \sw_{36} \sw_{78} +      \frac{1}{2} \sw_{12} \sw_{36} \sw_{58} + \frac{1}{2} \sw_{12} \sw_{36} \sw_{47} - \frac{1}{2} \sw_{12} \sw_{36} \sw_{45}\\
& + \frac{1}{2} \sw_{12} \sw_{34} \sw_{78} -      \frac{1}{2} \sw_{12} \sw_{34} \sw_{67} - \frac{1}{2} \sw_{12} \sw_{34} \sw_{58} + \frac{1}{2} \sw_{12} \sw_{34} \sw_{56}.
 \end{align*}
 }

To form a basis $\cB_4$
for the linear space spanned by products of at most four swap matrices, one takes $\cB_3$ and adds
\begin{enumerate}[\rm(a)]
\item[(Types 7)]
For each 7-tuple $i<j<k<\ell<p<q<r$ a select 36 products
of four swap matrices involving these seven indices;
\item[(Type 8)]
a certain selection of 91 out of the 105 distinct products of
four swap matrices on distinct indices $i<j<k<\ell<p<q<r<s$.
\end{enumerate}
Thus
\[
\begin{split}
\#\cB_4& =\#\cB_3+ 36{n\choose 7}+ \frac{91}{105} \frac{1}{4!}
{n\choose 2}{n-2\choose 2}{n-4\choose 2} {n-6\choose 2} \\
& = 
\frac1{40320}\big(91 n^8-2260 n^7+24094 n^6-138544 n^5+460579
   n^4\\
   & \phantom{\frac1{40320}\big(- } -869260 n^3+865956 n^2-340656 n+40320\big).
\end{split}
\]

\section{Gr\"obner basis for the Swap algebra \texorpdfstring{$\salg_4$}{A4}}
\label{app:GB34}

\begin{ex}[$n=4$]
The GB for $\sidl_4$ has 34 elements, namely
{\scriptsize
\begin{gather*}
-1 + s_{12}^2,\;  1 - s_{12} - s_{13} - s_{23} + s_{12} s_{13} + s_{12} s_{23},\;  
 1 - s_{12} - s_{14} - s_{24} + s_{12} s_{14} + s_{12} s_{24},\\  1 - s_{12} - s_{13} - s_{23} + s_{12} s_{13} + s_{13} s_{12},\;  
 -1 + s_{13}^2,\;  -s_{12} s_{13} + s_{13} s_{23},\;  1 - s_{13} - s_{14} - s_{34} + s_{13} s_{14} + s_{13} s_{34},\\  
 1 - s_{12} - s_{14} - s_{24} + s_{12} s_{14} + s_{14} s_{12},\;  1 - s_{13} - s_{14} - s_{34} + s_{13} s_{14} + s_{14} s_{13},\;  
 -1 + s_{14}^2,\\  -s_{12} s_{14} + s_{14} s_{24},\;  -s_{13} s_{14} + s_{14} s_{34},\;  -s_{12} s_{13} + s_{23} s_{12},\\  
 1 - s_{12} - s_{13} - s_{23} + s_{12} s_{13} + s_{23} s_{13},\;  -s_{14} s_{23} + s_{23} s_{14},\;  -1 + s_{23}^2,\\ 
 1 - s_{23} - s_{24} - s_{34} + s_{23} s_{24} + s_{23} s_{34},\;  -s_{12} s_{14} + s_{24} s_{12},\;  -s_{13} s_{24} + s_{24} s_{13},\\  
 1 - s_{12} - s_{14} - s_{24} + s_{12} s_{14} + s_{24} s_{14},\;  1 - s_{23} - s_{24} - s_{34} + s_{23} s_{24} + s_{24} s_{23},\;  
 -1 + s_{24}^2,\\  -s_{23} s_{24} + s_{24} s_{34},\;  -s_{12} s_{34} + s_{34} s_{12},\;  -s_{13} s_{14} + s_{34} s_{13},\\  
 1 - s_{13} - s_{14} - s_{34} + s_{13} s_{14} + s_{34} s_{14},\;  -s_{23} s_{24} + s_{34} s_{23},\;  
 1 - s_{23} - s_{24} - s_{34} + s_{23} s_{24} + s_{34} s_{24},\  -1 + s_{34}^2,\\  
 -1/2 + s_{12}/2 + s_{14}/2 + s_{23}/2 + s_{34}/2 - s_{12} s_{13}/2 - s_{12} s_{14}/2 - s_{12} s_{34}/2 - 
  s_{13} s_{14}/2 + s_{13} s_{24}/2 - s_{14} s_{23}/2 - s_{23} s_{24}/2 + s_{12} s_{13} s_{14},\\  
 1/2 - s_{12}/2 - s_{14}/2 + s_{23}/2 - s_{34}/2 - s_{12} s_{13}/2 + s_{12} s_{14}/2 + s_{12} s_{34}/2 + 
  s_{13} s_{14}/2 - s_{13} s_{24}/2 - s_{14} s_{23}/2 - s_{23} s_{24}/2 + s_{12} s_{13} s_{24},\\  
 1/2 - s_{12}/2 + s_{14}/2 - s_{23}/2 - s_{34}/2 + s_{12} s_{13}/2 - s_{12} s_{14}/2 + s_{12} s_{34}/2 - 
  s_{13} s_{14}/2 - s_{13} s_{24}/2 - s_{14} s_{23}/2 + s_{23} s_{24}/2 + s_{12} s_{14} s_{23},\\  
 -1/2 + s_{12}/2 + s_{14}/2 + s_{23}/2 + s_{34}/2 - s_{12} s_{13}/2 - s_{12} s_{14}/2 - s_{12} s_{34}/2 - 
  s_{13} s_{14}/2 + s_{13} s_{24}/2 - s_{14} s_{23}/2 - s_{23} s_{24}/2 + s_{13} s_{14} s_{23},\\  
 -1/2 + s_{12}/2 + s_{14}/2 + s_{23}/2 + s_{34}/2 - s_{12} s_{13}/2 - s_{12} s_{14}/2 - s_{12} s_{34}/2 - 
  s_{13} s_{14}/2 + s_{13} s_{24}/2 - s_{14} s_{23}/2 - s_{23} s_{24}/2 + s_{14} s_{23} s_{24}
\end{gather*}
}Notice that unlike in the $n=3$ case of \Cref{prop:swapGB3} also higher degree polynomials, namely cubics, appear.
\end{ex}

\begin{ex}
The following table gathers some data about GBs for $\salg_n$ for small values of $n$.

\begin{center}
\begin{tabular}{|c|c|c|c|c|c|}
\hline
& & highest degree  & number of highest \\
$n$ & size(GB) & polynomial in GB &  degree polynomials in GB\\
\hline
5 & 110 & 3 & 35\\
6 & 305 & 4 & 10 \\
7 & 620 & 4 & 102 \\
8 & 1665 & 5 & 37\\
\hline
\end{tabular}
\end{center}
\end{ex}

\newpage

\ssec{Gr\"obner bases for irreps \texorpdfstring{$[4-k,k]$}{[4-k,k]}}
Under the irrep $[3,1]$ the image of $\hampol {\clique 4}$ 
is $8\ I$. Under the irrep $[2,2]$
the image is $12\ I$. With this we can easily compute GB$_{3,1}$ 
and GB$_{2,2}$ with the help of a computer algebra system.
The former is given in \cref{eq:GB31}, and the latter in
\cref{eq:GB22}.
{\footnotesize
\begin{align}\label{eq:GB31}
\begin{autobreak}
\phantom{=}
-2 + s_{12} + s_{13} + s_{14} + s_{23} + s_{24} + s_{34},\;  -1 + s_{12}^2,\;  
 1 - s_{12} - s_{13} - s_{23} + s_{12} s_{13} + s_{12} s_{23},\;  1 - s_{12} - s_{14} - s_{24} + s_{12} s_{14} + s_{12} s_{24},\;  
 1 - s_{12} - s_{13} - s_{23} + s_{12} s_{13} + s_{13} s_{12},\;  -1 + s_{13}^2,\;  -s_{12} s_{13} + s_{13} s_{23},\;  
 1 - s_{13} - s_{24} + s_{13} s_{24},\;  1 - s_{12} - s_{14} - s_{24} + s_{12} s_{14} + s_{14} s_{12},\;  
 -1 + s_{12} + s_{23} + s_{24} + s_{13} s_{14} + s_{14} s_{13},\;  -1 + s_{14}^2,\;  1 - s_{14} - s_{23} + s_{14} s_{23},\;  
 -s_{12} s_{14} + s_{14} s_{24},\;  -s_{12} s_{13} + s_{23} s_{12},\;  1 - s_{12} - s_{13} - s_{23} + s_{12} s_{13} + s_{23} s_{13},\;  
 1 - s_{14} - s_{23} + s_{23} s_{14},\;  -1 + s_{23}^2,\;  s_{13} - s_{24} - s_{12} s_{13} + s_{12} s_{14} - s_{13} s_{14} + 
  s_{23} s_{24},\;  -s_{12} s_{14} + s_{24} s_{12},\;  1 - s_{13} - s_{24} + s_{24} s_{13},\;  
 1 - s_{12} - s_{14} - s_{24} + s_{12} s_{14} + s_{24} s_{14},\;  -1 + s_{12} + s_{14} + s_{24} + s_{12} s_{13} - 
  s_{12} s_{14} + s_{13} s_{14} + s_{24} s_{23},\;  -1 + s_{24}^2,\;  s_{13} - s_{12} s_{13} - s_{13} s_{14} + s_{12} s_{13} s_{14}
\end{autobreak}
\end{align}
}

{\footnotesize
\begin{align}\label{eq:GB22}
\begin{autobreak}
\phantom{=}
s_{12} + s_{13} + s_{14},\; s_{12} + s_{13} + s_{23},\; -s_{13} + s_{24},\; -s_{12} + s_{34},\; -1 + s_{12}^2,\; 
 1 + s_{12} s_{13} + s_{13} s_{12},\; -1 + s_{13}^2
\end{autobreak}
\end{align}
}

\section{\texorpdfstring{\Cref{ex:sym_las}$^2$}{Example 4.8 Squared}}\label{appd:Ex38}
We present the second relaxation of \Cref{ex:sym_las}.
Let $n=3$ and $d=2$. Then
\[
V_2(3)=(1,\, s_{12},\, s_{13},\, s_{23},\, s_{12}^2,\, s_{12}s_{13},\, s_{12}s_{23},\, 
s_{13}s_{12},\,  s_{13}^2,\, s_{13}s_{23},\, s_{23}s_{12},\, s_{23}s_{13},\, s_{23}^2
)^*.
\]
Thus $\mmp_2^3$ is the following $13\times 13$ matrix
\[
\resizebox{\linewidth}{!}{$
\left[
\begin{array}{cccccccccccccccccccccc}
1 & s_{12} & s_{13} & s_{23} & s_{12}^2 & s_{12} s_{13} & s_{12} s_{23} & s_{13} s_{12} & s_{13}^2 & s_{13} s_{23} & s_{23} s_{12} & s_{23} s_{13} & s_{23}^2 \\[1mm]  s_{12} & s_{12}^2 & s_{12} s_{13} & s_{12} s_{23} & s_{12}^3 & s_{12}^2 s_{13} & s_{12}^2 s_{23} & s_{12} s_{13} s_{12} & s_{12} s_{13}^2 & s_{12} s_{13} s_{23} & s_{12} s_{23} s_{12} & s_{12} s_{23} s_{13} & s_{12} s_{23}^2 \\[1mm]  s_{13} & s_{13} s_{12} & s_{13}^2 & s_{13} s_{23} & s_{13} s_{12}^2 & s_{13} s_{12} s_{13} & s_{13} s_{12} s_{23} & s_{13}^2 s_{12} & s_{13}^3 & s_{13}^2 s_{23} & s_{13} s_{23} s_{12} & s_{13} s_{23} s_{13} & s_{13} s_{23}^2 \\[1mm]  s_{23} & s_{23} s_{12} & s_{23} s_{13} & s_{23}^2 & s_{23} s_{12}^2 & s_{23} s_{12} s_{13} & s_{23} s_{12} s_{23} & s_{23} s_{13} s_{12} & s_{23} s_{13}^2 & s_{23} s_{13} s_{23} & s_{23}^2 s_{12} & s_{23}^2 s_{13} & s_{23}^3 \\[1mm]  s_{12}^2 & s_{12}^3 & s_{12}^2 s_{13} & s_{12}^2 s_{23} & s_{12}^4 & s_{12}^3 s_{13} & s_{12}^3 s_{23} & s_{12}^2 s_{13} s_{12} & s_{12}^2 s_{13}^2 & s_{12}^2 s_{13} s_{23} & s_{12}^2 s_{23} s_{12} & s_{12}^2 s_{23} s_{13} & s_{12}^2 s_{23}^2 \\[1mm]  s_{13} s_{12} & s_{13} s_{12}^2 & s_{13} s_{12} s_{13} & s_{13} s_{12} s_{23} & s_{13} s_{12}^3 & s_{13} s_{12}^2 s_{13} & s_{13} s_{12}^2 s_{23} & s_{13} s_{12} s_{13} s_{12} & s_{13} s_{12} s_{13}^2 & s_{13} s_{12} s_{13} s_{23} & s_{13} s_{12} s_{23} s_{12} & s_{13} s_{12} s_{23} s_{13} & s_{13} s_{12} s_{23}^2 \\[1mm]  s_{23} s_{12} & s_{23} s_{12}^2 & s_{23} s_{12} s_{13} & s_{23} s_{12} s_{23} & s_{23} s_{12}^3 & s_{23} s_{12}^2 s_{13} & s_{23} s_{12}^2 s_{23} & s_{23} s_{12} s_{13} s_{12} & s_{23} s_{12} s_{13}^2 & s_{23} s_{12} s_{13} s_{23} & s_{23} s_{12} s_{23} s_{12} & s_{23} s_{12} s_{23} s_{13} & s_{23} s_{12} s_{23}^2 \\[1mm]  s_{12} s_{13} & s_{12} s_{13} s_{12} & s_{12} s_{13}^2 & s_{12} s_{13} s_{23} & s_{12} s_{13} s_{12}^2 & s_{12} s_{13} s_{12} s_{13} & s_{12} s_{13} s_{12} s_{23} & s_{12} s_{13}^2 s_{12} & s_{12} s_{13}^3 & s_{12} s_{13}^2 s_{23} & s_{12} s_{13} s_{23} s_{12} & s_{12} s_{13} s_{23} s_{13} & s_{12} s_{13} s_{23}^2 \\[1mm]  s_{13}^2 & s_{13}^2 s_{12} & s_{13}^3 & s_{13}^2 s_{23} & s_{13}^2 s_{12}^2 & s_{13}^2 s_{12} s_{13} & s_{13}^2 s_{12} s_{23} & s_{13}^3 s_{12} & s_{13}^4 & s_{13}^3 s_{23} & s_{13}^2 s_{23} s_{12} & s_{13}^2 s_{23} s_{13} & s_{13}^2 s_{23}^2 \\[1mm]  s_{23} s_{13} & s_{23} s_{13} s_{12} & s_{23} s_{13}^2 & s_{23} s_{13} s_{23} & s_{23} s_{13} s_{12}^2 & s_{23} s_{13} s_{12} s_{13} & s_{23} s_{13} s_{12} s_{23} & s_{23} s_{13}^2 s_{12} & s_{23} s_{13}^3 & s_{23} s_{13}^2 s_{23} & s_{23} s_{13} s_{23} s_{12} & s_{23} s_{13} s_{23} s_{13} & s_{23} s_{13} s_{23}^2 \\[1mm]  s_{12} s_{23} & s_{12} s_{23} s_{12} & s_{12} s_{23} s_{13} & s_{12} s_{23}^2 & s_{12} s_{23} s_{12}^2 & s_{12} s_{23} s_{12} s_{13} & s_{12} s_{23} s_{12} s_{23} & s_{12} s_{23} s_{13} s_{12} & s_{12} s_{23} s_{13}^2 & s_{12} s_{23} s_{13} s_{23} & s_{12} s_{23}^2 s_{12} & s_{12} s_{23}^2 s_{13} & s_{12} s_{23}^3 \\[1mm]  s_{13} s_{23} & s_{13} s_{23} s_{12} & s_{13} s_{23} s_{13} & s_{13} s_{23}^2 & s_{13} s_{23} s_{12}^2 & s_{13} s_{23} s_{12} s_{13} & s_{13} s_{23} s_{12} s_{23} & s_{13} s_{23} s_{13} s_{12} & s_{13} s_{23} s_{13}^2 & s_{13} s_{23} s_{13} s_{23} & s_{13} s_{23}^2 s_{12} & s_{13} s_{23}^2 s_{13} & s_{13} s_{23}^3 \\[1mm]  s_{23}^2 & s_{23}^2 s_{12} & s_{23}^2 s_{13} & s_{23}^3 & s_{23}^2 s_{12}^2 & s_{23}^2 s_{12} s_{13} & s_{23}^2 s_{12} s_{23} & s_{23}^2 s_{13} s_{12} & s_{23}^2 s_{13}^2 & s_{23}^2 s_{13} s_{23} & s_{23}^3 s_{12} & s_{23}^3 s_{13} & s_{23}^4
\end{array}
\right]
$}
\]
leading to  the second relaxation which would be a $13\times 13$ SDP. However, the Veronese  $V_2(3)$ has redundancies. For instance, $s_{12}^2=1$, etc. Eliminating redundancies (more precisely, picking a basis for the image of $\free{\RR} s_2$ in 
$\free{\RR} s/\cI^{S_3}$) leads to the reduced Veronese, which we shall by abuse of notation still call $V_2(3)$:
\[
V_2(3)=(1,\; s_{12},\; s_{13},\; s_{23},\; s_{12}  s_{13},\; s_{12}  s_{23})^*.
\]
This yields a significantly smaller moment matrix pattern,
\[
\mmp_2^3=
\bem
1 & s_{12} & s_{13} & s_{23} & s_{12} s_{13} & s_{12} s_{23}\\[1mm] s_{12} & s_{12}^2 & s_{12} s_{13} & s_{12} s_{23} & s_{12}^2 s_{13} & s_{12}^2 s_{23}\\[1mm] s_{13} & s_{13} s_{12} & s_{13}^2 & s_{13} s_{23} & s_{13} s_{12} s_{13} & s_{13} s_{12} s_{23}\\[1mm] s_{23} & s_{23} s_{12} & s_{23} s_{13} & s_{23}^2 & s_{23} s_{12} s_{13} & s_{23} s_{12} s_{23}\\[1mm] s_{13} s_{12} & s_{13} s_{12}^2 & s_{13} s_{12} s_{13} & s_{13} s_{12} s_{23} & s_{13} s_{12}^2 s_{13} & s_{13} s_{12}^2 s_{23}\\[1mm] s_{23} s_{12} & s_{23} s_{12}^2 & s_{23} s_{12} s_{13} & s_{23} s_{12} s_{23} & s_{23} s_{12}^2 s_{13} & s_{23} s_{12}^2 s_{23}.
\eem
\]
Applying the replacement rules of \Cref{ex:sym_las} simplifies this matrix further,
\[
\mmp_2^3=
\bem
1 & s_{12} & s_{13} & s_{23} & s_{12} s_{13} & s_{12} s_{23}\\[1mm] s_{12} & 1 & s_{12} s_{13} & s_{12} s_{23} & s_{13} & s_{23}\\[1mm] s_{13} & s_{12} s_{23} & 1 & s_{12} s_{13} & s_{23} & s_{12}\\[1mm] s_{23} & s_{12} s_{13} & s_{12} s_{23} & 1 & s_{12} & s_{13}\\[1mm] s_{12} s_{23} & s_{13} & s_{23} & s_{12} & 1 & s_{12} s_{13}\\[1mm] s_{12} s_{13} & s_{23} & s_{12} & s_{13} & s_{12} s_{23} & 1
\eem
\]
We thus obtain a smaller second level hierarchy SDP:
\beq
\begin{aligned}\label{eq:wasteSDP2}
\nuflip_2(h)= 
\max & \;  \ell_{12} + \ell_{13} + \ell_{23}\\
\text{s.t. } & \; \\[-.5cm]
& \bem 
1 & \ell_{12} & \ell_{13} & \ell_{23} & \ell_{12,13}& \ell_{12} \ell_{23}\\[1mm] \ell_{12} & 1 & \ell_{12,13}& \ell_{12,23}& \ell_{13} & \ell_{23}\\[1mm] \ell_{13} & \ell_{12,23}& 1 & \ell_{12,13}& \ell_{23} & \ell_{12}\\[1mm] \ell_{23} & \ell_{12,13}& \ell_{12,23}& 1 & \ell_{12} & \ell_{13}\\[1mm] \ell_{12,23}& \ell_{13} & \ell_{23} & \ell_{12} & 1 & \ell_{12} \ell_{13}\\[1mm] \ell_{12,13}& \ell_{23} & \ell_{12} & \ell_{13} & \ell_{12,23}& 1
\eem \succeq 0.\\
\end{aligned}
\eeq
Since the matrix in \cref{eq:wasteSDP2} must be symmetric, we obtain $\ell_{12,23}=\ell_{12,13}$, reducing the number of unknowns by one,
\beq
\begin{aligned}\label{eq:wasteSDP22}
\nuflip_2(h)= 
\max & \;  \ell_{12} + \ell_{13} + \ell_{23}\\
\text{s.t. } & \; \\[-.5cm]
& \bem 
1 & \ell_{12} & \ell_{13} & \ell_{23} & \ell_{12,13}& \ell_{12} \ell_{23}\\[1mm] \ell_{12} & 1 & \ell_{12,13}& \ell_{12,13}& \ell_{13} & \ell_{23}\\[1mm] \ell_{13} & \ell_{12,13}& 1 & \ell_{12,13}& \ell_{23} & \ell_{12}\\[1mm] \ell_{23} & \ell_{12,13}& \ell_{12,13}& 1 & \ell_{12} & \ell_{13}\\[1mm] \ell_{12,13}& \ell_{13} & \ell_{23} & \ell_{12} & 1 & \ell_{12} \ell_{13}\\[1mm] \ell_{12,13}& \ell_{23} & \ell_{12} & \ell_{13} & \ell_{12,13}& 1
\eem \succeq 0\\
 = 3. \phantom{=} &
\end{aligned}
\eeq
Finally, since $\dim\C[S_3]=6$ is equal to the size of the reduced Veronese, the second relaxation $\nuflip_2(h)$ is automatically exact.

\begin{ex}\label{ex:S4}
Now consider $n=4$, and $\cI=\cI^{S_4}$. 
In this case the replacement rules are\footnote{In order to obtain a Gr\"obner basis (cf.~\Cref{sec:GB}) for the grlex order w.r.t.~$s_{12}< s_{13}< s_{14}< s_{23}< s_{24}< s_{34}$, one needs to add three additional replacement rules to the ones given in \cref{eq:S4preGB},
namely $s_{13} s_{14} s_{23} \to s_{12} s_{13} s_{14} ,\; 
s_{14} s_{23} s_{24} \to s_{12} s_{13} s_{14} ,\; 
s_{14} s_{23} s_{34} \to s_{12} s_{14} s_{23}$.}
\begin{align}\label{eq:S4preGB}
\begin{autobreak}
\phantom{=}
s_{12} ^2\to 1,\; 
s_{13} s_{12} \to s_{12} s_{23} ,\; 
s_{13} ^2\to 1,\; 
s_{13} s_{23} \to s_{12} s_{13} ,\; 
s_{14} s_{12} \to s_{12} s_{24} ,\; 
s_{14} s_{13} \to s_{13} s_{34} ,\; 
s_{14} ^2\to 1,\; 
s_{14} s_{24} \to s_{12} s_{14} ,\; 
s_{14} s_{34} \to s_{13} s_{14} ,\; 
s_{23} s_{12} \to s_{12} s_{13} ,\; 
s_{23} s_{13} \to s_{12} s_{23} ,\; 
s_{23} s_{14} \to s_{14} s_{23} ,\; 
s_{23} ^2\to 1,\; 
s_{24} s_{12} \to s_{12} s_{14} ,\; 
s_{24} s_{13} \to s_{13} s_{24} ,\; 
s_{24} s_{14} \to s_{12} s_{24} ,\; 
s_{24} s_{23} \to s_{23} s_{34} ,\; 
s_{24} ^2\to 1,\; 
s_{24} s_{34} \to s_{23} s_{24} ,\; 
s_{34} s_{12} \to s_{12} s_{34} ,\; 
s_{34} s_{13} \to s_{13} s_{14} ,\; 
s_{34} s_{14} \to s_{13} s_{34} ,\; 
s_{34} s_{23} \to s_{23} s_{24} ,\; 
s_{34} s_{24} \to s_{23} s_{34} ,\; 
s_{34} ^2\to 1.
\end{autobreak}
\end{align}
Then
$V_1(4)= ( 1, s_{12}, s_{13}, s_{14}, s_{23}, s_{24}, s_{34})^*$,
and
\[
\mmp_1^4=
\bem
1 & s_{12} & s_{13} & s_{14} & s_{23} & s_{24}\\[1mm] s_{12} & 1 & s_{12}  s_{13} & s_{12}  s_{14} & 
  s_{12}  s_{23} & s_{12}  s_{24}\\[1mm] s_{13} & s_{12}  s_{23} & 1 & s_{13}  s_{14} & 
  s_{12}  s_{13} & s_{13}  s_{24}\\[1mm] s_{14} & s_{12}  s_{24} & s_{13}  s_{34} & 1 & 
  s_{14}  s_{23} & s_{12}  s_{14}\\[1mm] s_{23} & s_{12}  s_{13} & s_{12}  s_{23} & s_{14}  s_{23} & 
  1 & s_{23}  s_{24}\\[1mm] s_{24} & s_{12}  s_{14} & s_{13}  s_{24} & s_{12}  s_{24} & 
  s_{23}  s_{34} & 1
\eem
\]
leading to the SDP constraint
\[
\bem
1 & \ell_{12} & \ell_{13} & \ell_{14} & \ell_{23} & \ell_{24}\\[1mm] \ell_{12} & 1 & \ell_{12,13} & \ell_{12,14} & 
  \ell_{12,23} & \ell_{12,24}\\[1mm] \ell_{13} & \ell_{12,23} & 1 & \ell_{13,14} & 
  \ell_{12,13} & \ell_{13,24}\\[1mm] \ell_{14} & \ell_{12,24} & \ell_{13,34} & 1 & 
  \ell_{14,23} & \ell_{12,14}\\[1mm] \ell_{23} & \ell_{12,13} & \ell_{12,23} & \ell_{14,23} & 
  1 & \ell_{23,24}\\[1mm] \ell_{24} & \ell_{12,14} & \ell_{13,24} & \ell_{12,24} & 
  \ell_{23,34} & 1
  \eem \succeq0.
\]
As before, symmetry of this matrix yields a few additional linear constraints, namely
\[
\ell_{12,13}=\ell_{12,23},\; \ell_{12,14} =\ell_{12,24},\;  \ell_{13,14}= \ell_{13,34}.
\]

The full $V_2(4)$ has 43 entries, and its reduced form is
\begin{multline*}
V_2(4)=
\big(1,\; s_{12},\; s_{13},\; s_{14},\; s_{23},\; s_{24},\; s_{34},\; s_{12} s_{13},\; s_{12} s_{14},\; s_{12} s_{23},\; s_{12} s_{24},\; s_{12} s_{34},\; s_{13} s_{14},\\ \; s_{13} s_{24},\; s_{13} s_{34}, \; s_{14} s_{23},\; s_{23} s_{24},\; s_{23} s_{34}\big)^*,
\end{multline*}
leading to the $18\times18$ moment matrix pattern
\[
\resizebox{\linewidth}{!}{$
\left[
\begin{array}{ccccccccccccccccccccccccccc}
1 & s_{12} & s_{13} & s_{14} & s_{23} & s_{24} & s_{34} & s_{12}  s_{13} & s_{12}  s_{14} & s_{12}  s_{23} &
   s_{12}  s_{24} & s_{12}  s_{34} & s_{13}  s_{14} & s_{13}  s_{24} & s_{13}  s_{34} & 
  s_{14}  s_{23} & s_{23}  s_{24} & s_{23}  s_{34}\\[1mm] s_{12} & 1 & s_{12}  s_{13} & 
  s_{12}  s_{14} & s_{12}  s_{23} & s_{12}  s_{24} & s_{12}  s_{34} & s_{13} & s_{14} & s_{23} & s_{24} & 
  s_{34} & s_{12}  s_{13}  s_{14} & s_{12}  s_{13}  s_{24} & s_{12}  s_{13}  s_{34} & 
  s_{12}  s_{14}  s_{23} & s_{12}  s_{23}  s_{24} & s_{12}  s_{23}  s_{34}\\[1mm] s_{13} & 
  s_{12}  s_{23} & 1 & s_{13}  s_{14} & s_{12}  s_{13} & s_{13}  s_{24} & s_{13}  s_{34} & s_{23} & 
  s_{12}  s_{14}  s_{23} & s_{12} & s_{12}  s_{23}  s_{24} & s_{12}  s_{23}  s_{34} & s_{14} & 
  s_{24} & s_{34} & s_{12}  s_{13}  s_{14} & s_{12}  s_{13}  s_{24} & 
  s_{12}  s_{13}  s_{34}\\[1mm] s_{14} & s_{12}  s_{24} & s_{13}  s_{34} & 1 & s_{14}  s_{23} & 
  s_{12}  s_{14} & s_{13}  s_{14} & s_{12}  s_{13}  s_{24} & s_{24} & s_{12}  s_{23}  s_{34} & 
  s_{12} & s_{12}  s_{23}  s_{24} & s_{34} & s_{12}  s_{13}  s_{34} & s_{13} & s_{23} & 
  s_{12}  s_{13}  s_{14} & s_{12}  s_{14}  s_{23}\\[1mm] s_{23} & s_{12}  s_{13} & s_{12}  s_{23} &
   s_{14}  s_{23} & 1 & s_{23}  s_{24} & s_{23}  s_{34} & s_{12} & s_{12}  s_{13}  s_{14} & s_{13} &
   s_{12}  s_{13}  s_{24} & s_{12}  s_{13}  s_{34} & s_{12}  s_{14}  s_{23} & 
  s_{12}  s_{23}  s_{24} & s_{12}  s_{23}  s_{34} & s_{14} & s_{24} & s_{34}\\[1mm] s_{24} & 
  s_{12}  s_{14} & s_{13}  s_{24} & s_{12}  s_{24} & s_{23}  s_{34} & 1 & s_{23}  s_{24} & 
  s_{12}  s_{13}  s_{34} & s_{12} & s_{12}  s_{14}  s_{23} & s_{14} & s_{12}  s_{13}  s_{14} & 
  s_{12}  s_{23}  s_{24} & s_{13} & s_{12}  s_{13}  s_{24} & s_{12}  s_{23}  s_{34} & s_{34} & 
  s_{23}\\[1mm] s_{34} & s_{12}  s_{34} & s_{13}  s_{14} & s_{13}  s_{34} & s_{23}  s_{24} & 
  s_{23}  s_{34} & 1 & s_{12}  s_{13}  s_{14} & s_{12}  s_{13}  s_{34} & 
  s_{12}  s_{23}  s_{24} & s_{12}  s_{23}  s_{34} & s_{12} & s_{13} & s_{12}  s_{14}  s_{23} & 
  s_{14} & s_{12}  s_{13}  s_{24} & s_{23} & s_{24}\\[1mm] s_{12}  s_{23} & s_{13} & s_{23} & 
  s_{12}  s_{14}  s_{23} & s_{12} & s_{12}  s_{23}  s_{24} & s_{12}  s_{23}  s_{34} & 1 & 
  s_{13}  s_{14} & s_{12}  s_{13} & s_{13}  s_{24} & s_{13}  s_{34} & s_{14}  s_{23} & 
  s_{23}  s_{24} & s_{23}  s_{34} & s_{12}  s_{14} & s_{12}  s_{24} & 
  s_{12}  s_{34}\\[1mm] s_{12}  s_{24} & s_{14} & s_{12}  s_{13}  s_{24} & s_{24} & 
  s_{12}  s_{23}  s_{34} & s_{12} & s_{12}  s_{23}  s_{24} & s_{13}  s_{34} & 1 & 
  s_{14}  s_{23} & s_{12}  s_{14} & s_{13}  s_{14} & s_{23}  s_{24} & s_{12}  s_{13} & 
  s_{13}  s_{24} & s_{23}  s_{34} & s_{12}  s_{34} & s_{12}  s_{23}\\[1mm] s_{12}  s_{13} & s_{23} & 
  s_{12} & s_{12}  s_{13}  s_{14} & s_{13} & s_{12}  s_{13}  s_{24} & s_{12}  s_{13}  s_{34} & 
  s_{12}  s_{23} & s_{14}  s_{23} & 1 & s_{23}  s_{24} & s_{23}  s_{34} & s_{12}  s_{14} & 
  s_{12}  s_{24} & s_{12}  s_{34} & s_{13}  s_{14} & s_{13}  s_{24} & 
  s_{13}  s_{34}\\[1mm] s_{12}  s_{14} & s_{24} & s_{12}  s_{13}  s_{34} & s_{12} & 
  s_{12}  s_{14}  s_{23} & s_{14} & s_{12}  s_{13}  s_{14} & s_{13}  s_{24} & s_{12}  s_{24} & 
  s_{23}  s_{34} & 1 & s_{23}  s_{24} & s_{12}  s_{34} & s_{13}  s_{34} & s_{12}  s_{13} & 
  s_{12}  s_{23} & s_{13}  s_{14} & s_{14}  s_{23}\\[1mm] s_{12}  s_{34} & s_{34} & 
  s_{12}  s_{13}  s_{14} & s_{12}  s_{13}  s_{34} & s_{12}  s_{23}  s_{24} & 
  s_{12}  s_{23}  s_{34} & s_{12} & s_{13}  s_{14} & s_{13}  s_{34} & s_{23}  s_{24} & 
  s_{23}  s_{34} & 1 & s_{12}  s_{13} & s_{14}  s_{23} & s_{12}  s_{14} & s_{13}  s_{24} & 
  s_{12}  s_{23} & s_{12}  s_{24}\\[1mm] s_{13}  s_{34} & s_{12}  s_{23}  s_{34} & s_{14} & s_{34} & 
  s_{12}  s_{13}  s_{24} & s_{12}  s_{13}  s_{34} & s_{13} & s_{14}  s_{23} & s_{23}  s_{34} & 
  s_{12}  s_{24} & s_{12}  s_{34} & s_{12}  s_{23} & 1 & s_{12}  s_{14} & s_{13}  s_{14} & 
  s_{23}  s_{24} & s_{12}  s_{13} & s_{13}  s_{24}\\[1mm] s_{13}  s_{24} & s_{12}  s_{14}  s_{23} &
   s_{24} & s_{12}  s_{23}  s_{24} & s_{12}  s_{13}  s_{34} & s_{13} & s_{12}  s_{13}  s_{24} & 
  s_{23}  s_{34} & s_{12}  s_{23} & s_{12}  s_{14} & s_{13}  s_{14} & s_{14}  s_{23} & 
  s_{12}  s_{24} & 1 & s_{23}  s_{24} & s_{12}  s_{34} & s_{13}  s_{34} & 
  s_{12}  s_{13}\\[1mm] s_{13}  s_{14} & s_{12}  s_{23}  s_{24} & s_{34} & s_{13} & 
  s_{12}  s_{13}  s_{14} & s_{12}  s_{14}  s_{23} & s_{14} & s_{23}  s_{24} & s_{13}  s_{24} & 
  s_{12}  s_{34} & s_{12}  s_{23} & s_{12}  s_{24} & s_{13}  s_{34} & s_{23}  s_{34} & 1 & 
  s_{12}  s_{13} & s_{14}  s_{23} & s_{12}  s_{14}\\[1mm] s_{14}  s_{23} & s_{12}  s_{13}  s_{24} &
   s_{12}  s_{23}  s_{34} & s_{23} & s_{14} & s_{12}  s_{13}  s_{14} & s_{12}  s_{14}  s_{23} & 
  s_{12}  s_{24} & s_{23}  s_{24} & s_{13}  s_{34} & s_{12}  s_{13} & s_{13}  s_{24} & 
  s_{23}  s_{34} & s_{12}  s_{34} & s_{12}  s_{23} & 1 & s_{12}  s_{14} & 
  s_{13}  s_{14}\\[1mm] s_{23}  s_{34} & s_{12}  s_{13}  s_{34} & s_{12}  s_{14}  s_{23} & 
  s_{12}  s_{23}  s_{34} & s_{24} & s_{34} & s_{23} & s_{12}  s_{14} & s_{12}  s_{34} & 
  s_{13}  s_{24} & s_{13}  s_{34} & s_{12}  s_{13} & s_{12}  s_{23} & s_{13}  s_{14} & 
  s_{14}  s_{23} & s_{12}  s_{24} & 1 & s_{23}  s_{24}\\[1mm] s_{23}  s_{24} & 
  s_{12}  s_{13}  s_{14} & s_{12}  s_{23}  s_{24} & s_{12}  s_{13}  s_{24} & s_{34} & s_{23} & 
  s_{24} & s_{12}  s_{34} & s_{12}  s_{13} & s_{13}  s_{14} & s_{14}  s_{23} & s_{12}  s_{14} & 
  s_{13}  s_{24} & s_{12}  s_{23} & s_{12}  s_{24} & s_{13}  s_{34} & s_{23}  s_{34} & 1
\end{array}
\right]$}
\]
We leave the construction of the corresponding SDP constraint to the interested reader.
\end{ex}

\section{Comparison with \texorpdfstring{\cite{jun2023hierarchy}}{[TRZ+]}}\label{app:trz}

An independent and simultaneously released work (\cite{jun2023hierarchy}) gives results that share some commonalities with those presented in this paper. These include construction of a new semidefinite programming hierarchy based on the swap matrices and exact solutions to \qmaxcut on specific graphs. Here we briefly compare and contrast these two papers. 

First we outline key differences in notation between the two papers.
Before defining a new semidefinite programming hierarchy, both papers introduce some additional formalism in order to focus on the algebraic structure of the swap matrices. However, the authors of~\cite{jun2023hierarchy} state their results primarily in terms of \df{operator programs}, while we work with \df{$*$-algebras} and \df{positive linear functionals (operator algebraic states)} defined on them. Moving back and forth between these two concepts is straightforward. In the language of operator programs, one considers maximizing (or minimizing) an expression of the form 
\begin{align}
    \ev{\theta(\{a_i\})}{\psi} \label{eq:operator_system_eig}
\end{align}
where the $\{a_i\}$ are non-commuting and self-adjoint operator variables, the $\psi$ are arbitrary vectors satisfying $\ketbra{\psi} = 1$, and the $\{a_i\}$ satisfy constraints 
\begin{align}
    \eta_j(\{a_i\}) = 0 \;\; \forall j
\end{align}
for some set of constraint polynomials $\{\eta_j\}$. In the language of $*$-algebras, one first defines a $*$-algebra $\mathcal{A}$ to be the algebra generated by self adjoint nc variables $\{a_i\}$ which satisfy relations 
\begin{align}
    \eta_j(\{a_i\}) = 0 \;\; \forall j 
\end{align}
then computes the maximum over positive linear functionals $L : \mathcal{A} \rightarrow \mathbb{C}$ with $L(1) = 1$ of the expression 
\begin{align}
    L(\theta(\{a_i\})). \label{eq:linear_functional_eig}
\end{align}
That these two formulations are equivalent 
follows 
from the GNS construction. 

Both this paper and~\cite{jun2023hierarchy} give a set of constraints $\{\eta_j\}$ which characterize the swap matrices. Formally, the papers prove that maximizing an expression of the form given in~\Cref{eq:operator_system_eig} (equivalently \Cref{eq:linear_functional_eig}) subject to these constrains is equivalent to computing the max eigenvalue of the matrix $\theta(\text{SWAP})$ obtained by replacing the $\{a_i\}$ variables in $\theta(\{a_i\})$ with explicit swap matrices. In~\cite{jun2023hierarchy} the operator program obtained when the $\{a_i\}$ variables (instead denoted $\{p_{ij}\}$) are subject to these constraints is denoted $\mathscr{Perm}(V,w)$. In this paper the $*$-algebra obtained when the $\{a_i\}$ variables (instead denoted $\{s_{ij}\}$) are subject to these constraints is call the \df{\sswap}.

Both papers also introduce hierarchies of semidefinite programs which give a converging series of upper bounds on \Cref{eq:operator_system_eig} (equivalently \Cref{eq:linear_functional_eig}). At each finite level both hierarchies construct a linear ``pseudoexpectation'', denoted here by $\tilde{\mathbb{E}}$, which assigns values 
(``pseudomoments'')
to all polynomials of degree $\leq 2d$ in the variables $\{a_i\}$. In both papers, these pseudoexpectations satisfy
\begin{align}
    \tilde{\mathbb{E}}(1) = 1 \qquad 
    \text{ and } \qquad
    \tilde{\mathbb{E}}(q^*q) \geq 0
\end{align}
for any nc polynomial $q$ of degree at most $d$.
However, there is an important difference between how strictly the pseudoexpectations constructed in the two papers enforce the constraints $\{\eta_j\}$. 
The pseudoexpectation for the $d$-th level hierarchy constructed in this paper -- which we call the $d$-th swap relaxation to \qmaxcut -- satisfies $\tilde{\mathbb{E}}[p] = 0$ for any degree $\leq 2d$ polynomial in the two sided ideal generated by $\{\eta_j\}$, so,
\begin{align}
\tilde{\mathbb{E}}[p] = 0 &\text{ if } \deg(p) \leq 2d \text{ and } \exists\; \text{monomials} \ \{\beta_{ij}\}, \{\gamma_{ij}\} \text{ s.t. } p = \sum_{i,j} \beta_{ij} \eta_j \gamma_{ij}. \label{eq:ideal_constraints}
\end{align}
Contrastingly, the pseudoexpectation for the $d$-th level hierarchy constructed in~\cite{jun2023hierarchy} only enforces the constraints in the \df{truncated ideal} generated by $\{ \eta_j\}$, meaning 
\begin{align}
\tilde{\mathbb{E}}[p] = 0 \text{ if } \deg(p) \leq 2d \text{ and } \exists\; \text{monomials}\ \{\beta_{ij}\}, \{\gamma_{ij}\}  \text{ s.t. } p = \sum_{ij} \beta_{ij} \eta_j \gamma_{ij} \;\;\;\; \nonumber \\
\textbf{with } \boldsymbol{\deg(\beta_{ij} \eta_j \gamma_{ij}) \leq 2d} \textbf{ for all } \boldsymbol{i, j}.
\label{eq:NPA_constraints}
\end{align}

In the case of swaps the polynomials $p$ defined in \Cref{eq:ideal_constraints} above could also be defined to be those $p$ of degree $\leq 2d$ which annihilate all of the swap matrices, namely the set of polynomials $p$ in the variables $s_{ij}$ such that
\begin{align}
\label{eq:Svariety}
p(\{\sw_{ij}\}) = 0 
\end{align}
where $\sw_{ij}$ is the matrix obtained by replacing each $s_{ij}$ variable with the corresponding Swap matrix. This follows from \Cref{prop:swap} in our paper or \cite[Theorem 3.8]{jun2023hierarchy}. 
The ideal corresponding to the swaps
can be defined by different sets of constraint
polynomials. 
 \Cref{eq:Svariety} 
implies 
the value obtained by our $d$th relaxation depends only on the ideal and does not depend on the choice of defining constraints 
$\{\eta_j\}$. 
 On the other hand, 
 this property may not be shared by 
  $d$-th relaxed value of~\cite{jun2023hierarchy}.

Further discussion of this difference is given in \Cref{rem:Id} in this paper, where the bound obtained at level $d$ of our hierarchy is denoted $\nu_d(h)$ and the bound obtained by the hierarchy in~\cite{jun2023hierarchy} is denoted $\bunderline{\nu}_d$. 
While both hierarchies converge after finitely many steps, this difference means that at each level the hierarchy presented in this paper gives an upper bound that may be tighter than the one presented in~\cite{jun2023hierarchy}.

We found  numerically (up to SDP accuracy in Mathematica, $10^{-7}$)
for graphs with up to 8 vertices and uniform edge weights 
the value of our second swap relaxation equals
the \exactn max/min eigenvalue of the \qmc Hamiltonian.
In contrast, \cite{jun2023hierarchy} do an impressive numerical study on a sampling of these same graphs and find 
the performance of their first relaxation in swaps 
is not \exactn (to the same tolerance) 
on numerous graphs with 6, 7 and 8 vertices, for details see \cite[Figure 4 of Section 5]{jun2023hierarchy}.

As mentioned in \cref{ssec:overview_techniques}, computing all constraints of the form given in \Cref{eq:ideal_constraints} at higher levels $d$ requires considerable technical work. 
This is the content of the later part of \Cref{sec:swap_ncSoS} of our paper (which gives an explicit linear algebraic basis for the degree $d$ monomials in the \sswap when $d \leq 4$) and of \Cref{sec:GB} (which applies Gr\"obner Basis to swap polynomials). These techniques for simplifying swap polynomials are not present in~\cite{jun2023hierarchy}. One consequence of our swap algebra theory,
see 
\Cref{eq:lassconverges},
is that the level
$\lceil \frac n 2 \rceil $ 
swap relaxation in our hierarchy solves the quantum max cut problem exactly.
The corresponding result for the hierarchy present in~\cite{jun2023hierarchy} is stated as an open question after 
Proposition 3.13 in \cite{jun2023hierarchy},
which gives the bound $\binom{n}{2}$.

The present paper and \cite{jun2023hierarchy} both theoretically analyze specific instances where the highest eigenvalue of \qmaxcut can be computed exactly, but with diverging aims. 
We provide a method to compute (in exact arithmetic) the maximum eigenvalue of the \qmc Hamiltonian on a certain family of graphs with uniform edge-weights, extending previously known results on exact solutions to \qmaxcut \cite{lieb1962ordering}. This result builds on the characterization of the \qmc Hamiltonian in terms of irreducible representations of the symmetric group via Schur-Weyl duality, and is largely independent of our construction of the swap relaxations. 

\cite{jun2023hierarchy}, on the other hand, is concerned with understanding instances where the SDP relaxations exactly solve the \qmaxcut problem as this can improve the analysis of approximation algorithms. Specifically, \cite{jun2023hierarchy} investigates whether either the first swap relaxation or the second quantum Lasserre relaxation over Paulis are exact or not on a number of graphs. 
Interestingly, there are cases where the SDP hierarchies considered by \cite{jun2023hierarchy} are proven to be inexact (see Section 4 of \cite{jun2023hierarchy}) but for which exact arithmetic solutions are known. An example of this is the \qmc Hamiltonian with uniform edge-weights on a clique with an odd number of vertices, which have exact solutions via representation theory (see, for example, \Cref{ssec:exact_arithmetic_solns_for_cliques} of this paper).
One special class of graphs solved by our methods are 
complete $k$-partite graphs
for any constant $k$, 
special cases of which include 
the clique and the crown graph.
These latter graphs are 
shown 
to be exactly computable by the SDPs in \cite{jun2023hierarchy}. In other cases, such as a uniformly weighted double-star graph and a star graph with positive weights, the SDP relaxations are shown to be \exactn but our representation theoretic method is unable to provide an exact solution. 

\end{document}

%% file: packages.tex
\usepackage{amsmath}
\usepackage{amssymb}
\usepackage{amsthm}
\usepackage{amsfonts}
\usepackage[pagebackref]{hyperref}
\hypersetup{
    colorlinks=true, 
    linkcolor=black, 
    citecolor=[rgb]{0,0,0.5}, 
    urlcolor=[rgb]{0,0,.54}
}
\usepackage{microtype}

\usepackage{accents}
\usepackage[ruled,titlenumbered]{algorithm2e}
\usepackage[toc,page]{appendix}
\usepackage{array}
\usepackage[english]{babel}
\usepackage{braket}
\usepackage{booktabs}
\usepackage{bigdelim}
\usepackage{cancel}
\usepackage[shortlabels]{enumitem}
\usepackage{float}
\usepackage[bottom]{footmisc}
\usepackage[utf8]{inputenc}
\usepackage[margin=1in]{geometry}
\usepackage{graphicx}
\usepackage[capitalize,nameinlink]{cleveref} 
\usepackage{mathtools}
\usepackage{multirow}
\usepackage{physics}
\usepackage{tabularx}
\usepackage[dvipsnames]{xcolor}
\usepackage{tikz}
\usepackage[textsize=footnotesize]{todonotes}
\usepackage{qcircuit}
\usetikzlibrary{positioning}
\usetikzlibrary{decorations.pathreplacing}
\usetikzlibrary{cd}
\usetikzlibrary{arrows.meta}
\usetikzlibrary{calc}
\usepackage{bbold}

\usepackage{tikzsymbols}
\usepackage{complexity}

\usepackage{xspace}

\renewcommand{\backref}[1]{}

\renewcommand{\backrefalt}[4]{%
\ifcase #1 %
\or
[p.\ #2]%
\else
[pp.\ #2]%
\fi}

\makeatletter
\renewcommand{\paragraph}{%
 \@startsection{paragraph}{4}%
 {\z@}{2.25ex \@plus .5ex \@minus .3ex}{-1em}%
 {\normalfont\normalsize\bfseries}%
}
\makeatother

\makeatletter
\newcommand{\para}{%
 \@startsection{paragraph}{4}%
 {\z@}{2ex \@plus 3.3ex \@minus .2ex}{-1em}%
 {\normalfont\normalsize\bfseries}%
}
\makeatother

%% file: adam.tex
\newtheorem{lem}{Lemma}[section]
\newtheorem*{lem*}{Lemma}
\newtheorem{thm}[lem]{Theorem}
\newtheorem*{thm*}{Theorem}
\newtheorem{prop}[lem]{Proposition}

\newtheorem{cor}[lem]{Corollary}
\newtheorem{defn}[lem]{Definition}

\theoremstyle{definition}

\newtheorem{rmk}[lem]{Remark}
\newtheorem{ex}[lem]{Example}
\crefname{thm}{Theorem}{Theorems}
\crefname{lem}{Lemma}{Lemmas}
\crefname{defn}{Definition}{Definitions}

\makeatletter
\newtheorem*{rep@theorem}{\rep@title}
\newcommand{\newreptheorem}[2]{%
\newenvironment{rep#1}[1]{%
 \def\rep@title{#2 \ref{##1}}%
 \begin{rep@theorem}}%
 {\end{rep@theorem}}}
\makeatother

\newreptheorem{thm}{Theorem}
\newreptheorem{lem}{Lemma}

\def\beq{\begin{equation}}
\def\eeq{\end{equation}}

\def\ben{\begin{enumerate}}
\def\een{\end{enumerate}}

\def\bem{\begin{bmatrix}}
\def\eem{\end{bmatrix}}

\newcommand{\df}[1]{{\emph{#1}}{\index{#1}}}

\def\cI{{\mathcal I}}
\def\cM{{\mathcal M}}
\def\cA{{\mathcal A}}
\def\cH{{\mathcal H}}

 \numberwithin{equation}{section}